\documentclass[11pt]{article}
\pdfoutput=1
\usepackage{jheppub}
\usepackage{tabu}
\usepackage[vcentermath]{youngtab}
\usepackage[usenames,dvipsnames,table]{xcolor}
\usepackage{graphicx,amsmath,amssymb,amsthm,multirow,array,bm,bbm,esint}
\usepackage[mathscr]{eucal}
\usepackage[bbgreekl]{mathbbol}
\usepackage{epsf,amsfonts}
\usepackage{slashed}
\usepackage[numbers,sort&compress]{natbib}
\usepackage{minibox}
\usepackage{rotating}
\usepackage{pdflscape}
\usepackage{array,tikz-cd}

\usepackage{slashed}

\newcommand{\Ttwo}{{\cal T}_{_{(2)}}\,}

 \def\shadeB{\cellcolor{blue!5}}

\definecolor{rust}{rgb}{0.8,0.2,0.2}

\newcommand{\prn}[1]{\left ( #1 \right )}
\newcommand{\brk}[1]{\left [ #1 \right ]}
\newcommand{\bigbr}[1]{\left\{ #1 \right\} }

\newcommand{\half}{\frac{1}{2}}

\newcommand{\Tr}[1]{\hbox{Tr}\left(#1\right)}

\newcommand{\vev}[1]{\langle #1 \rangle}
\def\ket#1{\mid  \! #1  \rangle}
\def\bra#1{\langle  #1 \! \mid}

\newcommand{\rhoi}{\hat{\rho}_{\text{initial}}}
\newcommand{\rhoT}{\hat{\rho}_{_T}}
\newcommand{\Ki}{\hat{K}_\text{initial}}

\newcommand{\diffBi}{\delta_{\Kbeta_{_0}}}

\newcommand{\QSK}{\mathcal{Q}_{_{SK}}}
\newcommand{\QSKb}{\overline{\mathcal{Q}}_{_{SK}}}
\newcommand{\QKMS}{\mathcal{Q}_{_{KMS}}}
\newcommand{\QKMSb}{\overline{\mathcal{Q}}_{_{KMS}}}
\newcommand{\CPT}{{}^{\text{CPT}}}

\newcommand{\Qzero}{\mathcal{Q}^0_{_{KMS}}}
\newcommand{\Qbeta}{\mathscr{L}_{_{KMS}}}
\newcommand{\al}[2]{\alpha_{#1}^{_{(#2)}}}
\newcommand{\IKMS}{\SF{\cal I}^{\text{\tiny{KMS}}}}
\newcommand{\LKMS}{\SF{\cal L}^{\text{\tiny{KMS}}}}
\newcommand{\IKMSb}{\SF{\overline{\cal I}}{}^{^{\text{\tiny{KMS}}}}}
\newcommand{\IKMSzero}{\SF{\cal I}^{\text{\tiny{KMS}}}_0\,}

\newcommand{\Op}[1]{\mathbb{#1}}
\newcommand{\OpH}[1]{\widehat{\mathbb{#1}}}

\newcommand{\SKR}[1]{\mathbb{#1}_{\skR}}
\newcommand{\SKL}[1]{\mathbb{#1}_{\skL}}

\newcommand{\SKAv}[1]{\mathbb{#1}_{{av}}}
\newcommand{\SKRel}[1]{\mathbb{#1}_{{dif}}}
\newcommand{\SKDif}[1]{\mathbb{#1}_{{dif}}}

\newcommand{\SKAdv}[1]{\mathbb{#1}_{adv}}
\newcommand{\SKRet}[1]{\mathbb{#1}_{ret}}

\newcommand{\SKG}[1]{\mathbb{#1}_{_G}}
\newcommand{\SKGb}[1]{\mathbb{#1}_{_{\overline{G}}}}

\newcommand{\SKS}[1]{\mathring{\mathbb{#1}}}

\newcommand{\FSgn}[1]{(-1)^{F_{\mathbb{#1}}}}
\newcommand{\GSgn}[1]{(-1)^{G_\mathbb{#1}}}
\newcommand{\gh}[1]{\text{gh}(#1)}

\newcommand{\gradcomm}[2]{ \brk{ #1, #2 }_{\scriptscriptstyle \pm} }
\newcommand{\comm}[2]{ \brk{ #1, #2 }}
\newcommand{\gradAnti}[2]{ \bigbr{ #1, #2 }_{\scriptscriptstyle \pm} }
\newcommand{\KeldBrk}[2]{ ( #1\ , #2 )_{_{SK}} }
\newcommand{\stepFn}[1]{\Theta_{{#1}} }

\newcommand{\avA}{ {\sf a} }
\newcommand{\ghA}{ {\sf g} }
\newcommand{\ghbA}{ {\bar {\sf g}} }
\newcommand{\difA}{ {\sf d} }

\newcommand{\thb}{{\bar{\theta}} }
\newcommand{\thetab}{\bar{\theta} }
\newcommand{\SF}[1]{\mathring{#1}}

\newcommand{\deltaB}{\delta_{_ {\bm\beta}}}
\newcommand{\fbeta}{\mathfrak{f}_{_ {\bm\beta}}}
\newcommand{\fbetaB}[1]{\mathfrak{f}^{\scriptscriptstyle{B}}_{#1}}
\newcommand{\fbetaF}[1]{\mathfrak{f}^{\scriptscriptstyle{F}}_{#1}}
\newcommand{\delKMS}{{\Delta}_{_ {\bm\beta}} }

\newcommand{\Kref}{{\bm \beta}}

\makeatletter
  \newcommand\Ttiny{\@setfontsize\Ttiny{1pt}{2}}
\makeatother

\newcommand{\Cref}{{\sf C}}

\newcommand{\form}[1]{\bm{#1}}

\newcommand{\lieD}{\pounds}

\newcommand{\Kbeta}{{\bm{\beta}}}
\newcommand{\LambdaB}{\Lambda_{\bm{\beta}}}

\newcommand{\KEq}{K}

\newcommand{\skR}{\text{\tiny R}}
\newcommand{\skL}{\text{\tiny L}}

\newcommand{\smallT}{{\sf \!{\scriptscriptstyle{T}}}}

\newcommand{\UT}{U(1)_{\scriptstyle{\sf T}}}

\title{Schwinger-Keldysh formalism I:\\
 BRST symmetries and superspace}

\author[a]{Felix M. Haehl}
\author[b]{\!, R.\ Loganayagam}
\author[c]{\!, Mukund Rangamani}
\affiliation[\,a]{Department of Physics and Astronomy, University of British Columbia,\\
6224 Agricultural Road, Vancouver, B.C.\ V6T 1Z1, Canada.}
\affiliation[\,b]{International Centre for Theoretical Sciences (ICTS-TIFR), \\
Shivakote, Hesaraghatta Hobli, Bengaluru 560089, India.}
\affiliation[\,c]{
Center for Quantum Mathematics and Physics (QMAP)  \\
Department of Physics, University of California, Davis, CA 95616 USA.}
\emailAdd{f.m.haehl@gmail.com}
\emailAdd{nayagam@gmail.com}
\emailAdd{mukund@physics.ucdavis.edu}

\abstract{
We review the Schwinger-Keldysh, or in-in, formalism for studying quantum dynamics of  systems out-of-equilibrium.  The main motivation is to rephrase well known facts in the subject in a mathematically elegant setting, by exhibiting a set of BRST symmetries inherent in the construction. We show how these fundamental symmetries can be made manifest by working in a superspace formalism. We argue that this rephrasing is extremely efficacious in understanding low energy dynamics following the usual renormalization group approach, for the BRST symmetries are robust under integrating out degrees of freedom.  In addition we discuss potential generalizations of the formalism that allow us to compute out-of-time-order correlation functions that have been the focus of recent attention in the context of chaos and scrambling. We also outline a set of problems ranging from stochastic dynamics, hydrodynamics,  dynamics  of entanglement in QFTs, and the physics of black holes and cosmology,  where we believe this framework could play a crucial role in demystifying various confusions.
Our companion paper \cite{Haehl:2016uah} describes in greater detail the mathematical framework embodying the topological symmetries we uncover here.
}

\begin{document}
\maketitle


\newpage

\section{Introduction}
\label{sec:intro}

The study of quantum dynamics out of equilibrium and in open systems necessarily involves working with mixed states. We appreciate that a  crucial aspect of the quantum evolution entails understanding the dynamical evolution, all the while keeping track of entanglement between the system and the external environment. This is best done by writing down a suitable density matrix for the system and studying its Hamiltonian evolution in the presence of the external stimulus. Such a density matrix evolves via the usual Heisenberg evolution so long as the entanglement between the system and environment is unchanging; in the generic  situation where entanglement may be modified it would undergo a form of generalized Linblad type evolution.

While one may a-priori be concerned that the information about the environment is hard to encode in the process, the monogamy of  quantum entanglement comes to the rescue. We only need to consider a part of the environment that is at least as large as our system; this suffices to encode the evolution of the system keeping track of the entanglement.  As long as the dynamics does not modify the entanglement between the system and the environment, i.e.,  entanglement is treated as a scare resource, this gives a complete characterization of the system's evolution. In effect, all one needs to do is to double the degrees of freedom, using a second copy of our system as the proxy for the environment.

This idea of doubling the degrees of freedom to describe mixed states of a quantum system has been well understood for over five decades since the seminal works of Schwinger \cite{Schwinger:1960qe}, Keldysh \cite{Keldysh:1964ud}, and Feynman and Vernon \cite{Feynman:1963fq}. This is a well studied subject with some very good reviews written over the years \cite{Chou:1984es,Landsman:1986uw,Maciejko, Kamenev:2009jj}. Despite these we will argue that a reformulation of the standard construction is necessitated, should one wish  to focus, not on microscopic degrees of freedom, but rather investigate properties of the low energy theory obtained by integrating out irrelevant modes.

The primary aim of these notes is to provide a novel perspective on the Schwinger-Keldysh formalism, emphasizing the symmetries that are inherent in the construction, elaborating on our earlier discussion in \cite{Haehl:2015foa}.  Essentially we wish to argue that the standard presentation of the Schwinger-Keldysh construction is a gauge-fixed formulation, which, whilst useful for many questions, is imbued with some inherent limitations. As we are familiar with in other areas of physics, a completely covariant construction with fully manifest symmetries allows one to address question of greater generality, for it enables separation of gauge artifacts from more fundamental issues. We will argue that the doubling of Schwinger and Keldysh should be better understood in terms of a topological structure, with a set of BRST charges that can be used to control the structure of the theory effectively.

Let us understand the rationale behind the Schwinger-Keldysh formalism, and its attendant complications when we try to describe effective field theories. We will sketch the physical picture here for open quantum systems where it is easy to keep track of the system we are interested in and the environment that it interacts with. Much of what we say can also be applied directly to closed quantum systems in mixed states which are also interesting in their own right.

Suppose we are interested in analyzing the dynamical evolution of an open quantum system, which we denote as $\mathbb{Q}$. If we have access to the microscopic description of both this system and the environment it interacts with, we have no real issue for we simply write down a complete closed quantum system by considering the detailed coupling between the two. To wit, denoting the environment  by $\mathbb{E}$, we would write down a standard path integral for $\mathbb{Q} \cup \mathbb{E}$, which would schematically look like
\begin{align}
\mathscr{Z}_{\mathbb{Q}\cup\mathbb{E}} = \int [{\cal D} \Phi_\mathbb{Q}] \, [{\cal D} \Phi_\mathbb{E}] \; \exp\left(\frac{i}{\hbar} \, \bigg[
S_\mathbb{Q}  + S_\mathbb{E} + S_\text{ent}(\mathbb{Q}\leftrightarrow\mathbb{E}) \bigg]\right)\,.
\end{align}
We can, of course, recognize such structures in various physical systems, which have been studied all across physics. Some familiar examples  of system-environment pairs to help orient the reader are presented in Table ~\ref{tab:examples}.
As is clear from our examples we are happy to view any form of coupling between the system and its environment, be it actual quantum entanglement in the initial state wavefunction, or an explicit  coupling which affects the dynamics, as a form of interaction $S_\text{ent}$.
\begin{table}[h]
\centering
\begin{tabular}{|| c | c | c ||c|| }
\hline\hline
\multicolumn{3}{||c||}{\shadeB{Open quantum systems and their environments}}\\
\hline
 System $\mathbb{Q}$ & Environment $\mathbb{E}$& Entanglement/Interaction \\
 \hline
Single harmonic oscillator &  Quantum oscillator bath & Harmonic couplings   \cite{Caldeira:1982iu} \\
Subsystem ${\cal H}_\mathcal{A}$ & Purifying complement ${\cal H}_{\mathcal{A}^c}$& Entanglement  structure \\
QFT modes with $\omega \leq \Lambda_{UV}$ & High energy modes & UV/IR interactions  \\
Open strings on D-branes &  Ambient closed string theory & Open-closed interactions  \\
Quantum mechanical system & Measurement apparatus & Projections \\
\hline\hline
\end{tabular}
\label{tab:examples}
\caption{A few examples of familiar open quantum systems which we encounter frequently. These are intended to orient the reader to the issues that we wish to emphasize in the bulk of our discussion.}
\end{table}

The essential thesis of working with an open system is however that we are agnostic of the environment's dynamics, and would prefer to integrate it out, so as to focus on the degrees of freedom of $\mathbb{Q}$ alone.  This is a useful thing to do when there exists a whole class of states and observeables where  environment plays a universal role, so that within that class one expects an effective autonomous description. We can imagine carrying out the path integral over the environment variables in order to get to this description. However, due to the interactions present in $S_\text{ent}(\mathbb{Q}\leftrightarrow\mathbb{E})$ we recognize immediately that the state of the system $\mathbb{Q}$ is necessarily mixed. While we oftentimes are interested in situations where this mixing is relatively weak and can be ignored, we emphasize that generically this is far from the case.  For instance in the standard renormalization group picture, one demonstrates that the irrelevant high energy modes are naturally suppressed when one considers the low energy degrees of freedom -- a statement that is sometimes referred to as {\em color transparency}. This however relies on the underlying quantum dynamics being suitably conventional, and is known to fail in systems where there is non-trivial mixing between UV and IR modes, e.g., non-commutative field theories \cite{Minwalla:1999px}, hydrodynamics \cite{Haehl:2015foa}, and gravity.

A natural consequence of integrating out the environment variables is that their role in setting up the underlying entanglement pattern has to be recorded somewhere at the end of the process. This is accomplished by two distinct elements in the path integral: \cite{Schwinger:1960qe,Feynman:1963fq,Keldysh:1964ud} realized that one first has to double the system variables $\Phi_\mathbb{Q} \mapsto \{ \Phi_\mathbb{Q}^\skL, \Phi_\mathbb{Q}^\skR\}$. This may be best understood by noting that a density matrix of the system $\mathbb{Q}$ is an operator on the Hilbert space and hence requires both a space of states (kets or the right Hilbert space) ${\cal H}_\mathbb{Q}^\skR$ and a space of conjugate states (bras or the left Hilbert space) ${\cal H}_\mathbb{Q}^\skL$ for its definition.

Thus, to begin with, the discussion of mixed states of a QFT necessarily involves a doubling. One writes down in lieu of the single-copy effective action for our system, the \emph{Schwinger-Keldysh action}
\begin{equation}
S_{SK} =S_\mathbb{Q}^\skR  - S_\mathbb{Q}^\skL \,.
\label{eq:skfactor}
\end{equation}
The relative sign can be easily understood by recalling that the Hamiltonian evolution of states and their conjugates is accomplished by the unitary evolution operator and its conjugate respectively, leading to a relative sign in the action. This Schwinger-Keldysh action suffices in circumstances where the role of the environment is to set up the correct entanglement structure in the system $\mathbb{Q}$, which then remains invariant in the subsequent evolution. Strictly speaking, in our above example we should first set up the appropriate entanglement between $\mathbb{Q}$ and $\mathbb{E}$ and evolve  the joint system with factorized unitary $U_{\mathbb{Q} \cup\mathbb{E}} =U_{\mathbb{Q} } \; U_\mathbb{E}$ to ensure that the evolution doesn't change this initial entanglement pattern.

 Per se, the discussion could thus be simply applicable to impure states of a closed quantum system. We emphasize that in such cases, given an initial density matrix, the  left-right factorized form \eqref{eq:skfactor} is true in the microscopic path integral description. This is sufficient for a path integral evolution of the system given appropriate initial and final boundary conditions, and it forms the basic object in the Schwinger-Keldysh theory.

This doubling is however insufficient in accounting for the interactions engendered into the Schwinger-Keldysh path integral. This point was the focus of \cite{Feynman:1963fq}, who argued that the process of integrating out necessarily leads to new terms in addition to \eqref{eq:skfactor}, which they christened {\em influence functionals}.  The precise statement is that the process of passing from the microscopic variables to the macroscopic ones necessarily induces  some interaction between the two Schwinger-Keldysh copies of the system.  To wit, the generating functional for the system after tracing out the environment takes the generic form:
\begin{equation}
\mathscr{Z}_\mathbb{Q} =
	\int [{\cal D} \Phi_\mathbb{Q}^\skL] \, [{\cal D} \Phi_\mathbb{Q}^\skR]
 	\; \exp\left(\frac{i}{\hbar} \, \bigg[
	S_\mathbb{Q}^\skR  - S_\mathbb{Q}^\skL + S_\text{IF}(\Phi^\skR; \Phi^\skL)
	\bigg]\right)\,.
\label{}
\end{equation}
We emphasize that the  Feynman-Vernon influence functionals, contained in $S_\text{IF}$, present new conceptual issues. Many of these issues are related to the fact that they are absent in a UV description and arise in IR only after various irrelevant modes have been integrated out for the class of states one is interested in. They are thus related to various phenomena that are unique to IR physics such as entropy, dissipation, decoherence, long-range entanglement, state dependent observables etc.  One might ask what is the
general non-linear structure of the  influence functionals, the rules they should obey in any quantum system, their renormalization and running.
The answer to these questions is unknown at present. As we will see in the sequel, the structure of  Feynman-Vernon terms is closely related to various fundamental questions about non-equilibrium systems and to the theory of open quantum systems. The astute reader will also recognize that the questions are broadly valid in discussions of gravity, either in cosmology or in studies of black holes, owing to the fact that the causal structure of the semiclassical spacetime precludes full knowledge of the degrees of freedom. Thus while the basic formalism of ascertaining the Schwinger-Keldysh action is clear from a microscopic perspective, things are much more murky when we have to deal with the low energy description.

While in the above we motivated the issues of interest using open quantum systems, we also alluded to the fact that similar statements ought to  apply when we consider mixed states of a closed quantum system (or pure states which behave  effectively as mixed states for the relevant observables). For the latter we have in mind Gibbsian density matrices describing the thermal state of a system, which show up in the discussion of thermodynamics and hydrodynamics. A useful way to demarcate the two situations is to realize that mixed states of a closed system undergo canonical Heisenberg evolution. As a result the transformation $\rho(t) = U \rho(0) U^\dagger$, with $U = e^{-i\, H\, t}$, is a unitary evolution which does not change the von Neumann entropy $S(\rho(t)) = -\Tr{\rho(t) \, \log \rho(t) } = -\Tr{\rho(0) \log \rho(0)} =S(\rho(0))$.\footnote{ A density matrix evolves oppositely to a Heisenberg picture operator
${\cal O}(t) = U^\dagger\, {\cal O}(0)\, U$ by virtue of the fact that the ordering is different. One can  infer the transformations directly from $\rho = \sum_\alpha \, c_\alpha \, \ket{\psi_\alpha} \bra{\psi_\alpha}$ and the fact that states undergo evolution via $\ket{\psi(t)} = e^{-i\,H\,t} \, \ket{\psi(0)}$.} We will refer to this as the invariance of the fine grained entropy of the system. On the contrary for open systems, the fine grained entropy may change owing to the interaction between $\mathbb{Q}$ and $\mathbb{E}$. It is conceivable that one needs to impose some restrictions on the entanglement patterns thus generated if we want to employ the Schwinger-Keldysh formalism.  We do not address here the full range of possibilities for open systems, pausing just to note that the discussion above applies, at the very least in circumstances where the environment can be treated classically. In such situations the interactions in $S_\text{ent}(\mathbb{Q} \leftrightarrow \mathbb{E})$ can be treated as classical sources for operators in $\mathbb{Q}$ with a probability distribution dictated by the semi-classical approximation to $S_\mathbb{E}$.

Our  motivation for getting intrigued by the problem of constructing effective Schwinger-Keldysh theories was primarily to understand the general structure of such effective actions in the fluid dynamical regime and  beyond \cite{Haehl:2015pja,Haehl:2015foa}.\footnote{ There are other groups that have thought about this issue, see for example  \cite{Endlich:2012vt,Grozdanov:2013dba,Kovtun:2014hpa,Harder:2015nxa}. Closer in spirit to our considerations is the recent work of \cite{Crossley:2015evo}, who take inspiration from the Schwinger-Keldysh formalism. In non-thermal states they argue for a single BRST supercharge to encompass the constraints of microscopic unitary. We will argue that there is a more natural structure involving two supercharges, which are CPT conjugates of each other.\label{fn:cgl1}}
  These are qualitatively similar to the classic problem of the Brownian oscillator which motivated \cite{Schwinger:1960qe}, or linear dissipative systems which inspired \cite{Feynman:1963fq}, albeit with a necessary upgrade to non-Gaussian interactions. What the above discussion emphasizes is that one needs to gain control over the unfactorized part contained in the influence functionals. We first encountered non-trivial influence functionals in the process of constructing an effective action for anomaly induced transport in hydrodynamics \cite{Haehl:2013hoa}.\footnote{ Initial attempts to understand anomalous transport from an effective action were made in \cite{Dubovsky:2011sk} who were successful in obtaining a single copy effective action for abelian flavour anomalies in two dimensions. Higher dimensions and non-abelian flavour symmetries necessarily involve doubling and influence functionals. In \cite{Kovtun:2014hpa} a hydrodynamic effective action was derived by exponentiating the classical equations of motion along the lines of \cite{Martin:1973zz}. As discussed in \cite{Haehl:2015pja} while this gives the general structure, it fails in general to account for all the constraints arising from microscopic unitarity. }  While quantum anomalies are sufficiently constraining and robust, and thus the influence functionals necessary to reproduce their effects are under sufficient control, our early construction did not provide a hint of why these terms were necessary.

In an attempt to understand influence functionals in hydrodynamics, we undertook a detailed analysis of hydrodynamic transport, which culminated in an eightfold classification of constitutive relations compatible with the phenomenological axioms of hydrodynamics \cite{Haehl:2014zda,Haehl:2015pja}. The major surprise was that most hydrodynamic transport is adiabatic, which by virtue of entropy non-production ought to admit a simple Lagrangian description (as for any conservative system). Here we encountered a second puzzle: an attempt to eschew the lessons of Schwinger-Keldysh effective actions only gives rise to actions encompassing two of the seven adiabatic classes (the eighth class is the dissipative transport). Taking seriously the lessons learnt from our explorations of anomalous transport, we were able to write down a Schwinger-Keldysh effective action which reproduced our eightfold classification.

We noticed however that the construction necessitated influence functions generically. By thinking about the microscopic structures in the Schwinger-Keldysh construction, we argued that such terms can be controlled, should one posit the existence of an emergent abelian gauge symmetry, which we called as $\UT$ KMS gauge symmetry. The underlying gauge invariance allowed us to forbid precisely those terms that were in tension with the hydrodynamic axiom that requires entropy to be produced (and not destroyed). Understanding the emergence of this gauge symmetry led us to revisiting the essentials of the Schwinger-Keldysh formalism, which as we outlined in \cite{Haehl:2015foa} are best viewed by extracting the topological invariances inherent in the construction. This suffices to reproduce the effective actions of \cite{Haehl:2014zda,Haehl:2015pja} for fluid dynamics as we recently explained in \cite{Haehl:2015uoc}.\footnote{ The paper \cite{Crossley:2015evo} which appeared around the same time as ours, also constructs  effective actions for dissipative hydrodynamics. As indicated in footnote \ref{fn:cgl1}  they posit a single supercharge as arising from the Schwinger-Keldysh construction, and argue for an emergent supercharge which enforces the KMS condition. They implement the latter as a discrete ${\mathbb Z}_2$ transformation. Despite these seeming differences, it turns out that  there is a close connection between the superalgebras that constrain the low energy dynamics in the two formalisms; we will explain this elsewhere. }

Our goal here is to elaborate on the statements made in \cite{Haehl:2015foa} and provide a perspective on the Schwinger-Keldysh construction that transcends the application we initially intended for it. We will therefore review the standard formalism from the viewpoint of computing out-of-equilibrium real time correlation functions -- we give a brief and heuristic overview in \S\ref{sec:skreview} and a detailed review of the important technical features in \S\ref{sec:skbasics}. In \S\ref{sec:skthemal} we review in standard language the KMS condition and its consequences for the special case of thermal dynamics. Some simple examples of Schwinger-Keldysh correlation functions with vacuum and thermal initial conditions are given in \S\ref{sec:examples}. This will conclude the review part of this paper. We then proceed in \S\ref{sec:skghosts} to explain abstractly the symmetries which are present in the doubled theory. This will lead to a reformulation of the standard formalism in terms of BRST symmetries, which can be neatly encapsulated in a superspace language. The extension of the formalism to include further BRST symmetries due to a KMS condition will be given in \S\ref{sec:KMScharges}. After a technical interlude on discrete symmetries in \S\ref{sec:cpt}, we proceed to a superspace analysis of Schwinger-Keldysh correlation functions in \S\ref{sec:superrules}. We will demonstrate how the necessity of soaking up ghost zero modes leads to a prescription for determining all ghost correlators, and we analyze the ambiguities in doing so. We will finally outline in \S\ref{sec:applications} a series of questions (and further generalizations) where we hope this viewpoint will be of use in demystifying various puzzles.

\section{A lighting review of standard Schwinger-Keldysh formalism}
\label{sec:skreview}

We review some details of the Schwinger-Keldysh (SK) technique in general, thus setting the stage for our reformulation to follow in the sequel. The idea here is to motivate an alternate and more detailed rationale for considering the doubled system. We examine real-time evolution in a relativistic QFT and remind the reader of  salient facts in the computation of real-time-ordered Green's functions. This discussion complements  the conceptual reasons given in the Introduction \S\ref{sec:intro}, which was primarily concerned with open quantum systems.  The discussion below largely follows the presentation in \cite{Chou:1984es, Maciejko, Kamenev:2009jj}.

Let us consider computing in a QFT the two-point Green's function for some (generically complex) Heisenberg operator ${\OpH O}(x)$ in some pure state\footnote{ We are assuming without loss of generality that we can purify mixed states by reintroducing the environment variables. }
\begin{align}
 G(x,x') = -i \langle{\Omega} | \, {\cal T}  \brk{ {\OpH O}(x) \, {\OpH O}^\dagger(x') } | \Omega \rangle \,,
\end{align}
where ${\cal T}$ will henceforth denote the standard time ordering. We use $x$ to denote the spacetime coordinates and will differentiate the temporal and spatial coordinates as $x = (t,{\bf x})$ when necessary.
For definiteness we will take $|\Omega\rangle$ to be  the ground state of the full interacting theory. Usually one chooses to work perturbatively by separating the interaction part from the full Hamiltonian, ${\OpH H} = {\OpH H}_0 + {\OpH H}_{int}$; one thence switches to the interaction picture, where the evolution operator
\begin{equation}
U(t_0,t) = {\cal T} \, \exp \left( -i \int_{t_0}^t dt' \, {\OpH H}_{int}(t')\right) ,
\label{}
\end{equation}
 defines temporal evolution of the interaction picture states. Using this expansion one then finds an expression for the two-point Green's function:
\begin{align}
 G(x,x') = -i \langle 0 | S^\dagger\,  {\cal T}\brk{{\OpH O}(x)\, {\OpH O}^\dagger(x') } \, S | 0 \rangle
         = -i \frac{\langle 0| {\cal T}\brk{ S \; {\OpH O}(x) \, {\OpH O}^\dagger(x')} |0\rangle}{\langle 0|S  |0\rangle} \,,
\label{eq:Gcalc}
\end{align}
where we introduced the $S$-matrix  $S \equiv U(-\infty, \infty)$ and the initial non-interacting ground state  of the Hamiltonian $\OpH{H}_0$, denoted $|0\rangle$. The r.h.s.\ of \eqref{eq:Gcalc}  is the starting point for the standard perturbation theory.

In writing the second equality we expressed the instantaneous late  time ground state in terms of the early time, assuming an adiabatic evolution of the system expressed as a property of the $S$-matrix. Namely, the phase picked up by acting on the final state with $S^\dagger $ is the same as the one accumulated during the evolution, i.e.,
$\langle 0 | S^\dagger = \langle 0 | e^{i\alpha}$ while $ \langle 0 | S |0 \rangle=e^{i\alpha}$, for some phase $\alpha$. One thus is assuming that the physical content of the ground state remains unchanged during the evolution, up to a phase rotation. This fails in non-equilibrium situations, where adiabatic evolution is not justified.

\begin{figure}[t!]
\begin{center}
\begin{tikzpicture}
\draw[thin,color=black,->] (-4.5,0) -- (4.5,0);
\draw[very thick,color=rust,dotted,->] (-5,1) -- (-4,1) node [above] {$\color{rust}\mathcal{C}$} -- (-3.5,1);
\draw[very thick,color=rust,->] (-3.5,1) -- (3.5,1);
\draw[very thick,color=rust,dotted,->] (3.75,1) node [above left] {${\color{black}\Op{O}_\skR}$} -- (5,1);
\draw[very thick,color=rust,dotted,<-] (-5,-1) -- (-3.5,-1);
\draw[very thick,color=rust,<-] (-3.5,-1) -- (3.5,-1);
\draw[very thick,color=rust,dotted,<-] (3.6,-1) node [below left] {${\color{black}\Op{O}_\skL}$} -- (5,-1);
\draw[thick,dotted, color=blue,<->] (3.5,0.05) -- (3.5,0.5) node [right] {$+i\varepsilon$} -- (3.5,0.92);
\draw[thick, dotted,color=blue,<->] (3.5,-0.05) -- (3.5,-0.5) node [right] {$-i\varepsilon$} -- (3.5,-0.92);
\draw[thick,color=blue,fill=blue] (3.5,0) circle (0.3ex);
\draw[thick,color=black,fill=black] (3.5,1) circle (0.3ex);
\draw[thick,color=black,fill=black] (3.5,-1) circle (0.3ex);
\end{tikzpicture}
\caption{Illustration of the generic Schwinger-Keldysh complex time contour. Every operator $\OpH{O}$ in the original theory has two representations in the Schwinger-Keldysh path integral, viz., $\Op{O}_\skR$ and $\Op{O}_\skL$, which can be thought of as the distinction as to what part of the contour the operator is inserted on. Right operators are time-ordered, while left operators are anti-time ordered.}
\label{fig:contour1}
\end{center}
\end{figure}
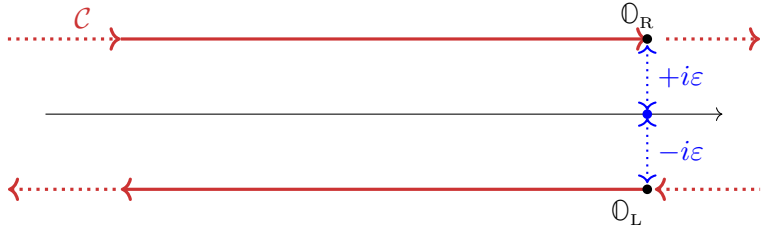

The Schwinger-Keldysh formalism deals with non-equilibrium dynamics by only ever making reference to the initial state,\footnote{ This explains why Schwinger-Keldysh formalism is sometimes also referred to as {\it in-in formalism}.} which may be taken w.l.o.g. to be an equilibrium configuration, the instantaneous vacuum state of ${\OpH H}_0$ at $t=-\infty$. We want to ensure now that should we evolve the system, that we make no assumption about what it would end up at late times. To this end, we should revert back after allowing the interactions to influence the system, to the initial state. In a path integral this can be done by introducing a SK-evolution operator, which evolves the system in a complex time contour. Let $\mathcal{C}$ be a contour in the complex time plane, that starts out at $t= -\infty+i\varepsilon$, follows the real axis, and then retraces its trajectory back with a small imaginary displacement by $-2i\varepsilon$, cf., Fig.~\ref{fig:contour1}. We have chosen to orient the contour, so that the direction of traversal is clockwise (about the origin say). We will also find it useful to label the forward leg of the contour, as the right $\text{R}$ part and the backward leg, the left $\text{L}$ part.

Given such a contour, we can work with operators which live in this complexified domain, and define the Schwinger-Keldysh S-matrix by working with contour-ordering prescription, viz.,
\begin{equation}
U_\mathcal{C}\equiv {\cal T}_\mathcal{C} \, \exp\left( -i \int_\mathcal{C} \,dt' \, {\OpH H}_{int}(t') \right) \,.
\end{equation}
There is  a sensible time-ordering prescription  inherited from this contour ordering. It is often however useful not to work with a single contour, but rather, work with fields and operators labeled by which part of the contour they appear on. This makes it clear that there needs to be a doubling of the degrees of freedom. We have left and right fields indexed by their position on the Schwinger-Keldysh contour $\mathcal{C}$. Furthermore, as illustrated, the operators on the right/forward leg are time-ordered, those on the left/backward leg are anti-time ordered, and the right operators precede those on the left leg of the contour.

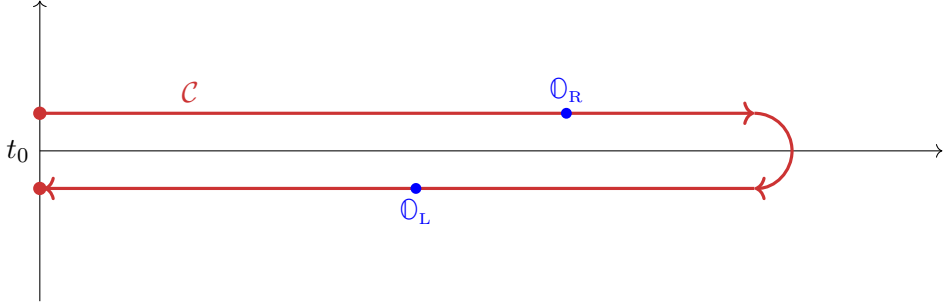
\begin{figure}[t!]
\begin{center}
\begin{tikzpicture}
\draw[thin,color=black,->] (-5,0) -- (7,0);
\draw[thin, color=black,->] (-5,-2) -- (-5,0) node [left] {$t_0$} --  (-5,2);
\draw[very thick,color=rust,->] (-5,0.5) -- (-3,0.5) node [above] {$\color{rust}\mathcal{C}$} -- (2,0.5) node [above] {${\color{blue}\Op{O}_\skR}$} -- (4.5,0.5);
\draw[very thick,color=rust,->] (4.5,-0.5) -- (0,-0.5) node [below] {${\color{blue}\Op{O}_\skL}$} -- (-4.95,-0.5);
\draw[very thick,color=rust,<-] (4.5,-0.5) arc (-90:90:0.5);
\draw[thick,color=rust,fill=rust] (-5,0.5) circle (0.45ex);
\draw[thick,color=rust,fill=rust] (-5,-0.5) circle (0.45ex);
\draw[thick,color=blue,fill=blue] (2,0.5) circle (0.35ex);
\draw[thick,color=blue,fill=blue] (0,-0.5) circle (0.35ex);
\end{tikzpicture}
\caption{SK time contour in the case where the initial state at time $t_0$ is known and the latest operator insertion happens at time $t$. The indicated operator insertions correspond to a real-time correlator $G_<(x,x')$.}
\label{fig:contour2}
\end{center}
\end{figure}

It is useful at this juncture to note a few salient facts about the integration contours, cf., \cite{Maciejko}. Should we have complete knowledge of the density matrix of the full
system at some finite time $t_0$, then we do not need to follow the contour all the way from $t=-\infty$ to $+\infty$ and back. It suffices to focus on the part of the contour from $t_0$ to $\text{max}(t,t')$ which corresponds to the future-most operator insertion before retracing back to the initial configuration, c.f., Fig.~\ref{fig:contour2}. Intuitively, all this is saying is that the knowledge of the density matrix can be treated as initial conditions for the subsequent evolution and that for finite time computations, details of how the system evolves to the future of all operator insertions are inessential.

We are now in a position to define the Schwinger-Keldysh Green's function. As an example, let us consider a complex (bosonic) operator $\OpH{O}$. Following standard discussion, we have
\begin{align}
 G_\mathcal{C}(x,x') = -i \langle \Omega | {\cal T}_\mathcal{C} \prn{ {\OpH O}(x) \, {\OpH O}^\dagger(x') }|\Omega\rangle
                     = -i \langle 0 | {\cal T}_\mathcal{C}\prn{ U_\mathcal{C} \,{\OpH O}(x)\, {\OpH O}^\dagger(x')} |0\rangle \,.
\label{eq:GSK}
\end{align}
We note that with the contour ordering we no longer have a normalizing denominator anymore, for the Schwinger-Keldysh S-matrix doesn't pick up a phase, $U_\mathcal{C} | 0\rangle = |0\rangle$. This contour prescription is sufficient to obtain the various Green's functions that one is usually interested in. Let us determine a prescription for these using the left-right basis of fields introduced above. Owing to the complexification of the contour, and doubling of the degrees of freedom we immediately see that we should have a $2 \times 2$ matrix of \emph{real-time}  Green's functions, corresponding to the choice of operator insertions on either segment. One thus defines:
\begin{align}
 G(x,x') = \begin{pmatrix}
 		G_{\skR \skR} & G_{\skR \skL} \\ G_{\skL \skR} & G_{\skL \skL}
 	\end{pmatrix}
         \equiv
         \begin{pmatrix}
         		G_{_F} & G_{_<} \\ G_{_>} & G_{_{\tilde F}}
         \end{pmatrix}
\,,
 \label{SK-CorrUnphys}
\end{align}
where we indicate the various Green's functions both by the contour positions of the operator insertions and the more familiar notation.
$G_F(x,x')$ is the well known Feynman propagator, $G_{\tilde F}(x,x')$ is an  anti-Feynman propagator with reversed time ordering
and we have in addition two new cross-contour correlators. These all have familiar definitions:
\begin{equation}
\begin{split}
 G_{_F}(x,x') &= -i \langle \Omega | {\cal T}\prn{{\OpH O}(x){\OpH O}^\dagger(x')}|\Omega\rangle \,, \\
 G_{_{\tilde F}}(x,x') &= -i \langle\Omega|  \bar{{\cal T}} \prn{{\OpH O}(x){\OpH O}^\dagger(x')}|\Omega\rangle \,,\\
 G_{_<} (x,x') &=  -i\langle \Omega| {\OpH O}^\dagger(x'){\OpH O}(x)|\Omega\rangle \,,\\
 G_{_>} (x,x') &= -i \langle \Omega| {\OpH O}(x){\OpH O}^\dagger(x') |\Omega\rangle \,.
\end{split}
\end{equation}
While it a-priori appears as though we have four non-trivial Green's functions, it is a simple matter to check the time-ordering prescriptions to note that they satisfy a simple linear relation,
\begin{equation}
G_{_F} + G_{_{\tilde F}} = G_{_>} + G_{_<} \,.
\label{eq:2ptrel}
\end{equation}

This is perhaps a lot more familiar if we adapt to a different basis and work with linear combinations of above to get the familiar
advanced, retarded and Keldysh functions, which are as usual defined as:
\begin{equation}
\begin{split}
 G_{ret}(x,x') &\equiv -i \,\Theta(t-t') \, \langle \Omega | \left[ \OpH{O}(x),{\OpH O}^\dagger(x')\right]  |\Omega\rangle
 	= G_{_F} - G_{_<} \,,
 \\
 G_{adv}(x,x') &\equiv i \Theta(t'-t) \, \langle \Omega |  \left[ \OpH{O}(x),{\OpH O}^\dagger(x')\right] |\Omega \rangle
 	= G_{_F}-G_{_>} \,,
 \\
 G_{K}(x,x') &\equiv -i \langle \Omega | \left\{{\OpH O}(x), {\OpH O}^\dagger(x')  \right\}|\Omega \rangle = G_{_F} + G_{_{\tilde F}} \,.
\end{split}
 \label{eq:advretSK}
\end{equation}
Note that the  Wightman two-point functions can easily be recovered from this basis.

We can continue in this vein to discuss higher point functions. An $n$-point Schwinger-Keldysh Green's function can be defined from the contour ordering and decomposes into $2^n$ component Green's functions as above. Generalizing \eqref{SK-CorrUnphys}, we can simply write $G_{_{I_1 I_2 \cdots I_n}}$ with $I_i \in\{\text{L},\text{R}\}$. The construction then naturally allows one to derive an identity, which implies the vanishing of a certain correlation function for every $n$ \cite{Chou:1984es} (see also \cite{Weldon:2005nr}).  We will see shortly that this statement, which is usually attributed to a normalization condition of $\Theta$ functions, is better interpreted as arising from a field redefinition topological invariance of the Schwinger-Keldysh construction.

Let us now try to rewrite the above discussion in terms of a path integral, introducing sources, so that we can start talking directly about the generating functions of correlators. Consider the Schwinger-Keldysh generating functional with suitable classical sources ${\cal J}(x)$ along the contour:
\begin{equation}
\begin{split}
 \mathscr{Z}_{SK}[{\cal J}(x)] &\equiv \langle \Omega | {\cal T}_\mathcal{C} \, \exp\left( i \int_\mathcal{C} \mathcal{L}[\Phi(x)] + {\cal J}(x) \Phi(x)\right) |\Omega\rangle  \\
   \mathscr{Z}_{SK}[{\cal J}_\skR(x),{\cal J}_\skL(x)]   &= \langle \Omega | {\cal T}_\mathcal{C} \, e^{i \int_{t=-\infty}^{t=\infty}\mathcal{L}[\Phi_\skR(x)]
       - \mathcal{L}[\Phi_\skL(x)] + {\cal J}_\skR(x) \Phi_\skR(x) - {\cal J}_\skL(x)\Phi_\skL(x) }|\Omega \rangle \,.
\end{split}
 \label{eq:SKGenFc}
\end{equation}
We started here with a single complex contour and thence  transformed into the representation involving left and right fields. The second line should be viewed as a \emph{single time representation} with the characteristic doubling of the sources and fields. The relative  sign in front of the part of the left
Lagrangian and sources corresponding to the left part of the contour makes explicit the fact that these terms should be integrated backwards in time.

Having the generating functional  $\mathscr{Z}_{SK}[{\cal J}_\skR(x),{\cal J}_\skL(x)]$ at hand, one may proceed to computing correlation functions $G_{_{I_1 I_2 \cdots I_n}}$ by functional differentiation
\begin{equation}
G_{_{I_1 I_2 \cdots I_n}}(x_1,x_2,\cdots, x_n) = \frac{\delta^n Z[{\cal J}_\skR(x), {\cal J}_\skL(x)]}{\delta {\cal J}_{_{I_1}}(x_1)\,\delta {\cal J}_{_{I_2}}(x_2)\, \cdots\,\delta {\cal J}_{_{I_n}}(x_n)} \bigg|_{{\cal J}_\skL(x) = {\cal J}_\skR(x) =0}
\label{eq:GnSK}
\end{equation}

To pass to from the left-right basis of correlation functions to one directly amenable to computing the time-ordered correlators as in \eqref{eq:advretSK} we can simply affect a basis change in the path integral. Performing a field redefinition we pass onto the average-difference basis:
\begin{align}
  \begin{pmatrix}
  \Phi_{_{av}} \\ \Phi_{_{dif}} \end{pmatrix} = \begin{pmatrix} \tfrac{1}{2} (\Phi_\skR+\Phi_\skL) \\ \Phi_\skR-\Phi_\skL
  \end{pmatrix}
   \,,\qquad
  \begin{pmatrix}
\SKAv{{\cal J}}\\ \SKRel{{\cal J}} \end{pmatrix} = \begin{pmatrix} \tfrac{1}{2} ({\cal J}_\skR + {\cal J}_\skL) \\ {\cal J}_\skR-{\cal J}_\skL \end{pmatrix} \,.
\end{align}
The generating functional then becomes
\begin{align}
 \mathscr{Z}[\SKAv{{\cal J}}(x), \SKRel{{\cal J}}(x)] =
 	\langle\Omega| {\cal T}_\mathcal{C}\,
 	e^{ i \int_{t=-\infty}^{t=\infty}
	    \mathcal{L}[\Phi_{_{av}} + \tfrac{1}{2} \,\Phi_{_{dif}}] -
	     \mathcal{L}[\Phi_{_{av}} - \tfrac{1}{2}\,\Phi_{_{dif}}]
    	+ \SKAv{{\cal J}}(x)\,\Phi_{_{dif}} +  \SKRel{{\cal J}}(x)\, \Phi_{_{av}}
    	}
    	 |\Omega\rangle \,.
\end{align}
The main fact we wish to highlight  is that the difference source $\SKRel{{\cal J}}(x)$ generates the response as a functional of the physical average field $\Phi_{_{av}}(x)$, while the average/common source $\SKAv{{\cal J}}(x)$ in turn does the same for the difference or fluctuation field $\Phi_{_{dif}}(x)$.

\begin{figure}[t!]
\begin{center}
\begin{tikzpicture}
\draw[thin,color=black,->] (-5,0) -- (7,0);
\draw[thin, color=black,->] (-5,-4) -- (-5,-3) node [left] {$t_0+i(\varepsilon-\beta_0)$} -- (-5,0) node [left] {$t_0$} --  (-5,0.5) node [left] {$t_0+i\varepsilon$} -- (-5,2);
\draw[very thick,color=rust,->] (-5,0.5) -- (-3,0.5) node [above] {$\color{rust}\mathcal{C}$} -- (2,0.5) node [above] {${\color{blue}\Op{O}_\skR}$} -- (4.5,0.5);
\draw[very thick,color=rust,->] (4.5,-0.5) -- (0,-0.5) node [below] {${\color{blue}\Op{O}_\skL}$} -- (-5,-0.5) -- (-5,-2.9);
\draw[very thick,color=rust,<-] (4.5,-0.5) arc (-90:90:0.5);
\draw[thick,color=rust,fill=rust] (-5,0.5) circle (0.45ex);
\draw[thick,color=rust,fill=rust] (-5,-3) circle (0.45ex);
\draw[thick,color=blue,fill=blue] (2,0.5) circle (0.35ex);
\draw[thick,color=blue,fill=blue] (0,-0.5) circle (0.35ex);
\end{tikzpicture}
\caption{SK time contour in thermal physics, where the initial state is a thermal state with an entanglement pattern encoded in a Euclidean partition function. The starting and end points of the contour are identified. The associated Euclidean (imaginary time) periodicity is set by the inverse temperature $\beta_0$.}
\label{fig:contour3}
\end{center}
\end{figure}
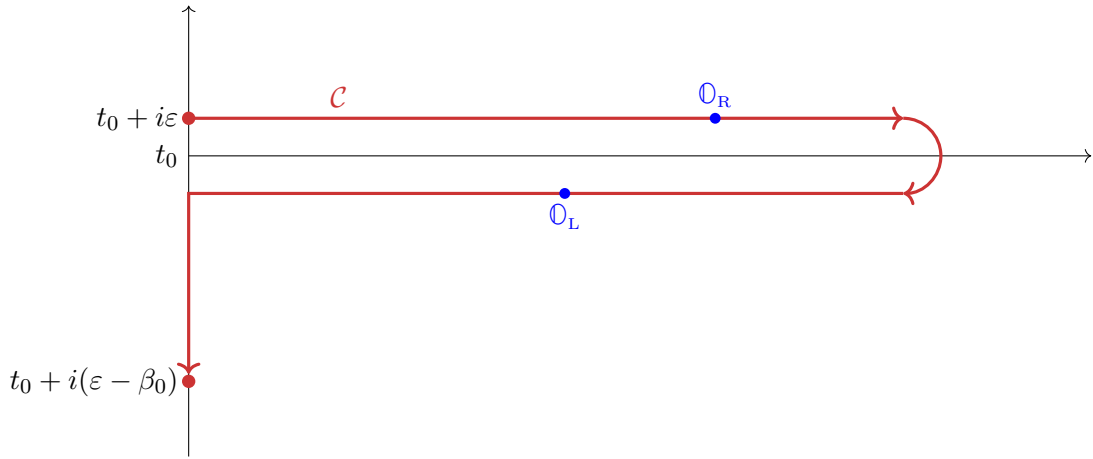

With future applications in mind, we briefly mention the special case of thermal initial conditions. For systems starting their evolution in a thermal state with inverse temperature $\beta_0$ at time $t_0$, the time contour can be illustrated as in Fig.~\ref{fig:contour3}. This presentation of the contour, which is necessary to consistently take into account initial state correlations, is sometimes referred to as Kadanoff-Baym contour.
That is, the thermal state generated by some Hamiltonian ${\OpH H}_0$ is described by an un-normalized initial density matrix $\rhoT = e^{-\beta_0 {\OpH H}_0}$. Such a state readily allows for a Euclidean description in terms of a partition function
\begin{equation}
\mathscr{Z}_{T}(\beta_0) = \text{Tr} \left( e^{-\beta_0 {\OpH H}_0} \right) \,.
\end{equation}
It is then clear that such a Euclidean path integral codifies the  correlations (or the entanglement pattern) of the initial state, and it corresponds to a Euclidean segment of evolution in the imaginary time direction as in Fig.\ \ref{fig:contour3}.
This special case is of significant interest if we wish to use the Schwinger-Keldysh formalism for the study of near-thermal correlations such as those underlying fluid dynamics. We will return to near-thermal physics in much more detail in \S\ref{sec:skthemal}.

We now have the essential features of the Schwinger-Keldysh construction in place. In the following we will try to  rephrase this discussion in more abstract terms and extract some useful lessons about the symmetries inherent in the construction.

\section{The basics of the Schwinger-Keldysh formalism}
\label{sec:skbasics}

We begin with an overview of some notational conventions that we will use in the rest of our discussion. Our task will then be to rewrite the discussion of \S\ref{sec:skreview} in an operator language that enables us to formulate an appropriate set of field redefinition BRST supercharges that are present in the Schwinger-Keldysh formalism. We first focus on arbitrary density matrices in a relativistic QFT and then subsequently discuss new features that arise when we consider thermal (or near-thermal) density matrices.

\subsection{Preliminaries: background and notation}


Consider a quantum system with a Hilbert space of states ${\cal H}$ (the space of `kets') and its dual space ${\cal H}^*$ (the space of `bras'). We will be agnostic for the present whether the quantum system is open or closed; our primary concern is to write down the Schwinger-Keldysh construction for dealing with density matrices $\hat{\rho}$ in this system.

To begin with we construct the Schwinger-Keldysh system for our quantum system: this is given by the tensor product Hilbert space
 ${\cal H}_\skR \otimes  {\cal H}_\skL^*$, where we have chosen to label the components as $\text{R}$ and $\text{L}$ for convenience. Following our earlier discussion we imagine that all the mixed states live in this extended space of states.
The main caveat is that not all elements of ${\cal H}_\skR \otimes {\cal H}_\skL^* $ can
 be normalized to give a mixed state whereas  any non-zero element of ${\cal H}$
can be normalized to a pure quantum state. Recall that density matrix $\hat{\rho}$ of an admissible mixed state should be
\begin{itemize}
\item Hermitian with non-negative eigenvalues, and
\item  should have non-zero but finite trace (which can then be normalized to unity).
\end{itemize}

 We adapt a notation where we denote standard operators on ${\cal H}$ (which are automatically elements of ${\cal H}_\skR \otimes {\cal H}_\skL^*$ )  by a hat. On the contrary there will be no hats on Schwinger-Keldysh operators which act on the entire space  ${\cal H}_\skR \otimes {\cal H}_\skL^*$. Operators in the extended system are sometimes referred to as  \emph{superoperators}.

Let $\hat{\mathbb{O}} \in {\cal H}_\skR \otimes {\cal H}_\skL^*$ be an operator acting on the state space  ${\cal H}$: we can then construct two corresponding superoperators acting on ${\cal H}_\skR \otimes {\cal H}_\skL^* $ of the form
\begin{equation}
\SKR{O} \equiv \OpH{O}  \otimes \Op{I} \,,\qquad
\SKL{O} \equiv \Op{I}   \otimes \OpH{O}\,.
\label{}
\end{equation}
As described in \S\ref{sec:skreview}, often one performs a \emph{Keldysh rotation} to instead work with the  difference and average operators defined via:
\begin{equation}\label{eq:KeldyshDef}
\begin{split}
\SKRel{O} \equiv  \SKR{O}-\SKL{O}\ , \qquad
\SKAv{O} &\equiv  \half \prn{ \SKR{O}+\SKL{O} }\,.
\end{split}
\end{equation}
We note that, after Keldysh rotation the average sources are associated with difference operators
and the difference sources are associated with average  operators. This a consequence of the following
relation relating right-left basis to Keldysh basis:
\begin{equation}\label{eq:KeldyshJ}
\begin{split}
\SKR{\mathcal{J}}\  \SKR{O} - \SKL{\mathcal{J}}\ \SKL{O}  =
\SKAv{\mathcal{J}}\ \SKRel{O} + \SKRel{\mathcal{J}}\ \SKAv{O}  \,.
\end{split}
\end{equation}
One may view the statement as saying that the Schwinger-Keldysh contour imparts a Lorentzian inner product between the left and right segments, and the passage to the Keldysh basis is akin to choosing light-cone variables.
In any event, varying the Schwinger-Keldysh action with respect to average sources gives the correlators with difference operators and vice-versa.

We should note here that in much of the literature the Keldysh basis introduced in \eqref{eq:KeldyshDef} is called the `$ra$' basis. The average operators are called the $r-$operators and the difference operators are called $a$-operators. We find this terminology less intuitive. Moreover, when we discuss thermal correlation functions, for $\rhoi$ being a thermal Gibbs density matrix, we will encounter the retarded-advanced basis (we use $ret-adv$ to denote them). To forestall any potential confusion, we propose to refer to the Keldysh basis as $av-dif$ operators.

With these preliminaries in place let us define the Schwinger-Keldysh generating functional $\mathscr{Z}_{SK}$ which is defined by the trace over the tensor product Hilbert space ${\cal H}_\skR \otimes {\cal H}_\skL^*$:
\begin{equation}
\mathscr{Z}_{SK}[{\cal J}_\skR,{\cal J}_\skL] \equiv \Tr{\ U[{\cal J}_\skR]\ \rhoi\ (U[{\cal J}_\skL])^\dag\ } \, ,
\label{eq:ZSKdef}
\end{equation}
where $\rhoi$ is the  initial density matrix of the system, $U$ denotes the unitary evolution operator of the QFT, and $U^\dag$ is its adjoint. We have allowed ourselves to deform the unitary evolution operators with a suitable sprinkling of both    right and left sources ${\cal J}_\skR$ and ${\cal J}_\skL$ respectively. These unitaries are defined in a standard way using time-ordered exponentials of the evolution operator deformed by the sources, viz.,
\begin{equation}
U[{\cal J}] =  {\cal T} \, \exp\left(  -i \int_{t_i}^{t}\, dt\, H[{\cal J}]   \right) \,, \qquad
(U[{\cal J}] )^\dagger=  \bar{{\cal T}} \, \exp\left(  i \int_{t_i}^{t}\, dt\, H[{\cal J}]   \right) \,.
\label{}
\end{equation}
We use the symbol $\mathcal{T}$ to denote time-ordering while $\bar{\mathcal{T}}$ denotes anti-time ordering.  In the absence of sources, the unitaries reduce to the standard Heisenberg operators for time independent Hamiltonians, viz., $U = e^{-i\,H \,t }$.

 Then, via functional differentiation with respect to the left and right sources, we can compute the Schwinger-Keldysh correlation functions, which are of the form:
\begin{equation}
\Tr{\ \rhoi \ \bar{\mathcal{T}} \prn{U^\dag \SKL{O} U^\dag \SKL{O}\ldots}\ \mathcal{T} \prn{U \SKR{O} U \SKR{O}\ldots}\ } \,.
\label{eq:SKCorr}
\end{equation}
We note that left operators are ordered to the \emph{left} of the right operators (thus justifying the terminology). We will have more to say about the Schwinger-Keldysh time-ordering prescription momentarily.

Note that the Schwinger-Keldysh description differs from the more familiar Feynman path-integral, which takes the form
\begin{equation}
\mathscr{Z}_{\text{Feynman}}[J] \equiv \langle \text{Vaccum}_{t=\infty} |\ U[{\cal J}]\  | \text{Vaccum}_{t=-\infty} \rangle \,.
\label{}
\end{equation}
This Feynman path integral only allows for the computation of time-ordered correlators of the form
\begin{equation}
\langle \text{Vaccum}_{t=\infty} |\ \mathcal{T} \prn{U \OpH{O} U \OpH{O}\ldots}\  | \text{Vaccum}_{t=-\infty} \rangle \, .
\label{}
\end{equation}
As noted in \S\ref{sec:skreview}, in a non-equilibrium or open quantum system we have an lack of knowledge of what the interacting final state of the system would be. The Schwinger-Keldysh construction cleverly avoids this issue, by reverting back to the initial state. This ensures that the entanglement built into the initial density matrix $\rhoi$, and the knowledge of the sources that one has turned on, suffices to compute the desired time ordered correlators.

\subsection{Schwinger-Keldysh time ordering}
\label{sec:torder}

We now introduce a notion of Schwinger-Keldysh time-ordering, which follows the contour ordering prescription introduced in \S\ref{sec:skreview}. To allow a general statement, let us first introduce the concept of mutual Grassmann parity of operators.
To do so, we first introduce the notion of a Grassmann number for an operator ${\Op{O}}$, which is defined to be
\begin{equation}
\GSgn{O}  =
\begin{cases}
+1 \,, \qquad  \Op{O}: \;\text{Grassmann even}  \\
-1 \,,\qquad \Op{O}:\; \text{Grassmann odd}
\end{cases}
\label{eq:GOdef}
\end{equation}

In addition to the Grassmann number it is also useful to keep track on occasion of the fermion number, which we denote $\FSgn{\Op{O}}$. We define this as
\begin{equation}
\FSgn{O}  =
\begin{cases}
+1 \,, \qquad  \Op{O}: \;\text{bosonic}  \\
-1 \,,\qquad \Op{O}:\; \text{fermionic}
\end{cases}
\label{eq:FOdef}
\end{equation}
On physical fields, $F_{\Op{O}}$ and $G_{\Op{O}}$ are the same. However, there is nevertheless an important distinction between the two (especially if one introduces ghosts in the description of the system, as we will do later on). The former cares about the statistics obeyed by the operator irrespective of its Grassmann parity, while the latter only cares about the Grassmann nature. In particular $F_\Op{O}$ is taken to be zero not only for the usual bosonic degrees of freedom one is used to, but also for Grassmann odd ghost particles; we will loosely refer to all such fields as `bosonic'. Similarly,
$F_\Op{O}$ is unity for Grassmann odd particles, as well as for  Grassmann even ghosts, both of which we refer to as `fermionic'.

Given the Grassmann number operator we can proceed to define the mutual Grassmann parity
$\GSgn{\Op{A}}{}^{G_\Op{B}}$ for two operators  $\OpH{A}(x)$ and $\OpH{B}(y)$ by multiplying the Grassmann numbers $G_\Op{A} \,G_\Op{B} $, which gives a relative sign when both operators are Grassmann odd. We will use this soon to define a generalized commutator that accounts for the Grassmann parity of the fields in question and to give the correct boundary conditions for thermal physics.

As the system evolves away from the initial density matrix $\rhoi$ prescribed at  $t=t_i$,  it generically is no more in equilibrium because of the external sources we turn on. In this case, as independently argued by Schwinger \cite{Schwinger:1960qe} and Keldysh \cite{Keldysh:1964ud}, we need to keep track of two copies of all the observables:  every operator $\OpH{O}$ is replaced by a right operator $\SKR{O}$ giving  time-ordered (or Feynman) correlators and a  left operator  $\SKL{O}$ giving  anti-time-ordered (or anti-Feynman) correlators. Thus
\begin{equation}\label{eq:RLCorr}
\begin{split}
\langle\ &\overline{\mathcal{T}} \brk{\OpH{O}^{(1)} \OpH{O}^{(2)} \ldots \OpH{O}^{(p)} } \ \mathcal{T}  \brk{\OpH{O}^{(p+1)} \OpH{O}^{(p+2)} \ldots \OpH{O}^{(p+q)} }\  \rangle \\
&\equiv \langle\ \mathcal{T}_{SK}\ \SKL{O}^{(1)} \SKL{O}^{(2)} \ldots \SKL{O}^{(p)} \ \SKR{O}^{(p+1)} \SKR{O}^{(p+2)} \ldots \SKR{O}^{(p+q)}\  \rangle \,,
\end{split}
\end{equation}
where the expectation value $\langle \ldots \rangle$ can be taken with an arbitrary initial condition and is just defined by the trace as in \eqref{eq:SKCorr}. The object  $\mathcal{T}_{SK}$ will henceforth be used to denote  Schwinger-Keldysh time ordering, deviating from the more conventional contour ordering used earlier in \S\ref{sec:skreview}. It can be easily understood by the mnemonic:  right operators are time ordered, left operators are anti-time ordered and left operators are ordered after the right operators.

To summarize, the nomenclature right vs left can be interpreted in various ways:
\begin{itemize}
\item The left operators are always ordered to the left of the right operators. Thus the right-left correlators can be used to compute the un-ordered correlators.
\item In the complex time plane, the right operators are placed on the time contour running to the right (increasing time) vs the left operators which are placed on the time contour running to the left (decreasing time), as in the contour ordering prescription described in \S\ref{sec:skreview}.
\item For thermal equilibrium states studied using the AdS/CFT correspondence cf., \cite{Maldacena:2001kr}, the right operators are placed at the right boundary of the eternal black hole whereas the left operators are placed at the left boundary.\footnote{ We will discuss in \S\ref{sec:applications} some rudimentary aspects of the Schwinger-Keldysh formalism applied to gravitational systems, drawing a distinction between the more familiar thermofield double construction which is employed in \cite{Maldacena:2001kr} and the Schwinger-Keldysh complex time contour.}
\end{itemize}

Returning back to our discussion of the Keldysh rotation, we note that we can transform the correlation functions from the
$\text{R}-\text{L}$ basis to the Keldysh ($av-dif$) basis. The prescription we seek is given for two-point functions (of mutually Grassmann even operators) quite simply to be
\begin{equation}
\begin{split}
\langle\ \mathcal{T}_{SK} \,\SKAv{A}(x)\ \SKAv{B}(y) \ \rangle
&=
	 \langle\   \gradAnti{\OpH{A}(x)}{\OpH{B}(y)}\  \rangle \,,
 \\
 \langle\ \mathcal{T}_{SK} \,\SKAv{A}(x)\ \SKDif{B}(y)\ \rangle
 &=
 	\stepFn{\Op{A}\Op{B}}\,  \langle\  \gradcomm{\OpH{A}(x)}{\OpH{B}(y)}\ \rangle \,,
 \\
 \langle\ \mathcal{T}_{SK}\, \SKDif{A}(x)\ \SKAv{B}(y) \  \rangle
 &=
 -\stepFn{\Op{B}\Op{A}} \ \langle\  \gradcomm{\OpH{A}(x)}{\OpH{B}(y)} \ \rangle \,,
 \\
\langle\ \mathcal{T}_{SK}\bigbr{ \SKAv{A}(x)\ \SKDif{B}(y)
-\SKDif{A}(x)\ \SKAv{B}(y) } \  \rangle
&=
\langle\  \gradcomm{\OpH{A}(x)}{\OpH{B}(y)}\ \rangle \,,\\
\langle\ \mathcal{T}_{SK} \,\SKDif{A}(x)\ \SKDif{B}(y) \ \rangle &=
0 \,.\\
\end{split}
\label{eq:skavdif}
\end{equation}
In the process of writing \eqref{eq:skavdif} we encounter a few new pieces of notation. Firstly, the graded commutator $\gradcomm{\ }{\ }$ is introduced so as to avoid our having to write commutators and anti-commutators in a case-by-case basis.  Likewise it is also convenient to define a graded anti-commutator  $\gradAnti{\ }{\ }$. These objects are defined using the Grassmann number operator $G_\Op{O}$ introduced in \eqref{eq:FOdef}:
\begin{align}
\gradcomm{\Op{A}}{\Op{B}} &= \Op{A}\, \Op{B} - (-)^{G_{\Op{A}}\,G_{\Op{B}} }\; \Op{B}\, \Op{A}   \,,
\nonumber \\
\gradAnti{\Op{A}}{\Op{B}} &= \half\left(\Op{A}\, \Op{B} + (-)^{G_{\Op{A}}\,G_{\Op{B}} }\; \Op{B}\, \Op{A} \right) \,.
\label{eq:gradCA}
\end{align}
These definitions clearly satisfy:
\begin{equation}
 \gradcomm{\Op{A}}{\Op{B}}=-(-)^{G_{\Op{A}}\,G_{\Op{B}} } \ \gradcomm{\Op{B}}{\Op{A}}\ ,\qquad
 \gradAnti{\Op{A}}{\Op{B}}=(-)^{G_{\Op{A}}\,G_{\Op{B}} }\ \gradAnti{\Op{B}}{\Op{A}} \,.
\label{eq:gradcac}
\end{equation}
One can also check that the graded commutators  obey a graded Jacobi identity of the form
\begin{equation}
\begin{split}
\gradcomm{\gradcomm{\Op{A}}{\Op{B}}}{\Op{C}}= \gradcomm{\Op{A}}{\gradcomm{\Op{B}}{\Op{C}}} - (-)^{G_{\Op{A}}\,G_{\Op{B}} } \gradcomm{\Op{B}}{\gradcomm{\Op{A}}{\Op{C}}} \,,
\end{split}
\end{equation}
which will prove useful when we study higher point correlation functions. A good discussion of these graded commutators and associated mathematical structures can be found in \cite{DeWitt:1992cy}.\footnote{ Since it becomes cumbersome to keep writing $\GSgn{\Op{A}}{}^{G_\Op{B}}$, we will often simplify this to $(-1)^{\Op{A}\Op{B}}$. Hopefully, it should be clear that the sign only cares about the mutual Grassmann parity of the operators in expressions. \label{fn:Grparity}}

We now turn to another notational aspect in \eqref{eq:skavdif}, concerning the step functions.

\subsection{Step function conventions}

The time ordering prescription comes as usual with a set of step functions, that appear when we want to compute certain physical correlation functions. There are various conventions used in the literature for these, so we will explain briefly our choice. We use $ \stepFn{\Op{A}>\Op{B}} =  \stepFn{\Op{A}\Op{B}}$ as a step function which is unity if the operator $\Op{A}$ lies in the causal future of the operator $\Op{B}$ and is zero if $\Op{A}$ lies in the
causal past of $\Op{B}$.  Similarly, $ \stepFn{\Op{A}<\Op{B}} = \stepFn{\Op{B}\Op{A}}$ is a step function
which is unity if the operator  $\Op{A}$ lies in the causal past of $\Op{B}$ and  is zero if $\Op{A}$ lies in the causal future of $\Op{B}$.  These step functions satisfy the identity:\footnote{
This normalization condition is what is usually invoked to argue that the two-point correlation function in the $\text{R}-\text{L}$ basis is exhausted by the advanced, retarded, and the Keldysh correlators. }
\begin{equation}
\stepFn{\Op{A}>\Op{B}} +  \stepFn{\Op{B}>\Op{A}} = 1 \,.
\label{eq:thetacon}
\end{equation}
 It is usual to give a regularizing prescription for what happens when the causal order is indeterminate. In addition it is natural to demand that whatever the prescription be, it  should continue to obey the identity above \eqref{eq:thetacon}.
Some of the commonly used regularizing prescriptions when the causal order is indeterminate are
\begin{equation}
\begin{split}
\text{ It\= o} &:  \quad  \stepFn{\Op{A}>\Op{B}}  = 1,\quad    \stepFn{\Op{A}<\Op{B}} = 0 . \\
\text{ (Fisk-) Stratonovich} &:  \quad  \stepFn{\Op{A}>\Op{B}}  = \half ,\quad     \stepFn{\Op{A}<\Op{B}}= \half . \\
\text{ Hanggi-Klimentovich} &:  \quad  \stepFn{\Op{A}>\Op{B}}  =0 ,\quad    \stepFn{\Op{A}<\Op{B}} = 1 . \\
\end{split}
\end{equation}
Since Stratanovich prescription is natural from the viewpoint of Fourier transforms and it is a CPT invariant regulator, we will employ it in what follows. We then have $\stepFn{\Op{A}>\Op{B}}  =  \stepFn{\Op{A}<\Op{B}} $ everywhere, when the causal order is indeterminate. Each of the three definitions above does respect the normalization condition \eqref{eq:thetacon}.\footnote{ Sometimes for  generalized Langevin theory in non-equilibrium physics and often in stochastic mathematics (including mathematical finance) the It\=o prescription is  preferred.  CPT exchanges It\=o and Hanggi-Klimentovich prescriptions and thus the CPT-violating  nature of  It\=o has to then be compensated by CPT-violating counter terms (as is usual with any symmetry violating regulator). The ghosts we will talk about later in this text often decouple in the It\=o prescription which is probably the reason it is preferred in fields which do not want to deal with ghosts.}

The discussion of the step functions generalizes clearly to multiple arguments, for we can simply iterate the definition pairwise for each insertion. For instance we can write:
\begin{equation}
\begin{split}
\stepFn{\Op{A}_1 > \Op{A}_2 > \cdots \Op{A}_n}  &\equiv  \stepFn{\Op{A}_1 \Op{A}_2\cdots \Op{A}_n}
 \,,  \\
&= \stepFn{\Op{A}_1 > \Op{A}_2 } \,\stepFn{\Op{A}_2 > \Op{A}_3 } \cdots
\stepFn{\Op{A}_{n-1} > \Op{A}_n }   = \stepFn{\Op{A}_1 \Op{A}_2 } \,\stepFn{\Op{A}_2  \Op{A}_3 } \cdots
\stepFn{\Op{A}_{n-1}  \Op{A}_n }
\end{split}
\end{equation}
In what follows we will always write the step functions with the explicit time ordering as indicted in the first line.
The analog of \eqref{eq:thetacon} is the generalized normalization condition:
\begin{equation}
\sum_{\text{permutations }\sigma} \stepFn{\Op{A}_{\sigma(1)} > \Op{A}_{\sigma(2)} > \cdots \Op{A}_{\sigma(n)}} = 1 \,.
\end{equation}
%

\subsection{Keldysh basis correlators}
\label{sec:keldysh}

We now have all the machinery to give  an explicit formula for the Keldysh basis correlators following \cite{Chou:1984es}. The simplest correlator is the one containing only difference operators and it vanishes identically, viz.,
\begin{equation}
\begin{split}
 \langle \mathcal{T}_{SK} \prod_k\ \SKRel{O}^{(k)} \rangle
 \equiv  \langle \mathcal{T}_{SK} \prod_k \prn{\SKR{O}^{(k)}-\SKL{O}^{(k)} }\rangle
  =  0 \,.
\end{split}
\label{eq:diff0}
\end{equation}
This is in fact easy to see directly from the definition of the generating function
$\mathscr{Z}_{SK}[{\cal J}_\skR,{\cal J}_\skL] $. First one notes that the difference operators $\SKRel{O}$ are sourced by the average sources ${\cal J}_{_{av}}$. This means that for computing \eqref{eq:diff0} we can w.l.o.g.\ set ${\cal J}_\skR = {\cal J}_\skL \equiv {\cal J}$ in the generating function before taking any functional derivatives. However, $\mathscr{Z}_{SK}[{\cal J},{\cal J}] = \Tr{\rhoi} $, owing to the cyclicity of the trace, c.f., \eqref{eq:ZSKdef}.
Thus we learn that the functional derivative of this result will vanish, simply  asserting that the  SK-path integral is unresponsive to a set of average sources, for it collapses to a statement of initial conditions. This proves \eqref{eq:diff0} and we conclude that one out of $2^n$ Schwinger-Keldysh $n$-point functions generically vanishes.

It must be emphasized that this fact holds independent of the dynamics, which after all, is contained in the unitary evolution operator $U$.  The universality of this statement, points to a fundamental symmetry principle. We will argue later that the Schwinger-Keldysh path integral behaves like a topological theory when restricted to this sector. In particular, the difference operators will be shown to be BRST exact, with the symmetry being traceable back to a set of field redefinitions inherent in the doubling from ${\cal H}$ to
${\cal H}_\skR \otimes {\cal H}_\skL^*$.

Since the correlation function with only difference operators vanishes, we can focus our attention on the most general Keldysh correlator with $p>0$ average operators and $q$ difference operators of the form
\begin{equation}
\begin{split}
 \langle \mathcal{T}_{SK} \SKAv{O}^{(1)} \SKAv{O}^{(2)} \ldots \SKAv{O}^{(p)} \SKRel{O}^{(p+1)} \SKRel{O}^{(p+2)} \ldots \SKRel{O}^{(p+q)} \rangle \,.
\end{split}
\end{equation}
Depending on the  relative time-ordering  of the operators in question, this correlator evaluates in general to a nested commutator or anti-commutator (depending on the statistics),  of the original operators $\{\OpH{O}^{(k)}\}$. To give an explicit formula, we begin by introducing some useful notation for   the  commutators/anti-commutators that
occur in Schwinger-Keldysh correlators. We introduce  the {\em Keldysh bracket} $\KeldBrk{\,\cdot}{\,\cdot\,}$ which \cite{Chou:1984es}
\begin{itemize}
\item takes a single copy (non-Schwinger-Keldysh) operator as its first entry,
\item takes  a Schwinger-Keldysh operator as the second entry, and
\item gives a right or a left product of the single copy operators as the result.
\end{itemize}
More precisely, we have\footnote{ We define the Keldysh bracket with an extra factor of half compared to  \cite{Chou:1984es}; this keeps the formulae simple and  saves us various powers of $2$ later on. }
\begin{equation}
\begin{split}
\KeldBrk{\OpH{A}}{\SKR{B}} &\equiv \OpH{A}\ \OpH{B} \,,\\
\KeldBrk{\OpH{A}}{\SKL{B}} &\equiv   (-)^{\Op{A}\Op{B}}\  \OpH{B}\ \OpH{A} \,.
\end{split}
\label{eq:kelbrk}
\end{equation}
Here  $(-)^{\Op{A}\Op{B}}$ is the extra relative sign that occurs  when both $\OpH{A}$ and $\OpH{B}$ are Grassmann odd, so we just keep track of the mutual Grassmann parity, c.f.,  footnote \ref{fn:Grparity}.

In the Keldysh basis \eqref{eq:KeldyshDef}, the Keldysh bracket evaluates to graded commutators and anti-commutators
\begin{equation}
\begin{split}
\KeldBrk{\OpH{A}}{\SKRel{B}} &\equiv \OpH{A}\ \OpH{B}-  (-)^{\Op{A}\Op{B}}\ \OpH{B}\ \OpH{A} \equiv \gradcomm{\OpH{A}}{\OpH{B}}\,, \\
\KeldBrk{\OpH{A}}{\SKAv{B}} &\equiv   \half\prn{\OpH{A}\ \OpH{B}+ (-)^{\Op{A}\Op{B}}\ \OpH{B}\ \OpH{A} } \equiv  \gradAnti{\OpH{A}}{\OpH{B}} \,.
\end{split}
\label{eq:kelbrkad}
\end{equation}
In particular, if $\OpH{I}$ is the identity operator then we have
\begin{equation}
\KeldBrk{\OpH{I}}{\SKRel{A}} = 0 \,, \qquad \KeldBrk{\OpH{I}}{\SKAv{A}} = \OpH{A}\,.
\label{eq:Iavdif}
\end{equation}

We can now expand  the most general Schwinger-Keldysh correlator  by writing it as a nested Keldysh bracket acting on identity operator and then applying Schwinger-Keldysh time-ordering. At every stage the Schwinger-Keldysh time-ordering is a particular choice of the step functions. One simply sums over all possible orderings of operators inside the nested Keldysh brackets and dresses each of them with the appropriate causal step function. For Schwinger-Keldysh operators $\Op{O}_{1}, \Op{O}_{2}, \cdots \Op{O}_{p}$ we can therefore write
\begin{equation}
\begin{split}
& \langle \mathcal{T}_{SK} \ \Op{O}_{1} \Op{O}_{2} \ldots \Op{O}_{p} \rangle \\
& \qquad = \sum_{\text{time orderings}} \Theta_{\sigma_1 \sigma_2  \cdots  \sigma_p}\;
  \langle  \KeldBrk{\cdots\KeldBrk{\KeldBrk{\OpH{I} }{ \Op{O}_{\sigma_1} } }{\Op{O}_{\sigma_2}}  \cdots}{\Op{O}_{\sigma_p}}
  \rangle  \,,
 \end{split}
 \label{eq:KeldBrkForm}
\end{equation}
where $\sigma_1 \sigma_2\,\cdots \sigma_p$ is a permutation of the $p$ indices. This expression be used to bring any correlation function to a standard form involving commutators and anti-commutators.

Note that we do not need to specify whether the Schwinger-Keldysh operators are averages or differences a-priori, since this is taken care of while evaluating the Keldysh brackets. Indeed, \eqref{eq:Iavdif} explicitly implements the \emph{largest time equation}, which says that the difference operator cannot be futuremost. We will discuss this in greater detail in \S\ref{sec:LargestTime}. For now let us try to see how these rules work in practice for low order correlation functions.

\paragraph{Two-point functions:}
As an easy example and illustration of this procedure, let us use the above algorithm for all four two-point correlators. First, we have
\begin{equation}
\begin{split}
\langle \mathcal{T}_{SK} \SKAv{A} \SKAv{B} \rangle
&=
	 \stepFn{\Op{A}\Op{B}}  \; \langle  \KeldBrk{ \KeldBrk{\OpH{I}}{\SKAv{A}} }{ \SKAv{B}  } \rangle
	 + (-1)^{\Op{A}\Op{B}}\ \stepFn{\Op{B}\Op{A}}  \; \langle
	  \KeldBrk{ \KeldBrk{\OpH{I}}{\SKAv{B}} }{ \SKAv{A}  } \rangle \\
&=
	\stepFn{\Op{A}\Op{B}}  \; \langle  \KeldBrk{ \OpH{A} }{ \SKAv{B}  } \rangle
	+ (-1)^{\Op{A}\Op{B}}\ \stepFn{\Op{B}\Op{A}}  \; \langle  \KeldBrk{ \OpH{B} }{ \SKAv{A}  } \rangle \\
&=
	\stepFn{\Op{A}\Op{B}}  \; \langle  \half \prn{\OpH{A}\, \OpH{B}
		+ (-1)^{\Op{A}\Op{B}}\ \OpH{B} \,\OpH{A} } \rangle
		+ (-1)^{\Op{A}\Op{B}}\ \stepFn{\Op{B}\Op{A}}  \; \langle
	\half \prn{\OpH{B}\,\OpH{A}+ (-1)^{\Op{A}\Op{B}}\ \OpH{A} \,\OpH{B} }\rangle
 \\
&=
	\; \langle  \gradAnti{\OpH{A}}{\OpH{B} } \rangle
\end{split}
\label{eq:2ptaa}
\end{equation}
where we used \eqref{eq:KeldBrkForm} and \eqref{eq:Iavdif}. This then gives the unordered Wightman two-point function.

Similarly, we may further deduce that
\begin{equation}
\begin{split}
\langle \mathcal{T}_{SK} \SKAv{A} \SKRel{B} \rangle
&=
	 (-1)^{\Op{A}\Op{B}} \ \langle \mathcal{T}_{SK}  \SKRel{B} \SKAv{A} \rangle \\
&=
	\stepFn{\Op{A} \Op{B}} \; \langle  \KeldBrk{ \KeldBrk{\OpH{I}}{\SKAv{A}} }{ \SKRel{B}  } \rangle \\
&=
	\stepFn{\Op{A} \Op{B}} \; \langle  \KeldBrk{ \OpH{A} }{ \SKRel{B}  } \rangle \\
&=
	 \stepFn{\Op{A} \Op{B}} \; \langle  \OpH{A}\,\OpH{B}- (-1)^{\Op{A}\Op{B}}\ \OpH{B} \, \OpH{A}  \rangle
=	  \stepFn{\Op{A} \Op{B}} \; \langle\  \gradcomm{\OpH{A}}{\OpH{B}}\ \rangle \,,
\end{split}
\label{eq:2ptad}
\end{equation}
where we obtained the second line by noting that difference operators can never be in the future of average operators.
Swapping the operators and using the step-function identity  \eqref{eq:thetacon}, we find the useful identity
\begin{equation}
\begin{split}
\langle \mathcal{T}_{SK} \prn{ \SKAv{A} \SKRel{B} -  \SKRel{A} \SKAv{B} }\rangle
=   \langle\  \gradcomm{\OpH{A}}{\OpH{B}}\ \rangle\,.
\end{split}
\label{eq:2ptadda}
\end{equation}
In this language the  last correlator, involving two difference operators, is identically zero,
\begin{equation}
\langle \mathcal{T}_{SK} \SKRel{A}\, \SKRel{B} \rangle =0 \,,
\label{eq:2ptdd}
\end{equation}
owing to the first identity in \eqref{eq:Iavdif}.

\paragraph{Three-point functions:}
We can similarly work out the three-point correlators  explicitly using the Keldysh brackets. For example, we have for all average operators a suitable time-ordered permutation of the symmetrized correlation, viz.,
\begin{equation}
\begin{split}
\langle \mathcal{T}_{SK} \SKAv{A}\ \SKAv{B}\ \SKAv{C} \rangle
&=
	 \stepFn{\Op{A}\Op{B}\Op{C}}\; \Big\langle \gradAnti{\gradAnti{\OpH{A}}{\OpH{B}}}{\OpH{C}} \Big\rangle
	+ (-)^{\Op{A}\,\Op{B}} \ \stepFn{\Op{B}\Op{A}\Op{C}}\; \Big\langle \gradAnti{\gradAnti{\OpH{B}}{\OpH{A}}}{\OpH{C}}
	\Big\rangle\\
&\quad
	+\; (-)^{\Op{B}\,\Op{C}} \ \stepFn{\Op{A}\Op{C}\Op{B}}  \; \Big\langle
	\gradAnti{ \gradAnti{\OpH{A}}{\OpH{C}} }{\OpH{B}} \Big\rangle \\
&\quad
	+\;  (-)^{(\Op{A}+\Op{B})\Op{C}} \ \stepFn{\Op{C}\Op{A}\Op{B}}
	\; \Big\langle  \gradAnti{ \gradAnti{\OpH{C}}{\OpH{A}} }{\OpH{B}} \Big\rangle\\
&\quad
	+\; (-)^{\Op{A} (\Op{B}+ \Op{C})} \ \stepFn{\Op{B}\Op{C}\Op{A}} \;
	 \Big\langle  \gradAnti{ \gradAnti{\OpH{B}}{\OpH{C}} }{\OpH{A}} \Big\rangle \\
&\quad
	 +\; (-)^{\Op{A}\, \Op{B}+\Op{A}\,\Op{C}+\Op{B}\,\Op{C}} \ \stepFn{\Op{C}\Op{B}\Op{A}}
	 \; \Big\langle  \gradAnti{ \gradAnti{\OpH{C}}{\OpH{B}} }{\OpH{A}} \Big\rangle
\end{split}
\label{eq:3ptCorrA}
\end{equation}
The correlators with difference operators give rise to commutators owing to \eqref{eq:kelbrkad}. We then have the response functions:
\begin{equation}
\begin{split}
\langle \mathcal{T}_{SK} \SKAv{A}\ \SKAv{B}\ \SKRel{C} \rangle
&=
	\stepFn{\Op{A}\Op{B}\Op{C}}  \Big\langle  \gradcomm{ \gradAnti{\OpH{A}}{\OpH{B}} }{\OpH{C}} \Big\rangle
	+ (-)^{\Op{B}\, \Op{C}} \ \stepFn{\Op{A}\Op{C}\Op{B}}  \Big\langle
	 \gradAnti{ \gradcomm{\OpH{A}}{\OpH{C}} }{\OpH{B}} \Big\rangle
\\
&
	\quad +\; (-)^{\Op{A} \, \Op{B}} \stepFn{\Op{B}\Op{A}\Op{C}}  \Big\langle  \gradcomm{ \gradAnti{\OpH{B}}{\OpH{A}} }{\OpH{C}} \Big\rangle \\
&\quad
	+\; (-)^{\Op{A}\, \Op{B}+\Op{A}\,\Op{C}} \ \stepFn{\Op{B}\Op{C}\Op{A}}  \Big\langle  \gradAnti{ \gradcomm{\OpH{B}}{\OpH{C}} }{\OpH{A}} \Big\rangle
\\
\langle \mathcal{T}_{SK} \SKAv{A}\ \SKRel{B}\ \SKRel{C} \rangle
&=
	 \stepFn{\Op{A}\Op{B}\Op{C}}  \Big\langle  \gradcomm{ \gradcomm{\OpH{A}}{\OpH{B}} }{\OpH{C}} \Big\rangle
	 + (-)^{\Op{B}\, \Op{C}} \ \stepFn{\Op{A}\Op{C}\Op{B}}  \Big\langle  \gradcomm{ \gradcomm{\OpH{A}}{\OpH{C}} }{\OpH{B}} \Big\rangle \\
\langle \mathcal{T}_{SK} \SKRel{A}\ \SKRel{B}\ \SKRel{C} \rangle
&= 0
\end{split}
\label{eq:3ptCorrAD}
\end{equation}
One can similarly work out higher point functions following the basic rule of the Keldysh bracket and its interplay with the Schwinger-Keldysh time-ordering.

\subsection{SK causality}
\label{sec:LargestTime}

By inspection of the Keldysh bracket algorithm presented in the previous subsection, we immediately  note the following causality property: if the future-most operator in the Schwinger-Keldysh correlator is a difference operator, then the correlator vanishes owing to the fact noted in \eqref{eq:Iavdif}:
\begin{equation}
\begin{split}
\text{If } t_n > t_k  \text{ for all } k=1,\ldots,n-1,  \text{ then:} \qquad
\langle \mathcal{T}_{SK} \; \SKRel{O}^{(n)} \,\prod_{k=1}^{n-1} \Op{O}_{I_k}^{(k)}   \rangle  = 0 \,.
\end{split}
\label{eq:LargestTime}
\end{equation}
for all choices $I_k \in \{av,\,dif\}$. We refer to this statement as the largest time equation following \cite{tHooft:1973pz}. In that discussion the largest time equation refers to a statement about cutting rules in computing Feynman amplitudes. As we explain in \S\ref{sec:applications} their discussion can also be efficiently worded in the current language.

The relation \eqref{eq:LargestTime} follows from the fact that for any operator $\OpH{A}$ we have the basic statement $\KeldBrk{\OpH{I}}{\SKRel{A}} = 0$.  Heuristically, the above `causality rule' can be thought of as the requirement that the state obtained by slicing the Schwinger-Keldysh contour at its turn-around point is annihilated by the difference operators.  The Schwinger-Keldysh construction ensures this by requiring that the future-most state be the maximally entangled (cat) state between the right and the left copies (see, e.g., \cite{Rangamani:2014isa} ).
As an aside, note that the basic identity \eqref{eq:diff0} is a special case of \eqref{eq:LargestTime} when all insertions are difference operators (i.e., $I_k=dif$ for all $k$).

The vanishing of correlators whose future-most insertion is a difference operator, Eq.~\eqref{eq:LargestTime}, can also easily be inferred from the defining Schwinger-Keldysh path integral \eqref{eq:ZSKdef}. To see this, let us assume that all operator insertions of a given correlation function lie in the time interval $[t_i,t_f]$ and also denote the time of the latest average operator insertion by $t_\text{top} \, (\leq t_f)$.We can always decompose the unitary implementing the evolution
as  an ordered sequence, i.e.,
\begin{equation}
U[{\cal J}] = U[{\cal J}, t_f - t_\text{top}] U[{\cal J}, t_\text{top}-t_i]  \,.
\label{}
\end{equation}
Thus one may write the generating functional \eqref{eq:ZSKdef} with the unitary evolution split across $t=t_\text{top}$ as
\begin{equation}
\mathscr{Z}_{SK}[{\cal J}_\skR,{\cal J}_\skL] = \Tr{\ U[{\cal J}_\skR, t_f - t_\text{top}] U[{\cal J}_\skR, t_\text{top}-t_i]\ \rhoi\  (U[{\cal J}_\skL, t_\text{top}-t_i])^\dag(U[{\cal J}_\skL, t_f - t_\text{top}])^\dag\ } \, ,
\label{eq:LargeTime}
\end{equation}

By assumption there are no average operator insertions after time $t_\text{top}$. Therefore,   for the purpose of calculating the a correlation function with a difference operator inserted at $t > t_\text{top}$ we can simply align the sources ${\cal J}_\skR = {\cal J}_\skL$ for times $t>t_\text{top}$. But in this alignment limit, the two outermost evolution operators in \eqref{eq:LargeTime} will cancel by cyclicity of the trace. As a result all dependence on sources drops out for times
$t>t_\text{top}$. Therefore any difference operator insertion for these late times (i.e., functional differentiation w.r.t.\ ${\cal J}_{av}$ for times $t>t_\text{top}$) will lead to a vanishing correlator.

This causality rule can be remembered by the mnemonic:
\begin{quote}
\emph{\textbf{R}elative (i.e., difference) operators should be \textbf{R}etarded in time and \textbf{A}verage operators  should be \textbf{A}dvanced in time.}
  \end{quote}
We remind the reader that the corresponding rule for sources is the opposite, owing to the Lorentzian inner product in the $\text{R}-\text{L}$ space. This leads to the following mnemonic for the sources
\begin{quote}
\emph{\textbf{A}verage sources should  be \textbf{R}etarded in time and  \textbf{R}elative sources should be \textbf{A}dvanced in time.}
\end{quote}

One immediate consequence of the above discussion is that for a $n$-point correlation function with one average and $(n-1)$ difference operators, all of which are inserted to the past of the  average  operator, i.e.,
\begin{equation}
\begin{split}
\langle \mathcal{T}_{SK}\ \SKAv{O}^{(n)}  \prod_{k=1}^{n-1} \SKRel{O}^{(k)}  \rangle
\end{split}
\end{equation}
gives the advanced Green's function.\footnote{ This is the  reason the difference operators are called advanced  operators in the retarded-advanced (RA) basis as we shall see in \S\ref{sec:skthemal}.}  Using the Keldysh bracket rules one can work out that this correlation function is a given by a sequence of nested commutators, see e.g., Eqs.~\eqref{eq:2ptad} and \eqref{eq:3ptCorrAD} for explicit expressions.

\section{Thermal correlation functions in Schwinger-Keldysh formalism}
\label{sec:skthemal}

Our discussion thus far has focused on an initial density matrix $\rhoi$ which was arbitrary. The initial state of the quantum system is mainly setting up for us an appropriate entanglement pattern for the degrees of freedom in ${\cal H}$. With this information we can only go as far as the discussion in \S\ref{sec:skbasics}.

However, not all density matrices are created equal, with some being more special than others. In what follows we will switch our focus on to thermal density matrices which enjoy some nice properties. To understand these, let us start by considering a QFT at finite temperature $T$. Should our theory contain some global symmetries we can also include some chemical potentials. One thus is considering the state of the system to be  a Gibbs density matrix, which gives the probabilities to find states with a given energy and charge:
\begin{equation}
\rhoT = e^{-\beta\, \left(\OpH{H} -\mu_{_I}\, \OpH{Q}^{\scriptsize{I}} \right)}
\label{}
\end{equation}
Here $\OpH{H}$ is the Hamiltonian for the quantum theory and $\OpH{Q}$ the flavour charge operator.
We have chosen not to normalize the density matrix;  the trace over the states then gives us the thermal partition function
\begin{equation}
\mathscr{Z}_{_T}(\beta, \mu_{_I}) = \Tr{\rhoT }
\label{eq:thermalZ}
\end{equation}

Usually one discusses thermal field theories in Minkowski spacetime ${\mathbb R}^{d-1,1}$. One furthermore, makes heavy use of the connection between thermal quantum field theories in $d$-spacetime dimensions and classical statistical mechanics in $(d-1)$ dimensions by realizing the operator $\rhoT$ as performing Hamiltonian evolution in imaginary time $t_{_E}$ by an amount set by the inverse temperature $\beta$. The role of the chemical potential then is to twist the charge fields by an  amount  set by the charge as they are taken around this imaginary Euclidean time.\footnote{ In classical statistical mechanics, the operator $\rhoT$ serves to determine the transfer matrix and the only information necessary to determine it are the Boltzmann weights, which give the relative probabilities for the occurrence of various energy levels.}

With this information we are now ready to understand the thermal boundary conditions implicit in $\rhoT$.  For any single-copy operator lying on the initial time slice $\Sigma_{\cal M}$ we require that the Kubo-Martin-Schwinger (KMS) periodicity condition \cite{Kubo:1957mj,Martin:1959jp}, be satisfied.\footnote{ This condition was first discussed independently in  papers by Kubo \cite{Kubo:1957mj} and by Martin-Schwinger \cite{Martin:1959jp}. However, the name was coined a bit afterward by Haag et.~al., \cite{Haag:1967sg} who applied this idea in the context of defining equilibrium configurations in axiomatic QFT.} The KMS condition says that bosonic operators are periodic under traversal of the thermal circle while fermionic operators are anti-periodic. We will now try to capture this information in a covariant form that will be useful in the sequel.

\subsection{Thermal equilibrium in  stationary curved spacetimes}
\label{sec:styT}

Insofar as thermal equilibrium is concerned, all one requires is that the system be stationary -- one does not require a globally constant temperature or chemical potentials. To allow for local temperature and chemical potential variations one can consider the system not on the flat Minkowski background, but rather on a curved spacetime, equipped with a timelike Killing vector field. Likewise the chemical potentials may be chosen to vary across space by turning on a background electromagnetic field. This idea of exploring thermal dynamics by turning on time-independent background sources has a rich history (see eg., \cite{Luttinger:1964zz}), but a systematic analysis of the subject has been undertaken recently in \cite{Banerjee:2012iz,Jensen:2012jh}.\footnote{ See also \cite{Banerjee:2012cr,Jensen:2012jy,Loganayagam:2012zg,Jensen:2012kj,Jensen:2013kka,Jensen:2013rga} for an application of these ideas to understand  anomaly induced effects in thermal physics. }

To explore this more general situation, it is efficacious to consider our quantum system residing on a background $\mathcal{M}_d $ with a non-trivial classical background metric and gauge field sources. Given the timelike Killing vector field $K^\mu$, we can adapt coordinates to it by choosing
\begin{equation}
K^\mu = \left(\frac{\partial}{\partial t}\right)^\mu
\label{}
\end{equation}
so that the background geometry can be brought to the Kaluza-Klein form:
\begin{equation}
ds^2 = -e^{2\,\sigma(x^m)} \left(dt + a_i(x^m)\, dx^i \right)^2 + \gamma_{ij}(x^m) \, dx^i\, dx^j\,, \qquad  \form{A} = A_0(x^m)\, dt + A_i(x^m)\, dx^i\,.
\label{}
\end{equation}
We have allowed the Killing field to not be hypersurface-orthogonal, as is the case with $a_i \neq 0$, but will demand that it be timelike globally (i.e., require that the background be free of ergosurfaces).

For the connection with the classical statistical mechanics, we can still exploit the analytic continuation to imaginary Euclidean time. To construct the corresponding manifold, we identify every point $p \in {\cal M}_d$  with a  point $p'$ in its
future, separated from it by a unit affine distance along the vector $K^\mu$. In other words points $p$ and $p'$ get identified if there exists a curve $x^\mu(\tau)$ parameterized by $\tau$ such that
\begin{equation}
\begin{split}
x^\mu(\tau=0) = p^\mu\ ,\qquad
x^\mu(\tau=1) = p'{}^\mu \ ,\qquad
\frac{dx^\mu}{d\tau} = \KEq^\mu \,.
\end{split}
\end{equation}
What this does is to construct from our original Lorentzian manifold ${\cal M}_d$, a corresponding Euclidean spacetime ${\cal M}_E$ which is endowed with a fibre bundle structure. The fibres are the Euclidean time circle, parameterized by $t_{_E} = i\, t$ which is fibered over a spacelike base $\Sigma_{\cal M}$. While we have discussed the construction for the geometry, a similar statement can be made for the flavour bundle, by identifying the fibres at $p$ and $p'$ up to a gauge transformation $\Lambda_K$.

Given this bundle structure in the Euclidean spacetime relevant for thermal equilibrium, it pays to work covariantly and  characterize the thermal Gibbs density matrix, not by a temperature and a chemical potential, but rather by a timelike inverse temperature vector $\Kbeta^\mu$ and a flavour gauge parameter $\LambdaB$. For the thermal equilibrium configurations that we have in mind here, these parameters can be identified with $\{\Kbeta^\mu,\LambdaB\} = \{K^\mu,\Lambda_K\}$, as constructed above.
Following  \cite{Haehl:2014zda} we refer to these quantities as the thermal vector and thermal twist respectively. The thermal vector sets up both the local inertial frame for the equilibrium configuration and the period of the local thermal (Euclidean time) circle. The thermal twist provides the correct boundary conditions for charged particles, ensuring that they pick up the right monodromy as they go around the thermal circle.

\subsection{The KMS condition}
\label{sec:kms}

We now have the necessary background to set up the KMS condition, which talks about the periodicity properties of operators (and thus their correlators) around the thermal circle in general. We will first discuss the KMS condition for a single copy theory and thence pass onto the  doubled Schwinger-Keldysh formalism.

The KMS condition asserts that  the thermal equilibrium correlation functions are periodic in Euclidean time. As such these  are non-local conditions on thermal Schwinger-Keldysh correlators which ensure that they are related to Euclidean correlators by analytic continuation. The scale of non-locality is simply the thermal scale, for we are essentially comparing operators that are related by a Euclidean time translation. While physically the periodicity in imaginary time is the essential content, it is useful to take a formal perspective, see e.g., \cite{Haag:1967sg,Araki:1968aa}.

Formally, the KMS conditions encode a certain analyticity property of Euclidean correlation functions. Usually they are stated as the requirement that the Green's functions obtained in the Euclidean theory are analytic in a strip in the complex time plane. Define $t_\mathbb{C} = t+ i \, t_\text{E}$ to be a complex time coordinate, which one may view as the coordinate in the complex plane on which the Schwinger-Keldysh contour is defined. The KMS condition requires that the Green's functions are analytic in the strip $\{ t_\mathbb{C } \in \mathbb{C}: \;\; 0 < \Im (t_\mathbb{C}  ) < \beta \} $.  This then requires that the two-point functions of  Heisenberg operators $\OpH{A}(t)$ and $\OpH{B}(t)$ which are elements of the algebra of observables   obey the periodicity condition
\begin{equation}
\Tr{\rhoT\, \OpH{A} (t -i\,\beta)\, \OpH{B}(0) } = \Tr{\rhoT\, \OpH{B}(0)\, \OpH{A} (t) }\,,
\label{eq:therkms}
\end{equation}
for bosonic operators $\OpH{A}$ and $\OpH{B}$ (below, we will also generalize to fermionic operators).
We used here conjugation of $\OpH{A}$ by the density matrix operator $\rhoT$, viz.,
\begin{align}
\OpH{A}(t-i\beta) = \rhoT^{-1} \, \OpH{A}(t) \, \rhoT
\label{eq:kmsconj}
\end{align}
 and cyclicity of the trace.  Motivated by this observation we will now define a notion of \emph{KMS conjugate} of an operator which will allow us to move the operator around the thermal circle.

These formal set of statements  are often used to define the notion of a KMS state from an algebraic QFT viewpoint. For a recent discussion see \cite{Gransee:2015aba}; these authors go on to discuss the notion of a local KMS state which meshes well with some of our earlier discussion of local equilibrium and hydrodynamics in \cite{Haehl:2015foa}. Another useful reference discussing the KMS condition and discrete symmetries which we found useful is \cite{Sieberer:2015hba}; we will comment on the relations with their definition of the KMS transformation in \S\ref{sec:Qkms}.

For simplicity, consider the case of thermal physics in ${\mathbb R}^{d-1,1}$, where we can relate the thermal shift along the Euclidean time circle as Hamiltonian evolution. One can then re-express \eqref{eq:kmsconj} using the explicit form of the density matrix:
\begin{equation}\label{eq:BetaShift}
\OpH{O}(t-i\beta) =
	e^{\beta\, \left(\OpH{H} -\mu_{_I}\, \OpH{Q}^{\scriptsize{I}} \right)} \; \OpH{O}(t) \,
	e^{-\beta\, \left(\OpH{H} -\mu_{_I}\, \OpH{Q}^{\scriptsize{I}} \right)} \,.
\end{equation}
We will find it convenient to think about this evolution as being achieved by a classical differential operator carrying out the time translation, viz., $\beta\, \frac{d}{dt}$, so that we can write
\begin{equation}
e^{-i \beta \frac{d}{dt}} \, \OpH{O}(t) \equiv \OpH{O}(t-i\beta)\,.
\label{eq:themalddt}
\end{equation}
We have chosen here to invert the order of conjugation to conform with the contour ordering we introduced earlier. The transformation above is appropriate for Schwinger-Keldysh left fields while the earlier definition is more natural for the right operators (see below).

This admits an easy generalization to the general situation of curved manifolds with timelike Killing field, as discussed in \S\ref{sec:styT}. We introduce a \emph{thermal time translation operator} $\deltaB$ whose task is to take any operator in the quantum system and translate it around the Euclidean circle. While we have indicated operators as $\OpH{O}$ without any decorations to indicate the Lorentz transformation properties, it should be clear that we need to keep track of the latter appropriately on curved spacetimes. To this end,  $\deltaB$ should Lie drag any quantum operator $\OpH{O}$ around the thermal circle, for such an action would be appropriately Lorentz covariant. We therefore declare
 $\deltaB$ be the Lie-derivative corresponding to the diffeomorphism and flavor transformation generated by $\{\Kbeta^\mu,\LambdaB\}$, so that
\begin{equation}
e^{-i\deltaB}\, \OpH{O}(t) \equiv e^{\beta\, \left(\OpH{H} -\mu_{_I}\, \OpH{Q}^{\scriptsize{I}} \right)} \; \OpH{O}(t) \,
	e^{-\beta\, \left(\OpH{H} -\mu_{_I}\, \OpH{Q}^{\scriptsize{I}} \right)} \,,
\label{eq:delBdef}
\end{equation}
continues to hold on curved backgrounds in global thermal equilibrium. Whilst as we have defined it, $\deltaB$ is a classical differential operator, we should  alert the reader not to view it  simply as a random diffeomorphism in the Euclidean time direction. It is a specific one that is solely determined by the boundary conditions given in the thermal density matrix $\rhoT$. Being thus a feature of the Gibbs density matrix, it is best thought of as a state-dependent (thermal) time translation.\footnote{ One necessary consequence of this fact is that the Noether charge associated with this thermal time translation is not the energy, but rather the entropy, cf., \cite{Haehl:2015pja} and comments in \cite{Haehl:2015foa,Haehl:2015uoc}.}

Now that we know how to relate operators around the Euclidean thermal circle, we can write down the KMS condition as the following statement:
\begin{equation}
\FSgn{O} e^{-i\deltaB} \OpH{O}\prn{t=t_i}\; \longrightarrow \; \OpH{O}\prn{t=t_i}  \,,
\label{eq:kms}
\end{equation}
where we now also account for the fermion number $\FSgn{O}$  of the operator ${\OpH O}$ introduced in \eqref{eq:FOdef}. Note that this replacement rule holds inside correlation functions since $t_i$ is the initial time. As explained in the discussion above,  $e^{-i\deltaB}$  time translates the operator in the negative imaginary time by $\Kbeta^\mu$ and then gives a flavor twist $\LambdaB$.
Hence at an operator level the KMS condition is simply saying that operators that are related by appropriate amount of Lie drag around the thermal circle are equivalent. This operator identity then asserts the thermal periodicity of the correlation functions envisaged in \cite{Kubo:1957mj,Martin:1959jp}.

It is convenient to further define  a derivative operator
\begin{equation}
i \delKMS \equiv 1-\FSgn{} e^{-i\deltaB} \,,
\label{eq:DelB}
\end{equation}
which measures the deviation from the KMS condition.
We can then  write the KMS condition as  a  differential statement, viz.,
\begin{equation}
\delKMS  \OpH{O} \; \longrightarrow \; 0 \,,
\label{eq:kmsDel}
\end{equation}
which, again, holds for initial time operator insertions in correlation functions. 

Let us further define a Grassmann-even thermal translation operator $\Qbeta$ via
\begin{equation}
\gradcomm{\Qbeta}{ \OpH{O}} = \delKMS{\OpH{O}}\,.
 \label{eq:Qbetadef}
\end{equation}
At present this formally defines an operator $\Qbeta$ which acts on the operator algebra as defined above. We will later find that this operator naturally fits into a larger algebraic structure as we shall unearth in due course.

As explained in \S\ref{sec:torder} the operator  $ \FSgn{O}  $ appearing in the KMS condition, Eq.~\eqref{eq:kms}  imposes the periodic or the anti-periodic boundary condition depending on whether the observable is bosonic or fermionic. By spin-statistics relation, the fermion parity  $ \FSgn{O}  $ is same as the Grassmann parity $\GSgn{O}$ for the physical fields. Below we will also work with ghost fields
whose fermion parity  $ \FSgn{O}  $ is opposite to their  Grassmann parity $\GSgn{O}$. This corresponds to the familiar fact
that the thermal boundary conditions for ghosts are determined by the physical operators they are related to, irrespective of their Grassmann parity. BRST ghosts of QCD, for example, are given periodic boundary conditions.

\subsection{Thermal sum rules from KMS}
\label{sec:thermalsum}

Having understood the KMS conditions for a single copy theory,  let us examine how KMS conditions appear in the Schwinger-Keldysh correlators. If we take the initial state to be a  Gibbs density matrix $\rho_i = \rhoT$, we can introduce a set of KMS conjugate operators
\begin{equation}
\SKL{\tilde{O}}\equiv \FSgn{O}  e^{-i\deltaB} \SKL{O} \,,
\label{eq:kmsconjL}
\end{equation}
which are thermal time-translates of the operator in question.
We could likewise also introduce in a similar fashion the analytically continued sources,
\begin{align}
\tilde{{\cal J}}_\skL \equiv \FSgn{\cal J} e^{-i\deltaB} {\cal J}_\skL \,,
\end{align}
and  similarly define analogous conjugations  for the Grassmann odd counterparts. In equilibrium the KMS condition guarantees us that we can replace $\{\SKL{O}, {\cal J}_\skL\} \;\to\; \{ \SKL{\tilde{O}}, \tilde{{\cal J}}_\skL \}$ and the physical correlation functions remain invariant.


The KMS conditions translate into a set of sum rules for the Schwinger-Keldysh theory \cite{Weldon:2005nr}:
\begin{equation}
\begin{split}
\langle \mathcal{T}_{SK} \prod_{k=1}^{n} \left(\SKR{O}^{(k)} -
\SKL{\tilde{O}}^{(k)}\right)\rangle = 0\,.
\end{split}
\label{eq:diff1}
\end{equation}
This says that correlation function of  differences  of right operators $\SKR{O} $ and the KMS conjugate of left operators $\SKL{O}$ (denoted $\SKL{\tilde{O}}$) vanish. 
One can check that this statement is compatible with our earlier statement phrased in terms of two-point functions \eqref{eq:therkms}. The general statement may of course be derived directly from there, but the cleanest statement is worded in terms of thermal sum rules.

Similarly, we have the analogue of the causality condition \eqref{eq:LargestTime}: a correlation function vanishes if the past-most insertion is a `time-twisted' difference operator $\SKR{O} - \SKL{\tilde{O}}$. This follows from the fact that $\SKR{O} - \SKL{\tilde{O}}$ annihilates the thermal density matrix.

\subsection{The retarded-advanced basis}
\label{sec:retadv}

One consequence of the KMS condition which relates operators related by a thermal translation, is that one expects the set of identities \eqref{eq:diff1} hold in correlation functions.
These sum rules which have been derived for example in \cite{Weldon:2005nr} can be succinctly stated by working in yet another basis of operators.  This new basis is called the \emph{retarded-advanced} basis, which is sometimes also referred to as the RA basis.\footnote{ As noted after Eq.~\eqref{eq:KeldyshJ}, the Keldysh basis itself in some circles is referred to as the $ra$ basis. We understand that this nomenclature originates from some historical confusion about the connections between the two bases. We will avoid this confusion altogether by sticking to the usage of `retarded-advanced' basis.} It is defined by the linear combination of the Schwinger-Keldysh operators , $\SKR{O}, \SKL{O}$ and their KMS shifted counterparts $\SKL{\tilde{O}}$.  Without loss of generality we make the choice:
\begin{equation}\label{eq:RADef}
\begin{split}
\SKAdv{O} \equiv  \SKR{O}-\SKL{O}\ , \qquad
\SKRet{O} &\equiv \frac{1}{1-\FSgn{O} e^{-i\deltaB} }  \prn{ \SKR{O}-\FSgn{O} e^{-i\deltaB} \SKL{O} } .
\end{split}
\end{equation}
Note that the retarded operator $\SKRet{O}$ is  defined with an inverse of $\delKMS$, so it should actually be thought  of as a solution to the differential equation
\begin{equation}
\begin{split}
 i\delKMS\SKRet{O} &=\SKR{O}-\FSgn{O} e^{-i\deltaB} \SKL{O} \,.
\end{split}
\end{equation}
which is solved with some initial  condition. We will choose our initial conditions to be
\begin{equation}
\begin{split}
\SKRet{O}(t=t_i) &=\SKR{O}(t=t_i)= \SKL{O}(t=t_i) = \OpH{O}(t=t_i) \,,\\
\SKAdv{O}(t=t_i) &=\SKR{O}(t=t_i)- \SKL{O}(t=t_i) = 0\ .
\end{split}
\end{equation}

It is a common practice to explicitly include the statistics of the operator in question in the definition. Recall that, for thermal correlation functions we should include the correct distribution function for bosons or fermions (which follows in turn from the periodicity conditions). This may be done by introducing another differential operator corresponding to Bose-Einstein or Fermi-Dirac distribution
\begin{equation}\label{eq:fDef}
\begin{split}
\fbeta \equiv \frac{1}{e^{i\deltaB}-\FSgn{}} \,.
\end{split}
\end{equation}
In terms of $\fbeta$ we can then write:
\begin{equation}\label{eq:RAfDef}
\begin{split}
\SKAdv{O} &\equiv  \SKR{O}-\SKL{O}\ , \\
\SKRet{O} &\equiv  \prn{1+\FSgn{O} \, \fbeta }\SKR{O}-\FSgn{O} \, \fbeta \, \SKL{O} \\
&= \SKAv{O}+\prn{\half+\FSgn{O} \fbeta }\SKRel{O}\ .
\end{split}
\end{equation}
These definitions can be then inverted to give right and left operators in terms of the retarded-advanced basis to be
\begin{equation}\label{eq:RLinRA}
\begin{split}
\SKR{O} &=  \SKRet{O}-\FSgn{O} \fbeta \SKAdv{O}\ , \\
\SKL{O} &= \SKRet{O}- \prn{1+\FSgn{O} \fbeta }\SKAdv{O}\,.
\end{split}
\end{equation}
Now that we have the explicit mapping, we can show that the Lorentzian inner product between the source and operator in the Schwinger-Keldysh construction still goes over into a non-diagonal (light-cone like) inner product between the $\SKRet{O}$ and $\SKAdv{O}$. A simple algebra leads to
\begin{equation}\label{eq:RAJ}
\begin{split}
\int \SKR{\mathcal{J}}\  \SKR{O} - \SKL{\mathcal{J}}\ \SKL{O}  =
\int \SKAdv{\mathcal{J}}\ \SKRet{O} + \SKRet{\mathcal{J}}\ \SKAdv{O} + \text{boundary contribution}
\end{split}
\end{equation}
up to some boundary contributions at the initial and final time slice. The extra contribution can be shown to take the form
\begin{equation}
\begin{split}
\text{boundary contribution}  = \int  &\prn{\frac{\FSgn{}e^{-i\deltaB}}{1-\FSgn{}e^{-i\deltaB}} \SKAdv{\mathcal{J}} } \prn{\frac{\FSgn{}e^{-i\deltaB}}{1-\FSgn{}e^{-i\deltaB}} \SKAdv{O} } \\
&- \int  \prn{ \frac{1}{1-\FSgn{}e^{-i\deltaB}} \SKAdv{\mathcal{J}} }\prn{ \frac{1}{1-\FSgn{}e^{-i\deltaB}} \SKAdv{O} }\,,
\end{split}
\end{equation}
which is zero by a  change of variables up to  contributions from the initial and the final time slice. Unfortunately, as far as we are aware,  the physical import of these boundary contributions (involving  $\delKMS^{-1}$ acting on advanced fields)  has not yet been studied much in the literature. We will ignore these boundary terms in what follows, but the reader should be alert to this fact.  Modulo this subtlety, as promised, in the retarded-advanced basis, the retarded source couples to the advanced operator and vice versa.

\subsection{Retarded-advanced correlators}
\label{sec:RetAdvCorr}

Using the definition \eqref{eq:RADef}, we can now compute correlation functions in the retarded-advanced basis by simply reverting back to our previous results in the average-difference basis, c.f., \S\ref{sec:keldysh}. For example, we find for two-point correlators the following relations:
\begin{equation}
\begin{split}
 \langle \mathcal{T}_{SK} \,\SKRet{A}\ \SKRet{B} \rangle
 &=
 	\Big\langle \gradAnti{ \OpH{A}}{\OpH{B}}\Big\rangle + \stepFn{\Op{A}\Op{B}} \,
 	\Big\langle \gradcomm{\OpH{A}}{\prn{\half+\FSgn{B} \fbeta }\OpH{B}} \Big\rangle
 \\
 &\qquad \qquad +
 	\; (-)^{\Op{A}\Op{B}}\, \stepFn{\Op{B}\Op{A}}  \, \Big\langle \gradcomm{\OpH{B}}{\prn{\half+\FSgn{A} \fbeta }\OpH{A}} \Big\rangle\,,
 \\
  \langle \mathcal{T}_{SK} \,\SKAdv{A}\ \SKRet{B} \rangle
  &=
  	(-)^{\Op{A}\Op{B}}\, \stepFn{\Op{B} \Op{A}}\, \Big\langle \gradcomm{\OpH{B}}{\OpH{A}} \Big\rangle\,,
  \\
  \langle \mathcal{T}_{SK} \,\SKRet{A}\ \SKAdv{B} \rangle
  &=
  	\stepFn{\Op{A} \Op{B}} \, \Big\langle \gradcomm{\OpH{A}}{\OpH{B}} \Big\rangle\,,
  \\
   \langle \mathcal{T}_{SK} \,\SKAdv{A}\ \SKAdv{B} \rangle &= 0\,,
\end{split}
\label{eq:retretCorr}
\end{equation}
where we use the fact that the Keldysh bracket \eqref{eq:kelbrkad} acts on $\fbeta  \SKRel{O}$ in the same way as on $\SKRel{O}$:
\begin{equation}
\KeldBrk{ \OpH{A}}{\fbeta  \SKRel{B}} = \gradcomm{\OpH{A}}{\fbeta \OpH{B}}\,.
\end{equation}
It is useful to write the expression for the fully retarded two-point correlator in terms of a thermally deformed anti-commutator. Let
\begin{equation}
\gradAnti{\Op{A}}{\Op{B}}^\beta  \equiv \gradAnti{\Op{A}}{\Op{B}} +
\stepFn{\Op{A}\Op{B}}  \gradcomm{\Op{A}}{\left(\frac{1}{2} + \FSgn{B} \, \fbeta \right) \Op{B}}+
(-)^{\Op{A}\, \Op{B}}\stepFn{\Op{B}\Op{A}}  \gradcomm{\Op{B}}{\left(\frac{1}{2} + \FSgn{A} \, \fbeta \right) \Op{A}} \,,
\label{}
\end{equation}
which continues to satisfy $\gradAnti{\Op{A}}{\Op{B}}^\beta = (-)^{\Op{A}\Op{B}} \gradAnti{\Op{B}}{\Op{A}}^\beta$.
Then we can write the first expression in \eqref{eq:retretCorr} as
\begin{equation}
\begin{split}
 \langle \mathcal{T}_{SK} \,\SKRet{A}\ \SKRet{B} \rangle
 &=
 	\Big\langle \gradAnti{ \OpH{A}}{\OpH{B}}^\beta \Big\rangle\,.
\end{split}
\end{equation}
that is, it takes the same form as $ \langle \mathcal{T}_{SK} \,\SKRet{A}\ \SKRet{B} \rangle$, with a simple replacement of commutators: $\{\,\cdot\,,\,\cdot\,\} \rightarrow\{\,\cdot\,,\,\cdot\,\}^\beta$.
The presence of the thermal anti-commutator implies that only the fluctuations over and above the thermal state are monitored by the above correlation function.\footnote{ It is easy to check for a harmonic oscillator: $\gradAnti{a}{a^\dag}^\beta = a^\dag\,a - \fbeta$.}

This pattern continues: all the higher $adv-ret$ correlation functions can be obtained from $dif-av$ correlators by replacing graded anti-commutators by graded {\it thermal} anti-commutators.
For example, one can check that the three-point functions involving advanced operators take the same form as \eqref{eq:3ptCorrA} and \eqref{eq:3ptCorrAD} with this particular replacement rule:
\begin{equation}
\begin{split}
\langle \mathcal{T}_{SK} \SKRet{A}\ \SKRet{B}\ \SKRet{C} \rangle
&=
	 \stepFn{\Op{A}\Op{B}\Op{C}}\; \Big\langle \gradAnti{\gradAnti{\OpH{A}}{\OpH{B}}^\beta}{\OpH{C}}^\beta \Big\rangle
	+ (-)^{\Op{A}\,\Op{B}} \ \stepFn{\Op{B}\Op{A}\Op{C}}\; \Big\langle \gradAnti{\gradAnti{\OpH{B}}{\OpH{A}}^\beta}{\OpH{C}}^\beta
	\Big\rangle\\
&\quad
	+\; (-)^{\Op{B}\,\Op{C}} \ \stepFn{\Op{A}\Op{C}\Op{B}}  \; \Big\langle
	\gradAnti{ \gradAnti{\OpH{A}}{\OpH{C}}^\beta }{\OpH{B}}^\beta \Big\rangle \\
&\quad
	+\;  (-)^{(\Op{A}+\Op{B})\Op{C}} \ \stepFn{\Op{C}\Op{A}\Op{B}}
	\; \Big\langle  \gradAnti{ \gradAnti{\OpH{C}}{\OpH{A}}^\beta }{\OpH{B}}^\beta \Big\rangle\\
&\quad
	+\; (-)^{\Op{A} (\Op{B}+ \Op{C})} \ \stepFn{\Op{B}\Op{C}\Op{A}} \;
	 \Big\langle  \gradAnti{ \gradAnti{\OpH{B}}{\OpH{C}}^\beta }{\OpH{A}}^\beta \Big\rangle \\
&\quad
	 +\; (-)^{\Op{A}\, \Op{B}+\Op{A}\,\Op{C}+\Op{B}\,\Op{C}} \ \stepFn{\Op{C}\Op{B}\Op{A}}
	 \; \Big\langle  \gradAnti{ \gradAnti{\OpH{C}}{\OpH{B}}^\beta }{\OpH{A}}^\beta \Big\rangle\\
\langle \mathcal{T}_{SK} \,\SKRet{A}\ \SKRet{B}\ \SKAdv{C} \rangle &=
	\stepFn{\Op{A}\Op{B}\Op{C}}  \Big\langle
	 \gradcomm{ \gradAnti{\OpH{A}}{\OpH{B}}^\beta }{\OpH{C}}  	 \Big\rangle
	+\; (-)^{\Op{B}\, \Op{C}} \ \stepFn{\Op{A}\Op{C}\Op{B}}  \Big\langle
	 \gradAnti{ \gradcomm{\OpH{A}}{\OpH{C}} }{\OpH{B}}^\beta
	 \Big\rangle
\\
&\quad
	 +\; (-)^{\Op{A} \, \Op{B}} \stepFn{\Op{B}\Op{A}\Op{C}}  \Big\langle
	  \gradcomm{ \gradAnti{\OpH{B}}{\OpH{A}}^\beta }{\OpH{C}}
	 \Big\rangle
	 +\; (-)^{\Op{A}(\Op{B}+\Op{C})} \ \stepFn{\Op{B}\Op{C}\Op{A}}  \Big\langle
	 \gradAnti{ \gradcomm{\OpH{B}}{\OpH{C}} }{\OpH{A}}^\beta
	 \Big\rangle \\
\langle \mathcal{T}_{SK} \,\SKRet{A}\ \SKAdv{B}\ \SKAdv{C} \rangle
&=
	\stepFn{\Op{A}\Op{B}\Op{C}}  \Big\langle
	 \gradcomm{ \gradcomm{\OpH{A}}{\OpH{B}} }{\OpH{C}}
	 \Big\rangle
	 + (-)^{\Op{B}\, \Op{C}} \ \stepFn{\Op{A}\Op{C}\Op{B}}  \Big\langle
	   \gradcomm{ \gradcomm{\OpH{A}}{\OpH{C}} }{\OpH{B}}
	   \Big \rangle\,,\\
\langle \mathcal{T}_{SK} \,\SKAdv{A}\ \SKAdv{B}\ \SKAdv{C} \rangle &= 0 \,.
\end{split}
\end{equation}
%

\subsection{The thermofield double}
\label{sec:thermofield}

The astute reader will note that we have so far refrained from discussing another commonly used framework for studying thermal correlation functions, viz., the thermofield double. As in the general Schwinger-Keldysh framework we use the fact that the thermal density matrix can be expressed as a pure entangled state in the ${\cal H}_\skR\otimes {\cal H}_\skL^*$ space. In particular, the thermofield double state, or the Hartle-Hawking state as it is known in some circles, has a very simple expression in the energy eigenbasis of the system. If we let $\{\ket{\mathfrak{r}_a}\}$ and $\{\ket{\mathfrak{l}_a}\}$ to be an energy eigenbasis of ${\cal H}_\skR$ and ${\cal H}_\skL$ respectively, then we have the state being given as\footnote{ We are schematically writing the state here with a notation that suggests a discrete spectrum of energy levels. The generalization to a continuum spectrum is straightforward.}
\begin{equation}
\ket{\Psi_\beta}  =\frac{1}{\sqrt{\mathscr{Z}_{_T}(\beta)}}  \;  \sum_{a}\, e^{-\frac{\beta}{2}\, E_a} \, \ket{ \mathfrak{r}_a \, \mathfrak{l}_a }\,.
\label{eq:hhpsi}
\end{equation}
Tracing out the left degrees of freedom leads to the desired thermal density matrix for the right fields as can be easily verified.

The fact that we have a relative Boltzmann weighting of the states by the energy $ e^{-\frac{\beta}{2}\, E_a} $	 in \eqref{eq:hhpsi} can be taken to literally mean that the left and right fields are separated by an evolution in imaginary time by an amount $\frac{\beta}{2}$. This would for instance be easily achieved by the Schwinger-Keldysh contour shown in Fig.~\ref{fig:contour3}. In operator language we can accommodate this shift by writing the following expression for the generating functional:
\begin{equation}
\mathscr{Z}_{_\text{TFD}}[{\cal J}_\skR, {\cal J}_\skL] = \Tr{(\rhoT)^\frac{1}{2} \, U[\mathcal{J}_\skR]\, (\rhoT)^\frac{1}{2} \, U^\dag[\mathcal{J}_\skL] }\,.
\label{eq:zthfld}
\end{equation}
The key difference from the earlier Schwinger-Keldysh construction described in \eqref{eq:ZSKdef} is that we have exploited the structure of the thermal density matrix as a Euclidean time evolution, and chosen to distribute this evolution into two independent parts. In the stationary global equilibrium state this is a consequence of the analytic structure of the thermal Gibbs density matrix.

The separation of the unitary evolution operators $U[{\cal J}_\skR]$ and $U^\dag[{\cal J}_\skL]$ with an insertion of the square root of the density matrix in \eqref{eq:zthfld} has drastic consequences for causality properties. Now with equal sources for the $\sf{L}$ and $\sf{R}$ fields, one no longer immediately discerns that the path integral reduces to a statement about the thermal partition function. Indeed, upon aligning ${\cal J}_\skR= {\cal J}_\skL = {\cal J}$ one ends up with
\begin{equation}
\mathscr{Z}_{_\text{TFD}}[{\cal J}] =  \Tr{(\rhoT)^\frac{1}{2} \, U[\mathcal{J}]\, (\rhoT)^\frac{1}{2} \, U^\dag[\mathcal{J}] }  \stackrel{?}{=} \Tr{\rhoT}\,.
\label{}
\end{equation}
On the other hand, the Schwinger-Keldysh path integral $\mathscr{Z}_{SK}[{\cal J}_\skR , {\cal J}_\skL]$ was engineered to precisely reproduce the equilibrium answer  $ \Tr{\rhoT}$ when the sources were set equal  (c.f., the discussion at the beginning of \S\ref{sec:keldysh}).

The thermofield double partition function does give the correct answer for equal sources, but to see this one has to invoke the analytic properties of the thermofield double construction. At an operator level, to pass from
$\mathscr{Z}_{SK}$ to $\mathscr{Z}_{_\text{TFD}}$ we have to use the fact that the unitary operator conjugates back to itself under a thermal evolution by half a period, i.e.,
\begin{equation}
U[{\cal J}] = (\rhoT)^{-\frac{1}{2}} \; U[{\cal J}]\; (\rhoT)^\frac{1}{2}\,.
\label{eq:rhoTFD}
\end{equation}
This is a consequence of the KMS condition since $\rhoT$ implements Hamiltonian evolution along the Euclidean time circle by half a thermal period.

The thermofield double treats the left and right degrees of freedom of the Schwinger-Keldysh construction symmetrically, but it does so at the price of obscuring the physical properties of the correlation functions. These get buried into the detailed analytic properties. It should be clear that such a description would only be useful to compute equilibrium correlation functions by turning on well-behaved (suitably analytic) sources.  If we are willing to accept the analyticity requirement, then we note that a-priori the only feature that singles out the thermofield double path integral \eqref{eq:zthfld} is this $\sf{L}\leftrightarrow \sf{R}$ symmetry. One could choose to consider a more general one-parameter family of generating functions $\mathscr{Z}_{_{\alpha-\text{TFD}}}$ by redistributing the Euclidean evolutions asymmetrically:
\begin{equation}
\mathscr{Z}_{_{\alpha-\text{TFD}}}[{\cal J}_\skR, {\cal J}_\skL] = \Tr{(\rhoT)^\alpha \, U[\mathcal{J}_\skR]\, (\rhoT)^{1-\alpha} \, U^\dag[\mathcal{J}_\skL] }\,.
\label{}
\end{equation}
The Schwinger-Keldysh path integral is obtained in the formal limit $\alpha \to 0$.

In a genuine non-equilibrium scenario one does not expect to see the sources being controllable to ensure the requisite analyticity. Therefore for the most part we will eschew the usage of the thermofield double construction. We will however return to it when we discuss aspects of the Schwinger-Keldysh formalism for holographic field theories in \S\ref{sec:grav}.

\section{Examples}
\label{sec:examples}

Thus far we have been discussing abstract quantum systems and have focused on generic operators therein.
While this has the advantage of  setting up the formalism in one swoop for any quantum system, it is also useful to record some of the salient properties in some simple examples.

To this end we now describe the basic results for three simple theories to get a taste for the formalism. We will discuss free scalars, fermions, and vectors in $d$ dimensions to exemplify the basic aspects.

Having fixed the theory, we still have the choice of the initial state to pick. Since arbitrary density matrices are still complicated, we will focus on two simple cases:
\begin{itemize}
\item Computing real time correlation functions in the vacuum state $\rhoi =\; \ket{0}\bra{0}$
\item Correlations in thermal density matrices
\end{itemize}
These situations have been well studied in the literature and we will mostly review the salient results.

When we discuss the situation where  fields are at finite temperature, $\beta$ will as usual denote  the inverse temperature. For scalars and fermions we will also make provision for an abelian $U(1)$ flavour symmetry, thus allowing the freedom to turn on a chemical potential  $\mu$. We will take the fields to have charge $q$ under this symmetry and turn on the chemical potential by a background gauge field coupling.

\subsection{Free scalar field}
\label{sec:freeB}

We wish to write down the vacuum and thermal correlation functions for a complex scalar field with charge $q$ under an Abelian flavour symmetry. The Schwinger-Keldysh construction instructs us to consider the scalar action:
\begin{equation}
-S_\text{scalar}  = \int \,d^dx\, \sqrt{-g}\, \left[\frac{1}{2}\, \partial_\mu \phi_\skR^\dagger \,\partial^\mu \phi_\skR -
\frac{1}{2}\, \partial_\mu \phi_\skL^\dagger \,\partial^\mu \phi_\skL \right]\,.
\label{}
\end{equation}

To write various formulae compactly  it is best to pass onto Fourier space. We find it convenient to introduce the Lorentz-invariant momentum space  integral
\begin{equation}
\int_p \, \mathfrak{I}   \equiv \int \frac{1}{(2E_p)} \frac{d^{d-1}p}{(2\pi)^{d-1}} \; \mathfrak{I} \,, \qquad
e^{ip.x} \equiv e^{i (\vec{p} \cdot \vec{x} -  E_p \,t) }\,.
\label{}
\end{equation}
with $E_p  =\sqrt{\vec{p} \cdot \vec{p} }$ as appropriate for a massless scalar. In the context of thermal correlators, we will have occasion to use the  Bose-Einstein  distribution function \eqref{eq:fDef} which for the problem of interest takes the form
\begin{equation}
\begin{split}
\fbetaB{q} \equiv \frac{1}{e^{\beta (E_p - q\,\mu)} -1}\,.
\end{split}
\label{eq:BEfnq}
\end{equation}
We have chosen to display the charge dependence explicitly since it helps simplify the presentation of the answers.

\paragraph{Vacuum correlation functions:} The zero temperature real-time vacuum correlators can be explicitly written down in momentum space for the free boson:
\begin{equation}
\begin{split}
\langle\ \mathcal{T}_{SK}\, \SKR{\phi}(x)\ \SKR{\phi}^\dag(y)\ \rangle &=
\int_p \bigbr{  \stepFn{xy} \;  e^{ip.(x-y)} + \stepFn{yx} \;  e^{-ip.(x-y)}   }\,
 \\
\langle\ \mathcal{T}_{SK}\, \SKR{\phi}(x)\ \SKL{\phi}^\dag(y)\ \rangle &=
\int_p \   e^{-ip.(x-y)}     \,
 \\
\langle\ \mathcal{T}_{SK} \,\SKL{\phi}(x)\ \SKR{\phi}^\dag(y)\ \rangle &=
\int_p \  e^{ip.(x-y)}   \,
 \\
\langle\ \mathcal{T}_{SK}\, \SKL{\phi}(x)\ \SKL{\phi}^\dag(y)\ \rangle &=
\int_p \bigbr{  \stepFn{yx} \; e^{ip.(x-y)} + \stepFn{xy} \;  e^{-ip.(x-y)}   }\,.
 \\
\end{split}
\label{eq:SvacLR}
\end{equation}
One can readily see that these four two-point correlators are not independent, for a linear combination of them vanishes. This feature becomes manifest if we switch to the average-difference basis. In order to get the physical two-point correlation functions, we perform the Keldysh rotation which leads to the expressions in the average-difference basis
 \begin{equation}
\begin{split}
\langle\ \mathcal{T}_{SK} \, \SKAv{\phi}(x)\ \SKAv{\phi}^\dag(y)\ \rangle &= \half \,
\int_p \brk{   e^{ip.(x-y)} +   e^{-ip.(x-y)}   }\,,
 \\
 \langle\ \mathcal{T}_{SK} \, \SKAv{\phi}(x)\ \SKRel{\phi}^\dag(y)\ \rangle
 &=  \stepFn{xy}\
\int_p \brk{  e^{ip.(x-y)} -  e^{-ip.(x-y)}   } \,, \\
\langle\ \mathcal{T}_{SK} \, \SKRel{\phi}(x)\ \SKAv{\phi}^\dag(y)\  \rangle &=
\stepFn{yx} \
\int_p \brk{  -e^{ip.(x-y)} +  e^{-ip.(x-y)}   }\,,\\
\langle\ \mathcal{T}_{SK} \, \SKRel{\phi}(x)\ \SKRel{\phi}^\dag(y)\ \rangle &=
0 \,,\\
\end{split}
\label{eq:Svacad}
\end{equation}
and we note in particular the combination
\begin{equation}
\langle\ \mathcal{T}_{SK}\bigbr{
 \SKAv{\phi}(x)\ \SKRel{\phi}^\dag(y)\
- \SKRel{\phi}(x)\ \SKAv{\phi}^\dag(y)\
}\  \rangle =
\int_p \brk{  e^{ip.(x-y)} -  e^{-ip.(x-y)}   }  \,.
\end{equation}

\paragraph{Thermal correlation functions:} The thermal correlation functions are similarly easy to obtain. Now the fact that we have populated the modes of the scalar with a thermal distribution results in terms involving the Bose-Einstein distribution function $\fbetaB{q}$ defined in \eqref{eq:BEfnq}.  The Schwinger-Keldysh propagators can be shown to take the Mills form \cite{Chou:1984es}:
\begin{equation}
\begin{split}
\langle\ \mathcal{T}_{SK}\, \SKR{\phi}(x)\ \SKR{\phi}^\dag(y)\ \rangle &=
\int_p \bigbr{ \brk{\fbetaB{+q} +\stepFn{xy} } e^{ip.(x-y)} +\brk{ \fbetaB{-q}+\stepFn{yx} }  e^{-ip.(x-y)}   }\,,
 \\
\langle\ \mathcal{T}_{SK}\, \SKR{\phi}(x)\ \SKL{\phi}^\dag(y)\ \rangle &=
\int_p \brk{ \fbetaB{+q} e^{ip.(x-y)} + \prn{\fbetaB{-q} + 1 } e^{-ip.(x-y)}   }\,,
 \\
\langle\ \mathcal{T}_{SK}\, \SKL{\phi}(x)\ \SKR{\phi}^\dag(y)\ \rangle &=
\int_p \brk{ \prn{\fbetaB{+q} +1} e^{ip.(x-y)} + \fbetaB{-q}  e^{-ip.(x-y)}   }\,,
 \\
\langle\ \mathcal{T}_{SK}\, \SKL{\phi}(x)\ \SKL{\phi}^\dag(y)\ \rangle &=
\int_p \bigbr{ \brk{\fbetaB{+q} +\stepFn{yx} } e^{ip.(x-y)} +\brk{ \fbetaB{-q}+\stepFn{xy} }  e^{-ip.(x-y)}   }\,.
 \\
\end{split}
\label{eq:StherLR}
\end{equation}
We observe that these correlators reduce to the vacuum correlators \eqref{eq:SvacLR} in the event of zero occupation number, $\fbetaB{+q} = \fbetaB{-q} = 0$. Further, the correlators \eqref{eq:StherLR} simplify after Keldysh rotation; one can check that the expressions  for the thermal correlators are essentially given by the expressions in \eqref{eq:Svacad} with the one exception
\begin{equation}
\begin{split}
\langle\ \mathcal{T}_{SK}\,  \SKAv{\phi}(x)\ \SKAv{\phi}^\dag(y)\ \rangle &=
\int_p \brk{ \prn{\fbetaB{+q} +\half } e^{ip.(x-y)} + \prn{\fbetaB{-q} + \half } e^{-ip.(x-y)}   } \,.
\end{split}
\label{eq:Stherad}
\end{equation}

\subsection{Free fermion field}
\label{sec:freeF}

As a second example, we consider now the Schwinger-Keldysh doubled theory of a free, massive fermion in four dimensions:
\begin{equation}
-S_\text{fermion}  = \int \,d^4x\, \sqrt{-g}\, \left[ \bar{\psi}_\skR\, (i\slashed{\partial} - m) \psi_\skR - \bar{\psi}_\skL (i\slashed{\partial} - m) \psi_\skL\right]\,.
\label{}
\end{equation}
We now give the vacuum and thermal correlators in analogy to the scalar case, see for instance \cite{Kamenev:2011aa} for details. Since we work in mostly positive signature our conventions for spin-$\frac{1}{2}$ fields are similar to those of
\cite{Srednicki:2007qs}.

\paragraph{Vacuum correlation functions:} The zero temperature real-time vacuum correlators take a form that is very similar to the case of free scalars:
\begin{equation}
\begin{split}
\langle\ \mathcal{T}_{SK} \,\SKR{\psi}(x)\ \SKR{\bar{\psi}}(y)\ \rangle &=
\int_p \bigbr{\stepFn{xy} \;  e^{ip.(x-y)} \,(-\slashed{p}+m) + \stepFn{yx} \;  e^{-ip.(x-y)}  \,(\slashed{p}+m) }\,,
 \\
\langle\ \mathcal{T}_{SK} \,\SKR{\psi}(x)\ \SKL{\bar{\psi}}(y)\ \rangle &=
 \int_p \   e^{-ip.(x-y)} \,(\slashed{p}+m)     \,,
 \\
\langle\ \mathcal{T}_{SK} \,\SKL{\psi}(x)\ \SKR{\bar{\psi}}(y)\ \rangle &=
\int_p \  e^{ip.(x-y)}   \,(-\slashed{p}+m)\,,
 \\
\langle\ \mathcal{T}_{SK}\, \SKL{\psi}(x)\ \SKL{\bar{\psi}}(y)\ \rangle &=
\int_p \bigbr{ \stepFn{yx} \; e^{ip.(x-y)}\,(-\slashed{p}+m) + \stepFn{xy} \;  e^{-ip.(x-y)} \,(\slashed{p}+m)  }\,.
 \\
\end{split}
\label{eq:FvacLR}
\end{equation}
As one can immediately confirm these can be obtained by acting with the operator $i\slashed{\partial}+m$ on the corresponding scalar result.

In analogy to the discussion of the scalar field, we can read off the correlation functions in Keldysh basis:
 \begin{equation}
\begin{split}
\langle\ \mathcal{T}_{SK}  \,\SKAv{\psi}(x)\ \SKAv{\bar{\psi}}(y)\ \rangle
&=
	\half \,\int_p \brk{ e^{ip.(x-y)} \,(-\slashed{p}+m)+   e^{-ip.(x-y)} \,(\slashed{p}+m)  }\,,
 \\
 \langle\ \mathcal{T}_{SK} \, \SKAv{\psi}(x)\ \SKRel{\bar{\psi}}(y)\ \rangle
 &=
	 \stepFn{xy}\ \int_p \brk{  e^{ip.(x-y)}\,(-\slashed{p}+m) - e^{-ip.(x-y)}  \,(\slashed{p}+m) }  \,,\\
\langle\ \mathcal{T}_{SK} \, \SKRel{\psi}(x)\ \SKAv{\bar{\psi}}(y)\  \rangle
 &=
	\stepFn{yx} \ \int_p \brk{ - e^{ip.(x-y)}\,(-\slashed{p}+m) +  e^{-ip.(x-y)} \,(\slashed{p}+m)  }\,,\\
\langle\ \mathcal{T}_{SK}\,  \SKRel{\psi}(x)\ \SKRel{\bar{\psi}}(y)\ \rangle
&=
	0\,. \\
\end{split}
\label{eq:Fvacad}
\end{equation}
We note in passing that the condensed matter literature prefers a different convention originating from \cite{Larkin:1969aa} for the average and difference operators for the Dirac conjugates $\bar{\psi}$ (see for instance \cite{Kamenev:2011aa}). Their definition involves declaring $\bar{\psi}_\skR - \bar{\psi}_\skL$ to be the average operator as opposed to the difference operator as we would have it. We find this extremely counter-intuitive especially in working out the constraints from the largest time equation. So we will adopt a homogeneous definition as in \eqref{eq:KeldyshDef} for both $\psi$ and $\bar{\psi}$.

\paragraph{Thermal correlation functions:} The thermal correlation functions for the free fermion can similarly be computed. To write this down we need the  Fermi-Dirac  distribution function
\begin{equation}
\begin{split}
\fbetaF{q} \equiv \frac{1}{e^{\beta (E_p - q\,\mu)} +1}\,.
\end{split}
\label{eq:FDfnq}
\end{equation}
With this in hand one can check that the correlation functions take  the following form:
\begin{equation}
\begin{split}
\langle\ \mathcal{T}_{SK}\, \SKR{\psi}(x)\ \SKR{\bar{\psi}}(y)\ \rangle &=
\int_p \bigbr{ \brk{-\fbetaF{+q} +\stepFn{xy} } e^{ip.(x-y)} \,(-\slashed{p}+m)+\brk{- \fbetaF{-q}+\stepFn{yx} }  e^{-ip.(x-y)} \,(\slashed{p}+m)  }\,,
 \\
\langle\ \mathcal{T}_{SK}\, \SKR{\psi}(x)\ \SKL{\bar{\psi}}(y)\ \rangle &=
\int_p \brk{ -\fbetaF{+q} \,e^{ip.(x-y)} \,(-\slashed{p}+m)+ \prn{1-\fbetaF{-q} } e^{-ip.(x-y)} \,(\slashed{p}+m)  }\,,
 \\
\langle\ \mathcal{T}_{SK}\, \SKL{\psi}(x)\ \SKR{\bar{\psi}}(y)\ \rangle &=
\int_p \brk{ \prn{1-\fbetaF{+q} } e^{ip.(x-y)} \,(-\slashed{p}+m)- \fbetaF{-q}\,  e^{-ip.(x-y)} \,(\slashed{p}+m)  }\,,
 \\
\langle\ \mathcal{T}_{SK}\, \SKL{\psi}(x)\ \SKL{\bar{\psi}}(y)\ \rangle &=
\int_p \bigbr{ \brk{-\fbetaF{+q} +\stepFn{yx} } e^{ip.(x-y)} \,(-\slashed{p}+m)+\brk{ -\fbetaF{-q}+\stepFn{xy} }  e^{-ip.(x-y)}\,(\slashed{p}+m)   }\,,
 \\
\end{split}
\label{eq:FtherLR}
\end{equation}
 The correlators in the Keldysh basis are the same as the vacuum correlators \eqref{eq:Fvacad}, with one exception:
\begin{equation}
\langle\ \mathcal{T}_{SK}  \,\SKAv{\psi}(x)\ \SKAv{\bar{\psi}}(y)\ \rangle =
\int_p \brk{ \left( -\fbetaF{+q} + \half \right) e^{ip.(x-y)} \,(-\slashed{p}+m)+  \left( -\fbetaF{-q} + \half \right) e^{-ip.(x-y)} \,(\slashed{p}+m)  }\,.
\end{equation}
%

\subsection{Vector field}
\label{sec:freeV}

Consider the doubled theory of a photon in the Feynman gauge:
\begin{equation}
 S_\text{photon} = \int d^4x \, \sqrt{-g} \, \left[\left( -\frac{1}{4} (F_{\mu\nu})_\skR \, F^{\mu\nu}_\skR - \frac{1}{2} \, (\partial_\mu A^\mu_\skR)^2 \right) - \left(- \frac{1}{4} (F_{\mu\nu})_\skL \, F^{\mu\nu}_\skL  - \frac{1}{2} \, (\partial_\mu A_\skL^\mu)^2\right)\right]\,.
\end{equation}
Correlators in this gauge have the advantage of being proportional to Klein-Gordon propagators; so one can immediately write down from the scalar result the answers for the photon correlation functions.

\paragraph{Vacuum correlation functions:} The zero temperature real-time vacuum correlators take a form that is very similar to the case of free scalars:
\begin{equation}
\begin{split}
\langle\ \mathcal{T}_{SK} \,A_\skR^\mu(x)\ A_\skR^\nu(y)\ \rangle
&=
	g^{\mu\nu} \; \int_p \bigbr{\stepFn{xy} \;  e^{ip.(x-y)} + \stepFn{yx} \;  e^{-ip.(x-y)}  }
 \\
\langle\ \mathcal{T}_{SK} \,A_\skR^\mu(x)\ A_\skL^\nu(y)\ \rangle
&=
	g^{\mu\nu}\; \int_p \   e^{-ip.(x-y)}     \,
 \\
\langle\ \mathcal{T}_{SK} \,A_\skL^\mu(x)\ A_\skR^\nu(y)\ \rangle
&=
	g^{\mu\nu}\; \int_p \  e^{ip.(x-y)}   \,
 \\
\langle\ \mathcal{T}_{SK} \,A_\skL^\mu(x)\ A_\skL^\nu(y)\ \rangle
&=
	g^{\mu\nu}\; \int_p \bigbr{  \stepFn{yx} \; e^{ip.(x-y)} + \stepFn{xy} \;  e^{-ip.(x-y)}   }\,.
 \\
\end{split}
\label{eq:VvacLR}
\end{equation}
One can obtain the results in the Keldysh basis by a quite easily from here.

We will not at the moment write down the thermal correlation functions. These expressions are a bit more involved in Feynman gauge since the Faddeev-Popov ghosts do not decouple quite so simply, as they do for the vacuum correlators. This should hopefully be clear following our discussion  in \S\ref{sec:skghosts}.

\section{BRST symmetries and ghosts}
\label{sec:skghosts}

Thus far we have confined our discussion to the standard presentation of the Schwinger-Keldysh construction whereby we double the degrees of  freedom in  the quantum system by taking two copies of the same. While sufficient for most purposes of computing time ordered correlation functions, we wish now to argue that this framework naturally admits some additional structure in the form of a BRST symmetry.

\subsection{Field redefinition BRST symmetries}
\label{sec:frbrst}

To motivate the statement of Schwinger-Keldysh BRST symmetry, let us first consider consequences of field redefinitions in a standard single-copy path integral.
Let $\varphi_i$ be the physical dynamical fields. Since the field values themselves are not important, we can freely redefine them by the replacement $\varphi_i \rightarrow \varphi_i(\xi_j, \vartheta_a)$. We have chosen here not only to have new fields $\xi_j$
 which coincide in number with the original set of fields, but also to introduce some more redundancy with the fields $\vartheta_a$.
 The fact that we do not need to preserve the total number of fields  is obvious for we can always trivially integrate out non-interacting modes.

It should help here to consider an explicit example. We can for instance consider as our system two decoupled free scalar fields $\varphi_1, \varphi_2$. For the redefinition we take these two to be functions of three new bosonic degrees of freedom, viz., $\varphi_1(\chi, \xi,\vartheta)$ and $\varphi_2(\chi, \xi,\vartheta)$. Explicitly, we could map the non-interacting scalar theory to some complicated interacting theory involving the extra variables thus introduced. For example let us imagine the transformation:
\begin{equation}
\begin{split}
{\cal L}[\varphi_1,\varphi_2] &= \frac{1}{2}\, \sum_{i=1}^2\, \partial_\mu \varphi_i \, \partial^\mu \varphi_i  \\
\longrightarrow \quad {\cal L}[\chi,\xi, \vartheta] &=
	\frac{1}{2}\, \partial_\mu \left(\chi +  {\cal F} (\chi, \xi , \nabla^2) \right)
	\partial^\mu \left(\chi +  {\cal F} (\chi, \xi,  \nabla^2)  \right)
	\\
	& \qquad \qquad +\; \frac{1}{2}
	\partial_\mu \left(\xi + \vartheta\right) \partial^\mu \left(\xi + \vartheta\right) .
\end{split}
\label{eq:fredef}
\end{equation}
Clearly the second line is obtained from the first by a complicated redefinition involving some function ${\cal F}$ of the fields and their derivatives, which obscures the simplicity of the theory. However, the physical observables one would compute in the two cases would agree. And they must, for all we have done is introduce redundant variables. We expect that there should be some constraints inherent in the path integral, which manifest themselves as operator identities. The simplest diagnostic of such statements is to work out the general Ward identities of the theory.

These statements are all very familiar in the context of gauge invariance. We are simply generalizing consideration from that case to situations where the redefinition is not in some symmetry direction. In particular, the Lagrangian by itself is not required to be invariant under the  general class of transformations envisioned above. But a key element from the context of gauge symmetry still plays a useful role: we can associate with the field redefinition freedom a BRST invariance \cite{Alfaro:1989rx,Alfaro:1992np}.
The derivation proceeds analogously to the standard discussion of BRST symmetries in gauge theories; we examine the constraints in the redundant description, introduce auxiliary variables and transformation laws etc..

We can be rather explicit here and can work out, for example, the story for the simpler case when the transformations in \eqref{eq:fredef} have ${\cal F} =0$. Then $\chi = \varphi_1$  which we henceforth ignore and concentrate on the shift $\varphi_2 = \xi + \vartheta$. The redundant transformations are simply $\delta \xi = \lambda $ and $\delta \vartheta = -\lambda$. One would naturally want to gauge fix  $\vartheta =0$.  What we expect is indeed to recover the original Lagrangian with some relabeled fields. But this is no longer the unique choice; other gauge choices are equally possible. To retain covariance, what we should do is to introduce Faddeev-Popov ghost fields enlarging the symmetry to include the ghosts and the Lagrange multiplier fields. Letting $c, {\bar c}$ be the ghost fields and $b$ the (bosonic) Nakanishi-Lautrup field, the transformations for the free field are
\begin{equation}
\delta \xi =c \,, \quad
\delta \vartheta = -c\,, \quad
\delta c = 0 \,, \quad
\delta {\bar c} = -i\, b\,,\quad
\delta b = 0
\label{eq:freeccb}
\end{equation}
We can obtain this by the standard Faddeev-Popov procedure by using the gauge fixing condition $\vartheta=0$ which has a unit determinant. The action gauge fixed this way, then reads (ignoring the field $\varphi_1 = \chi$):
\begin{equation}
-S =
	\int d^dx \left\{  \half \, \partial_\mu (\xi+\vartheta) \, \partial^\mu (\xi+\vartheta)  - \bar{c} \, c
	+ i \, b \, \vartheta \right\}
	+ \int d^dx \;  \frac{1}{2} \,\partial_\mu \chi \, \partial^\mu \chi  \,.
\end{equation}
A moment's thought should convince the reader that we can integrate out the $b$ and $\vartheta$ fields, leaving behind the free $\xi$ field with a decoupled ghost sector. One can readily get rid of the latter as well, but lets say that we choose to retain it for the present.

In the enlarged system, the presence of the BRST symmetry \eqref{eq:freeccb} allows us to infer a Ward identity
\begin{equation}
\langle \frac{\delta G(\xi)}{\delta \xi} +i\, G(\xi)\, \frac{\delta S(\xi)}{\delta \xi} \rangle =0\,,
\end{equation}
where $G(\xi)$ is an arbitrary function. This statement is easily verified for the free field, but what is useful is that the Ward identity will hold even if $\xi$  had self-interactions. It is then often referred to as the statement of general Schwinger-Dyson equations \cite{Alfaro:1989rx,Alfaro:1992np}. We can consider the replacement $\varphi_2 \mapsto \xi + \vartheta$ in any interacting scalar theory; the arguments above can be paralleled with ease.

  None of the above statements should come as a surprise. There are much simpler ways to arrive at these results. We have chosen here to perversely display some trivial consequences of introducing redundancy into a free field theory.  While clearly an overkill for this simple problem, the phrasing of the field redefinition freedom as a BRST symmetry is the main lesson we want to extract.
  Once we have this basic statement we can extend our considerations to other examples of interest.
  This  perspective has some useful implications in the Schwinger-Keldysh construction, as we now describe.

\subsection{The topological sector of Schwinger-Keldysh}
\label{sec:topsk}

Let us attempt to draw parallels between the Schwinger-Keldysh construction and the field redefinition story described in the preceding subsection. Recall that by definition the Schwinger-Keldysh path integral computes the generating functional $\mathscr{Z}_{SK}[{\cal J}_\skR,{\cal J}_\skL] $  introduced in Eq.~\eqref{eq:ZSKdef}. To define this object we have doubled the degrees of freedom by introducing a second copy of the physical Hilbert space.  Doing so necessarily introduces some redundancy whose effects we would now like to understand.

 One may in fact view \eqref{eq:diff0} (i.e., the statement that the correlation functions of difference operators vanish) as a consequence of the redundancy thus introduced.  Indeed, what we have is a Ward identity, which is independent of the dynamics of the theory. Irrespective of the unitary operator that evolves the system, we are finding that the Schwinger-Keldysh construction demands that a set of correlation functions vanish. As indicated earlier, turning on equal sources for the right and left degrees of freedom results in the theory localizing to the initial conditions built into $\rhoi$.

 It is rather remarkable to have a general vanishing theorem for correlation functions in a quantum system.  The one context where such a statement is natural is when we have a system with an underlying BRST symmetry. In such a situation the symmetry guarantees that correlators are trivial for  BRST exact operators.

The lesson we wish to draw is that  the difference operators $\SKRel{O}$ are BRST exact for the field redefinition BRST invariance which comes hand in hand with the Schwinger-Keldysh doubling.  Then the vanishing result \eqref{eq:diff0} would follow naturally as a consequence of the symmetries. Since the Schwinger-Keldysh Lagrangian is given by ${\cal L}[\Phi_\skR] - {\cal L}[\Phi_\skL]$, the change of variables in question is  schematically of the form $\Phi_\skR \mapsto \Phi_\skR + \Psi $   and  $\Phi_\skL \mapsto \Phi_\skL+ \Psi $, i.e., a correlated shift of the two sets of degrees of freedom. Closely following the discussion
in \S\ref{sec:frbrst} we can construct the BRST charges according to the usual rules.
We will leave this construction as an exercise for the reader and jump immediately to a set of algebraic statements about the operators in a generic Schwinger-Keldysh doubled theory. For a simple example illustrating this algebra, we refer the reader to \S\ref{sec:scalars}.

A useful way to think about the BRST symmetry in the Schwinger-Keldysh context is to view the standard construction involving left and right degrees of freedom as a gauge fixed version. While there is nothing wrong with this, in certain contexts the gauge fixing condition may prevent one from clearly seeing some of the underlying structure. Often one likes to work with a covariant presentation of the theory to avoid such issues. The Schwinger-Keldysh BRST construction we are about to give should be viewed in this light. The process of covariantizing involves perhaps working with a larger set of variables, but the price paid is worthwhile since all the symmetries are manifest.

Lets first see how this works in the Schwinger-Keldysh formalism. Since there are going to be ghost degrees of freedom associated with the BRST symmetries, let us introduce them at the outset. Given an operator $\OpH{O}$ in the single copy theory,
the Schwinger-Keldysh formalism demands to introduce a pair of left and right degrees of freedom. Furthermore, we have as a consequence of the BRST field redefinition symmetry a pair of ghosts denoted by $\SKGb{O}$ and $\SKG{O}$. The assignment of quantum numbers to these fields is inherited from $\OpH{O}$ with one crucial difference: $\SKGb{O}$ and $\SKG{O}$  have opposite Grassmann parity to $\OpH{O}$. They however have the  same fermionic parity (and hence the same  thermal boundary conditions). One consequence of this assignment is the
violation of the spin-statistics relation by  $\SKGb{O}$ and $\SKG{O}$ following from their ghostly nature.

 Thus, every operator $\OpH{O}$ in the single copy theory is  replaced in the Schwinger-Keldysh theory by a quadruplet  $\{\SKL{O},\SKGb{O},\SKG{O},\SKR{O}\}$ which we will collectively
refer to as the \emph{SK multiplet} associated with $\OpH{O}$. We will now explore the consequences of this quadrupling, fleshing out in the process the BRST symmetries we seek to explore.

\subsection{SK supercharges}
\label{sec:Qsk}

We have upgraded the Schwinger-Keldysh formalism to a quadruplet of fields $\{\SKL{O},\SKGb{O},\SKG{O},\SKR{O}\}$ which form a basic multiplet in the construction. We will soon introduce  a convenient way to package them, but for now let us work out how these should transform into each other under the aforementioned BRST symmetry. We will refer to the field redefinition BRST symmetry as the \emph{Schwinger-Keldysh} symmetry and correspondingly introduce a pair of
\emph{Schwinger-Keldysh (SK) supercharges} $\QSK$ and $\QSKb$ which implement the transformations. These supercharges are Grassmann odd BRST operators with zero fermion number. Without further ado, the considerations of \S\ref{sec:frbrst} and \S\ref{sec:topsk} lead us to the action of these charges defined by the following  graded commutators:
\begin{equation}\label{eq:QSKdefRL}
\begin{split}
\gradcomm{\QSK}{\SKL{O}}& = \gradcomm{\QSK}{\SKR{O}}  = \SKG{O},\quad
\gradcomm{\QSK}{\SKG{O} } = 0 ,\quad
\gradcomm{\QSK}{\SKGb{O}} = -\prn{\SKR{O}-\SKL{O}} , \\
\gradcomm{\QSKb}{\SKL{O}}& = \gradcomm{\QSKb}{\SKR{O}}  = \SKGb{O},\quad
\gradcomm{\QSKb}{\SKGb{O} } = 0 ,\quad
\gradcomm{\QSKb}{\SKG{O}} = \prn{\SKR{O}-\SKL{O}} .
\end{split}
\end{equation}

In what follows it will be useful to keep track of ghost number for various operators. The physical operators
$\OpH{O}$  and their Schwinger-Keldysh counterparts have zero ghost number. We will choose to assign ghost number $\pm1$ to
$\SKG{O}$ and $\SKGb{O}$ respectively. Ghost number conservation then demands a compatible assignment to the
supercharges; we make the following choice:
\begin{equation}
\gh{\SKG{O}} = \gh{\QSK} =+1 \,, \qquad \gh{\SKGb{O}} = \gh{\QSKb} =-1 \,.
\label{eq:ghnum}
\end{equation}
The action of the supercharges can be usefully captured in a diagrammatic form, viz.,
\begin{equation}
\begin{tikzcd}
&\SKR{O}, \SKL{O} \arrow{ld}{\QSK} \arrow{rd}[below]{\QSKb}    &   \\
\SKG{O}\arrow{rd}{\QSKb} & & \SKGb{O} \arrow{ld}[above]{\!\!\!\!\!\!\!\!\!\!\!\!\!\!-\QSK}\\
&   \SKR{O}- \SKL{O} &
\end{tikzcd}
\label{eq:qskaction}
\end{equation}
with the understanding that $\QSK$ and $\QSKb$ maps should be interpreted as commutator actions on the Hilbert space.

The one peculiarity of our ghost number assignment is that it increases right to left on this diagram. While we have denoted both $\SKL{O}$ and $\SKR{O}$ on the top row, it is clear that writing both of them is slightly redundant, and we could equivalently resort to the Keldysh basis of $av-dif$ operators. In the  Keldysh basis, the action of the supercharges can be checked to take the form
\begin{equation}\label{eq:QSKdefKeld}
\begin{split}
\gradcomm{\QSK}{\SKAv{O}}&=  \SKG{O},\quad
\gradcomm{\QSK}{\SKG{O} } = 0 ,\quad
\gradcomm{\QSK}{\SKGb{O}} = - \SKRel{O} ,\quad
\gradcomm{\QSK}{\SKRel{O}} = 0\ , \\
\gradcomm{\QSKb}{\SKAv{O}}&=  \SKGb{O},\quad
\gradcomm{\QSKb}{\SKGb{O} } = 0 ,\quad
\gradcomm{\QSKb}{\SKG{O}} =  \SKRel{O} ,\quad
\gradcomm{\QSKb}{\SKRel{O}} = 0 \,,
\end{split}
\end{equation}
or equivalently
\begin{equation}
\begin{tikzcd}
&\SKAv{O} \arrow{ld}{\QSK} \arrow{rd}[below]{\QSKb}    &   \\
\SKG{O}\arrow{rd}{\QSKb} & & \SKGb{O} \arrow{ld}[above]{\!\!\!\!\!\!\!\!\!\!\!\!\!\!-\QSK}\\
&   \SKRel{O} &
\end{tikzcd}
\label{eq:qskactionAvDif}
\end{equation}

The commutation relations make it clear in either case that $\SKRel{O}$ is both $\QSK$ and $\QSKb$ exact, thus assuring that their correlation functions vanish. In either presentation, it is easy to check that
\begin{equation}
\QSK^2=\QSKb^2= \gradcomm{\QSK}{\QSKb}=0\,.
\label{eq:qsksq}
\end{equation}
 We note that the  ghost operators $\SKGb{O}$ and $\SKG{O}$ occur naturally as the ghosts
corresponding to the right-left symmetric shift generated by the Schwinger-Keldysh supercharges.

It is worthwhile comparing the discussion above with the more familiar discussion of BRST symmetries in gauge theories. In that case we introduce the ghosts by upgrading the gauge transformation parameters. One usually defines a  single BRST charge  $\mathcal{Q}$ by requiring that it perform a gauge transformation of the  physical fields along the ghost. With the ghost number assignment as in \eqref{eq:ghnum} we have an alignment in the charge assignment of the BRST operator and the ghost field. The partner anti-ghost field comes with an opposite ghost charge, to ensure that we have a net vanishing of ghost number for terms that appear in the action. Equivalently, when we exponentiate the Jacobian arising from the gauge fixing condition, we have a pair of ghosts with equal and opposite ghost number; only one of them is chosen to be obtained by gauge transforming the physical fields. Clearly there is an analogous construction where we should invoke a BRST transformation in the anti-ghost direction, $\bar{\mathcal{Q}}$. The two pairs of BRST charges are individually nilpotent and should anti-commute among themselves. In either case the Lagrange multiplier or the Nakanishi-Lautrup field, which enters through the gauge fixing condition, is BRST exact  -- it is obtained as the $\mathcal{Q}$ action on the anti-ghost or the  $\bar{\mathcal{Q}}$ action on the ghost.

This is exactly the structure present in \eqref{eq:QSKdefRL} or \eqref{eq:qskaction}. The BRST charges $\QSK$  and $\QSKb$ perform field redefinitions of the Schwinger-Keldysh fields in the ghost and anti-ghost directions respectively. The difference operator $\SKRel{O}$ is the Nakanishi-Lautrup field of this redefinition redundancy.\footnote{ As noted, it has been suggested in \cite{Crossley:2015evo} that the Schwinger-Keldysh construction can implement the Ward identity \eqref{eq:diff0} by the presence of a single BRST supercharge. 
The field redefinition arguments, as well as the known constructions for path integrals to compute partition sums as opposed to indices (see \cite{Vafa:1994tf} for a clear physical discussion), suggest to us however that the correct formalism involves having both supercharges made manifest, as we have chosen to do. Indeed in our applications of this formalism to construct hydrodynamic effective actions \cite{Haehl:2015uoc},  we have found no obstructions to making both supercharges manifest. This lends support to the general structure proposed above,  and for the rest of the discussion we will simply work out the consequences of two Schwinger-Keldysh supercharges, without further qualifiers. 
}

It will also be convenient for us to consider the definition of the SK-supercharges in the advanced-retarded basis. This will prove useful in the sequel when we wish to specify to thermal correlation functions.  The transformation into this basis is defined in
\eqref{eq:RADef} and it involves the operator that implements thermal time translations $\deltaB$. However, the thermal translations commute with any field redefinition we perform in the Schwinger-Keldysh basis. With this understanding it is then clear that in the
advanced-retarded basis the SK-supercharges act as
\begin{equation}\label{eq:QSKdefRA}
\begin{split}
\gradcomm{\QSK}{\SKRet{O}}&=\SKG{O},\quad
\gradcomm{\QSK}{\SKG{O} } =0 ,\quad
\gradcomm{\QSK}{\SKGb{O}} = - \SKAdv{O} ,\quad
\gradcomm{\QSK}{\SKAdv{O}} = 0\ , \\
\gradcomm{\QSKb}{\SKRet{O}} &= \SKGb{O},\quad
\gradcomm{\QSKb}{\SKGb{O} } = 0 ,\quad
\gradcomm{\QSKb}{\SKG{O}} =  \SKAdv{O} ,\quad
\gradcomm{\QSKb}{\SKAdv{O}} = 0 .
\end{split}
\end{equation}
%

\subsection{Example: free scalar field}
\label{sec:scalars}

Let us flesh out this abstract discussion with an example. Consider the free scalar field described in \S\ref{sec:freeB}:
\begin{equation}
\begin{split}
-S_\text{scalar}  &= \int \,d^dx\, \sqrt{-g}\, \left(\frac{1}{2}\, \partial_\mu \phi_\skR^\dagger\, \partial^\mu \phi_\skR -
\frac{1}{2}\, \partial_\mu \phi_\skL^\dagger\, \partial^\mu \phi_\skL \right)\\
&=  \int \, d^dx\, \sqrt{-g} \, \left( \frac{1}{2} \, \partial_\mu \phi_{_{av}}^\dagger \, \partial^\mu \phi_{_{dif}} + \frac{1}{2} \, \partial_\mu \phi_{_{dif}}^\dagger \, \partial^\mu \phi_{_{av}}\right) \,.
\end{split}
\label{}
\end{equation}
We have the Schwinger-Keldysh fields $\phi_\skR$ and $\phi_\skL$ whose correlation functions in a thermal state have been described hitherto (see
\eqref{eq:StherLR}). We claim that this Schwinger-Keldysh doubled theory has a hidden field redefinition invariance of the form $\{\phi_\skR \rightarrow \phi_\skR + \chi,\;\phi_\skL \rightarrow \phi_\skL + \chi\}$, or equivalently in the Keldysh basis $\{\phi_{_{av}} \rightarrow \phi_{_{av}} + \chi,\; \phi_{_{dif}} \rightarrow \phi_{_{dif}}\}$.

In order to make this symmetry manifest, we now have to include the Schwinger-Keldysh ghost fields, which fill out the quartet of fields. According to our discussion in \S\ref{sec:frbrst} we expect that adding a decoupled ghost sector will make the field redefinition Schwinger-Keldysh BRST symmetry manifest.
Introduce Schwinger-Keldysh ghosts $c$ and $\bar{c}$ which are scalar fields with odd Grassmann parity. The fields $\{\phi_{_{av}}, c, \bar{c},\phi_{_{dif}}\}$ then form the Schwinger-Keldysh multiplet  for this theory. We take the ghost charge assignments to be
\begin{equation}
\gh{\phi_\skL} = \gh{\phi_\skR} = 0 \,, \qquad \gh{c} = +1 \,, \qquad \gh{\bar{c}} = -1 \,.
\label{eq:scalargh}
\end{equation}
The Schwinger-Keldysh supercharge action may be inferred from the diagram:
\begin{equation}
\begin{tikzcd}
&\phi_{_{av}} \arrow{ld}{\QSK} \arrow{rd}[below]{\QSKb}   &   \\
c \arrow{rd}{\QSKb} & & \bar{c} \arrow{ld}[above]{\!\!\!\!\!\!\!\!\!\!\!\!\!\!-\QSK}\\
&   \phi_{_{dif}} &
\end{tikzcd}
\label{eq:qskB}
\end{equation}
where we pass onto the Keldysh basis for convenience.

Adding the ghosts, the manifestly BRST invariant Schwinger-Keldysh Lagrangian for the scalar and the ghosts takes the form:
\begin{equation}
-S_{\text{scalar+ghosts}} = \int d^d x\,\sqrt{-g}\, \left( \frac{1}{2} \, \partial_\mu \phi_{_{av}}^\dagger \, \partial^\mu \phi_{_{dif}} + \frac{1}{2} \, \partial_\mu \phi_{_{dif}}^\dagger \, \partial^\mu \phi_{_{av}}+
c^\dagger\, \nabla^2  \bar{c} - \bar{c}^\dagger \, \nabla^2 c
\right)\,.
\label{eq:skBaction}
\end{equation}
The ghosts form a decoupled sector in this theory owing to the absence of interactions. It is easy to check that $\QSK$ and $\QSKb$ are symmetries of this action. In fact, the nilpotency of the supercharges allows us to make the symmetry manifest by writing the action in a BRST exact form:
\begin{equation}
-S_{\text{scalar+ghosts}} =  \gradcomm{\QSKb}{\gradcomm{\QSK}{\int d^d x\,\sqrt{-g}\, \left( \frac{1}{2} \, \partial_\mu \phi_{_{av}}^\dagger \, \partial^\mu \phi_{_{av}}\right) }} \,.
\label{eq:Sscalarghost2}
\end{equation}

The correlation functions of the various fields can be chosen to be:
\begin{equation}
\begin{split}
&\langle\ \mathcal{T}_{SK} \; c(x)\ c^\dag(y)\ \rangle =
\langle\ \mathcal{T}_{SK} \;\bar{c}(y)\ \bar{c}^\dag(y)\ \rangle =
0 \\
&  \langle\ \mathcal{T}_{SK}\;  c (x)\ \bar{c}^\dag(y)\  \rangle
=  -\langle\ \mathcal{T}_{SK}\;  c (x)\ \bar{c}^\dag(y)\  \rangle^\star  \\
& \hspace{3cm} =
\langle\ \mathcal{T}_{SK} \;  \phi_{_{av}}(x) \phi^\dag_{_{dif}}(y) -
\phi_{_{dif}}(y) \phi^\dag_{_{av}}(x) \  \rangle \\
& \hspace{3cm}=   \int_p \brk{ e^{ip.(x-y)} - e^{-ip.(x-y)}   } \\
&
\end{split}
\end{equation}
The correlation functions in the second line  are written down for a particular choice of the boundary conditions. As we explain in some detail in \S\ref{sec:superrules}, there is a certain amount of ambiguity in the ghost correlation functions owing to the choice of boundary conditions we can impose on the superspace path integral. The result above is for a particular choice which is simple and incorporates the results for the average difference correlation functions from
\eqref{eq:Svacad}.

\subsection{The Schwinger-Keldysh superfields}
\label{sec:superfields}

The interpretation of the Schwinger-Keldysh construction in terms of the quartet of fields including the ghosts can be succinctly summarized using a superfield language. This allows for a relative ease in constructing effective actions which admit a correct action of the $\QSK$ and $\QSKb$ charges.

To describe the superfields, we start by upgrading the background geometry where our quantum system resides to admit a supermanifold structure. To the coordinates $x^\mu$ of the geometry, we add two Grassmann valued super-coordinates $\theta$ and $\bar{\theta}$ which parameterize the superfield directions. We will take these super-coordinates to carry non-trivial ghost number, with the assignment:
\begin{equation}
\gh{\theta} = +1 \,, \qquad \gh{\bar{\theta}} = -1 \,.
\label{eq:ghththb}
\end{equation}

We can now view the quartet $\{\SKR{O}, \SKL{O}, \SKG{O}, \SKGb{O}\} $ as a single superfield $\SKS{O}$ with a superspace Taylor expansion. For the superfield we will make different choices depending on whether we discuss real-time vacuum correlation functions or thermal correlators. Define
\begin{align}
\SKS{O} \equiv
\begin{cases}
& \SKAv{O} + \theta\ \SKGb{O} +\bar{\theta}\  \SKG{O} + \bar{\theta} \theta\ \SKDif{O}  \,, \qquad \rhoi = \ket{0}\bra{0} \,, \\
&  \SKRet{O} + \theta\ \SKGb{O} +\bar{\theta}\  \SKG{O} + \bar{\theta} \theta\ \SKAdv{O}  \,, \qquad \rhoi = \rhoT\,.
\end{cases}\label{eq:KMSsup}
\end{align}
The choice is made such that the bottom component corresponds to the physical average/retarded field
$\SKAv{O}$ or $\SKRet{O}$ and the top component to the Nakanishi-Lautrup difference operator $\SKRel{O}$ (equivalently $\SKAdv{O}$).  We use the notation introduced in \cite{Haehl:2015uoc} to denote the Schwinger-Keldysh superfield corresponding to an operator $\OpH{O}$, by simply affixing an accent ``$\ \mathring{} \ $'' above it.

The expression for the superfield can be written equivalently in the other basis of operators introduced hitherto in a straightforward manner
\begin{equation}
\begin{split}
\SKS{O} &\equiv
	\prn{1+\FSgn{O} \fbeta+ \bar{\theta} \theta\ }\SKR{O} -
	\prn{\FSgn{O}\fbeta+ \bar{\theta} \theta\ }\SKL{O}
	+ \theta\ \SKGb{O} +\bar{\theta}\  \SKG{O}\\
&\equiv
	\SKAv{O} +\prn{\half+\FSgn{O} \fbeta + \bar{\theta} \theta\ }\SKRel{O} + \theta\ \SKGb{O} +\bar{\theta}\  \SKG{O}\,.
\end{split}
\end{equation}
From the superfield we can always recover the operator in question by projection -- we simply set
$\theta= \bar{\theta} =0$. We will denote this operation by ``$|$'', viz.,
\begin{equation}
\SF{\Op{O}} | \equiv \SF{\Op{O}} \big{|}_{\theta=\bar\theta=0} = \Op{O} \,.
\label{eq:sfproj}
\end{equation}

The advantage of working with superfields is that the operations of  $\QSK$ and $\QSKb$ can be equivalently understood in terms of super-derivations.
We now give two equivalent ways of thinking about the action of $\{\QSK,\QSKb\}$ on superfields. The first point of view simply takes the supercoordinates to anti-commute with the Schwinger-Keldysh supercharges:
\begin{equation}
\gradcomm{\QSK}{\theta} = \gradcomm{\QSK}{\bar{\theta}} = \gradcomm{\QSKb}{\theta}=\gradcomm{\QSKb}{\theta} = 0
\label{eq:qqbththb}
\end{equation}
With this understanding, $\{\QSK,\QSKb\}$ simply act on superfields componentwise (picking up signs when they pass through $\theta$ or $\thb$). This yields:
\begin{equation}
\begin{split}
\gradcomm{\QSK}{\SF{\Op{O}}} &= \gradcomm{\QSK}{\Op{O}} - \theta \, \gradcomm{\QSK}{\SKGb{O}} - \thb \, \gradcomm{\QSK}{\SKG{O}} + \thb\theta \, \gradcomm{\QSK}{\SKRel{O}} \\
&= \SKG{O} + \theta \, \SKRel{O} \,,
\end{split}
\end{equation}
and similarly for $\QSKb$. This encodes the action of the SK supercharges componentwise.

Alternately, there is a more straightforward way to obtain the same result. Namely, we simply take $\{\QSK,\QSKb\}$ to act as translation generators $\{\partial_\thb,\partial_\theta\}$ in the Grassmann-odd directions:
\begin{equation}
\gradcomm{\QSK}{\SF{\Op{O}}} =\frac{\partial \SF{\Op{O}}}{\partial \bar\theta} \,, \qquad
 \gradcomm{\QSKb}{\SF{\Op{O}}}  = \frac{\partial \SF{\Op{O}}}{\partial {\theta}} \,.
\label{eq:Qderact3}
\end{equation}
This expression makes it patently clear that the superspace geometrises the Schwinger-Keldysh differentials $\{\QSK,\QSKb\}$.

These expressions are indeed the ones we would write based on the construction of the basic Schwinger-Keldysh multiplet. To see this explicitly let us instead work this out using $\QSK, \QSKb$.  One finds using \eqref{eq:Qderact3} that the bottom component transforms as
\begin{equation}
\begin{split}
\gradcomm{\QSK}{\SKRet{O}} &= \frac{\partial\SKS{O}}{\partial\bar{\theta}} \bigg{|}= \left(\SKG{O}  +\theta\ \SKAdv{O}\right)| = \SKG{O} \,,\\
\gradcomm{\QSKb}{\SKRet{O}}  &= \frac{\partial\SKS{O}}{\partial\theta}\bigg{|}= \left(\SKGb{O}  -  \bar{\theta}\ \SKAdv{O}\right)| = \SKGb{O}\,.
\end{split}
\end{equation}
 In ascertaining this we are using the fact that the supercharges themselves do not have a non-trivial dependence on the supercoordinates. To compute the action on the superpartners of $\SKRet{O}$ such as $\SKG{O}, \SKGb{O}, \SKAdv{O}$ etc., we have to first construct superfields whose bottom component is the object of interest. For  example,
\begin{equation}
\gradcomm{\QSK}{\SKGb{O}} = \gradcomm{\QSK}{\frac{\partial \SF{\Op{O}}}{\partial\theta} \bigg{|}} =
\frac{\partial}{\partial\bar{\theta}} \left(\frac{\partial \SF{\Op{O}}}{\partial\theta}\right) \bigg{|} = -\SKAdv{O} \,,
\end{equation}
where the sign originates from the order of the superderivations performed. Working with these expressions one can check that \eqref{eq:QSKdefRA} (or equivalently \eqref{eq:QSKdefKeld}) are satisfied.

To summarize, we have just argued for the following elegant statement: \textit{by simply upgrading the single-copy QFT to a theory in superspace $\{x^\mu,\theta,\bar\theta\}$ automatically gives a Schwinger-Keldysh theory with a ghost sector that guarantees the universal symmetries inherent in the doubling structure of Schwinger-Keldysh formalism.} Let us briefly reconsider a simple example for illustration.

\paragraph{Free scalar in superspace:}
Another advantage of introducing the superfields is that the action can be written compactly in terms of a superspace integral. For example the scalar field theory discussed in \S\ref{sec:scalars} can be written in terms of the superfield
\begin{equation}
\SF{\phi} = \phi_{_{av}} + \bar{\theta} \, c + \theta \, \bar{c} + \bar{\theta} \theta \, \phi_{_{dif}} \,.
 \label{eq:phisf}
\end{equation}
The action \eqref{eq:Sscalarghost2} in superspace is simply
\begin{equation}
-S_\text{scalar+ghosts} = \int d^dx\,\sqrt{-g}\,\int d\theta\, d\bar{\theta} \, \left(\frac{1}{2}\, \partial_\mu \SF{\phi}^\dagger \,\partial^\mu \SF{\phi} \right) \,,
\label{}
\end{equation}
up to a total derivative. The integration over superspace is then just the statement that the action is $\QSK$ and $\QSKb$ exact. This way, working in superspace automatically ensures a formalism that is manifestly Schwinger-Keldysh field redefinition invariant.

\section{Thermal BRST symmetries and SK-KMS superalgebra}
\label{sec:KMScharges}

Our discussion of the BRST symmetries has thus far focused  on generic density matrices. We have already seen in \S\ref{sec:skthemal} that there are special features of thermal density matrices that imply an additional structure. The primary new ingredient is the interpretation of the thermal density matrix in terms of a Euclidean evolution and the associated KMS condition.
In this section we focus on the case of thermal density matrices and derive an additional BRST structure associated with  the KMS condition.

\subsection{The KMS supercharges}
\label{sec:Qkms}
To implement the KMS condition efficiently, we introduced in \eqref{eq:kmsconjL} the KMS conjugate of an operator and the corresponding source $\tilde{{\cal J}}$. 
The KMS condition is embodied in the sum rule Eq.~\eqref{eq:diff1}.
One consequence of this observation is that we can now repeat our arguments in \S\ref{sec:Qsk} with say ${\cal J}_\skL$ replaced by  $\tilde{{\cal J}}_\skL$ and $\SKL{O}$ replaced by $\SKL{\tilde{O}}$.\footnote{ We find this analogy useful to motivate the charges below, but we do not claim that this replacement is a symmetry of the correlation functions.} One then should encounter an entirely new set of topological charges that involve the KMS conjugates of our operators.

To be precise, let us invoke  a second set of mutually anti-commuting, nilpotent, Grassmann odd operators which we will call the \emph{KMS supercharges} $\QKMS$ and $\QKMSb$. They should induce an algebra completely analogous to the universal Schwinger-Keldysh field redefinitions \eqref{eq:QSKdefRL}, but now implementing the fact that in a thermal state we additionally have field redefinitions whose Nakanishi-Lautrup field is $\SKR{O}-\SKL{\tilde{O}}$. The latter {\it `KMS shifted'} difference operators should hence be the BRST and anti-BRST exact objects in the KMS symmetry algebra. Furthermore, this second algebra associated with the KMS condition should not introduce new ghost fields, since the involved Grassmann neutral fields are just the original $\SKR{O}$ and $\SKL{O}$ and their thermal time translations.
This motivates us to define the $\QKMS$ and $\QKMSb$ action on the Schwinger-Keldysh quadruplet  $\{\SKL{O},\SKGb{O},\SKG{O},\SKR{O}\}$ introduced in \S\ref{sec:topsk} as follows:
\begin{align}
& \gradcomm{\QKMS}{\SKL{O}} = i\,\SKG{O}\,, & &
\gradcomm{\QKMS}{\SKR{O}} =
    i\,\FSgn{O}e^{-i\deltaB}\SKG{O} \,, \nonumber \\
& \gradcomm{\QKMS}{\SKG{O}} = 0\, , & &
\gradcomm{\QKMS}{\SKGb{O}} = -i\,\left( \SKR{O}- \SKL{\tilde{O}}\right) \equiv -i\,\left(\SKR{O}-\FSgn{O}  e^{-i\deltaB} \SKL{O}\right) \,, \nonumber \\
& \gradcomm{\QKMSb}{\SKL{O}} = -i\,\SKGb{O}\,, & &
\gradcomm{\QKMSb}{\SKR{O}} =
   - i\,\FSgn{O}e^{-i\deltaB}\SKGb{O} \,, \nonumber \\
& \gradcomm{\QKMSb}{\SKGb{O}} = 0 \, , & &
\gradcomm{\QKMSb}{\SKG{O}} = -i\,\left(\SKR{O}- \SKL{\tilde{O}} \right)\equiv -i\,\left(\SKR{O}-\FSgn{O}  e^{-i\deltaB} \SKL{O}\right) \, .
\label{eq:QKMSdefRL}
\end{align}
One may write this in a diagrammatic notation making explicit the similarity to Eq.~\eqref{eq:qskaction} as:
\begin{equation}
\begin{aligned}
\begin{tikzcd}
&\quad\SKL{O}\quad \arrow{ld}{\!\!\!\QKMS} \arrow{rd}[below]{\!\!\!\!\QKMSb}   &   \\
i\,\SKG{{O}}\arrow{rd}{\!\!\!\QKMSb \quad\;-\QKMS} & & -i\,\SKGb{{O}} \arrow{ld}\\
&   \SKR{O}- \SKL{\tilde{O}} &
\end{tikzcd}
\end{aligned}
\label{eq:qkmsaction}
\end{equation}

Basically all we have done is to rewrite the field redefinition supercharges by working with the physical operators and realizing that the KMS invariance implies that $\SKR{O} - \SKL{\tilde{O}}$ should belong to the topological sector of the theory, for it has vanishing self-correlations as in \eqref{eq:diff1}. This explains how the algebra \eqref{eq:QKMSdefRL} can be derived: one starts by writing down the descendants of $\SKL{O}$, using the same ghosts as in the universal Schwinger-Keldysh field redefinition algebra and the Nakanishi-Lautrup field $\SKR{O}-\SKL{\tilde{O}}$ to describe the KMS condition. Then, the descendants of $\SKR{O}$ are immediately fixed by consistency (in particular by demanding that $\SKR{O}-\SKL{\tilde{O}}$ be closed under $\QKMS$ and $\QKMSb$). In defining the KMS supercharges in \eqref{eq:qkmsaction}, we have chosen to sneak in a few factors of  $i$, which is purely a choice of convention. This differs from the definition given in \cite{Haehl:2015foa}, but turns out to be more natural.\footnote{ The choice made here is natural from the superspace perspective. We have effectively chosen to do a ghost number rotation on the KMS supercharges relative to the choice made in \cite{Haehl:2015foa}:
\begin{equation}
 [\QKMS]_{_\text{here}} = i[\QKMS]_\text{\cite{Haehl:2015foa}} \,,\qquad
  [\QKMSb]_{_\text{here}} = -i[\QKMSb]_\text{\cite{Haehl:2015foa}} \,,\qquad
   [\Qzero]_{_\text{here}} = i[{\cal Q}_0]_\text{\cite{Haehl:2015foa}}\,,
 \end{equation}
 where $\Qzero$ will be defined below, c.f., \eqref{eq:Q0defOrig}.
 It also transpires that this choice corresponds to an anti-Hermitian representation for certain gauge fields in superspace which will prove useful in interpreting these structures from an equivariant cohomology perspective in \cite{Haehl:2016uah}.  \label{foot:conventions}}

The KMS supercharges thus defined are nilpotent $\QKMS^2 = \QKMSb^2 =0$. This follows once we realize that the supercharges commute with the thermal translation generator $\delKMS$.

One can make similar statements by choosing to replace  $\SKR{O}$ by its KMS conjugate instead. However, to respect the time ordering prescription the right operators should be conjugated slightly differently. One can for instance check that $\SKR{\tilde{O}} \equiv \FSgn{O}e^{i\deltaB}\SKR{O}$ has the same diagram as \eqref{eq:qkmsaction},  by noting that the Schwinger-Keldysh contour should be traversed in the opposite orientation to go from $\text{L} \to \text{R}$. The nicer way to circumvent this subtlety is by simply passing to a more convenient basis; as we will see below, the transformation properties are expressed most compactly in the $adv-ret$ basis (as was the case for the action of Schwinger-Keldysh supercharges).\footnote{ There are subtleties associated with the conjugation of $\SKR{O}$ owing to issues relating to convergence in the Euclidean domain.}

It is interesting to contrast this discussion with other implementations of the KMS condition. For instance, in \cite{Sieberer:2015hba} the authors define a transformation $\mathcal{K}_\beta$ which acts to map $\SKR{O}(t) \mapsto \SKL{O}(t-i \frac{\beta}{2})$ and $\SKL{O}(t) \mapsto \SKR{O}(t+i \frac{\beta}{2})$, motivated by the thermofield double construction. This acts as a ${\mathbb Z}_2$ transformation on the Schwinger-Keldysh fields, and modulo an overall parity and time-reversal corresponds to the dynamical KMS symmetry of \cite{Crossley:2015evo}. Supplementing this with an addition $i \frac{\beta}{2}$ time translation, we would see that the net effect is to replace 
$\SKR{O} \mapsto \SKL{\tilde{O}}$ and $\SKL{O} \mapsto \SKR{O}$, whence \eqref{eq:diff1} would follow from the basic identity of Schwinger-Keldysh \eqref{eq:diff0} (modulo at best an irrelevant overall sign). We are implementing the transformation somewhat differently at this stage, but it has the same intended effect on the correlation functions.

The action of the KMS supercharges in the other bases can be also readily ascertained.  For instance in Keldysh basis, we have
\begin{align}
& \gradcomm{\QKMS}{\SKAv{O}} = -\prn{i+\half \delKMS}\SKG{O}, & &
\gradcomm{\QKMS}{\SKGb{O}} = -\delKMS \SKAv{O} + \prn{i+ \half \delKMS}\SKRel{O} , \nonumber \\
&\gradcomm{\QKMS}{\SKG{O}} = 0 , & &
\gradcomm{\QKMS}{\SKRel{O}} = -\delKMS\SKG{O}\ ,  \nonumber \\
&\gradcomm{\QKMSb}{\SKAv{O}} = \prn{i+\half \delKMS}\SKGb{O}, & &
\gradcomm{\QKMSb}{\SKGb{O}} = 0 , \nonumber \\
&\gradcomm{\QKMSb}{\SKG{O}} = \delKMS \SKAv{O} - \prn{i+ \half \delKMS}\SKRel{O} , & &
\gradcomm{\QKMSb}{\SKRel{O}} = \delKMS\SKGb{O}\ .
\label{eq:QKMSdefKeld}
\end{align}
On the other hand in  the retarded-advanced basis, we obtain the most compact version of the KMS algebra:
\begin{align}
& \gradcomm{\QKMS}{\SKRet{O}} = 0, & &
\gradcomm{\QKMS}{\SKGb{O}} =  \delKMS\SKRet{O}  \,, \nonumber \\
& \gradcomm{\QKMS}{\SKG{O}} = 0 , & &
\gradcomm{\QKMS}{\SKAdv{O}} = \delKMS\SKG{O}\  \,, \nonumber \\
& \gradcomm{\QKMSb}{\SKRet{O}} = 0, & &
\gradcomm{\QKMSb}{\SKGb{O}} = 0   \,, \nonumber \\
& \gradcomm{\QKMSb}{\SKG{O}} = \delKMS\SKRet{O}, & &
\gradcomm{\QKMSb}{\SKAdv{O}} = -\delKMS\SKGb{O}\ .
\label{eq:QKMSdefRA}
\end{align}
In obtaining these expression we used the fact that $\QKMS$ and $\QKMSb$ commute with the Hamiltonian
$\OpH{H}$. This once again follows from the KMS invariance and can be viewed as a consequence of the field redefinition symmetry in the Schwinger-Keldysh construction combined with the Euclidean periodicity imposed by the KMS condition for thermal density matrices. In diagrammatic form, we find simply
\begin{equation}
\begin{aligned}
\begin{tikzcd}
&\quad\SKAdv{O}\quad \arrow{ld}{\!\!\!\QKMS} \arrow{rd}[below]{\!\!\!\!\QKMSb}   &   \\
\delKMS \SKG{{O}}\arrow{rd}{\!\!\!\QKMSb \quad\;-\QKMS} & & -\delKMS\SKGb{{O}} \arrow{ld}\\
&  \delKMS\delKMS \SKRet{O} &
\end{tikzcd}
\end{aligned}
\label{eq:qkmsactionAR}
\end{equation}

While formally similar to our discussion of the Schwinger-Keldysh supercharges, there is a very crucial distinction in the KMS algebra described above. Since $\QKMS$ and $\QKMSb$ involve thermal translations $e^{-i\deltaB}$  by a finite amount involving the inverse temperature, they relate fields which are physically separated along the Euclidean thermal circle. This implies that these charges are necessarily non-local. In global thermal equilibrium one can work with the Fourier modes of fields along the Euclidean time direction, viz., the Matsubara decomposition, and define the operators rather precisely. Beyond this special case however one expects that the strict definition of these supercharges comes with various associated subtleties. We will remark on these issues when we discuss the analog for generic density matrices later on.

\subsection{The quadruplet of thermal translations}
\label{sec:quadruplet}

 We have derived the existence of Grassmann-odd supercharges $\{\QSK,\QSKb\}$, which encode the field redefinition symmetry inherent in the Schwinger-Keldysh construction,  and have argued that the KMS condition leads to BRST supercharges $\{\QKMS,\QKMSb\}$ in the same vein. We now motivate the introduction of a new operator $\Qzero$, which together with the KMS charges, and the thermal translation operator $\Qbeta$ (which we recall acts as $\delKMS$) forms a quartet of super-KMS transformations.

 Firstly, realize that by construction $\{\QKMS,\QKMSb\}$ provide two Grassmann-odd generators of thermal translations, with $\gh{\QKMS} = +1$ and $\gh{\QKMSb} = -1$ respectively. We furthermore have a Grassmann-even thermal translation operator $\Qbeta$ defined in \eqref{eq:Qbetadef} which measures deviations from the KMS condition. Its action on the entire Schwinger-Keldysh multiplet which we reproduce here for convenience is simply
\begin{equation}
\gradcomm{\Qbeta}{ \Op{O}} = \delKMS{\Op{O}}\,, \qquad \qquad
 \Op{O} \in\{\SKR{O},\SKL{O},\SKG{O},\SKGb{O}\}
 \label{eq:Qbetadef2}
\end{equation}
which follows by virtue of $\gh{\Qbeta} =0$.  It is easy to check by explicit evaluation that the operations introduced so far satisfy
\begin{equation}
\gradcomm{\QSK}{\QKMSb} =  \gradcomm{\QSKb}{\QKMS} = \Qbeta \,,
\label{eq:}
\end{equation}
with all other graded commutators vanishing. In particular note that $\Qbeta$ has vanishing commutators with $\{\QSK,\QSKb,\QKMS,\QKMSb\}$.

But we now encounter a problem -- the three  KMS operators fail to generate a super-multiplet of actions. Based on the superfield construction  in \S\ref{sec:superfields} we might expect to find a fourth generator that completes them into a multiplet of super-transformations, on which $\QSK$ and $\QSKb$ act naturally as super-derivations along the lines of \eqref{eq:Qderact3}. This prompts us to ascertain a new Grassmann-even generator, $\Qzero$, which completes the KMS operations into a multiplet.

The easiest way to proceed is to intuit that $\Qzero$ action should only involve the KMS deviation differential operator $\delKMS$ and it should suitably intertwine with the other generators. Given that there is no passage from $\QSK$ to $\QKMS$ using $\Qbeta$ or likewise for their partners (the only ghost number conserving possibilities), we can ask if there is an operator $\Qzero$ that intertwines with $\QSK$ to produce $\QKMS$. One simple way to proceed is to require that the quartet of KMS operations fits into a diagram of the form \eqref{eq:qskaction}, viz.,
\begin{equation}
\begin{tikzcd}
&\Qzero \arrow{ld}{\QSK} \arrow{rd}[below]{\QSKb}    &   \\
\QKMS\arrow{rd}{\QSKb} & &  -\QKMSb \arrow{ld}[above]{\!\!\!\!\!\!\!\!\!\!\!\!\!\!-\QSK}\\
&   \Qbeta &
\end{tikzcd}
\label{eq:QzeroDiag0}
\end{equation}
where arrows indicate as before action via graded commutator, e.g., $\gradcomm{\QSK}{\Qzero} = \QKMS$ etc.. In a sense $\Qzero$ is the basic (`top component') thermal translation, with the other three generators appearing as its descendants in the Schwinger-Keldysh cohomology.

With this motivation in mind, it is now easy to write down the generator $\Qzero$, which has all the desired properties. It acts on the Schwinger-Keldysh quartet $\{\SKR{O}, \SKL{O}, \SKG{O}, \SKGb{O}\}$  as
 \begin{align}
   \gradcomm{\Qzero}{\SKL{O}} = \frac{i}{1-\FSgn{O} e^{-i\diffBi}} \left( \SKR{O} - \SKL{\tilde{O}} \right) \,,\qquad&
   \gradcomm{\Qzero}{\SKR{O}} = \frac{i\,\FSgn{O} e^{-i\diffBi}}{1-\FSgn{O} e^{-i\diffBi}} \left( \SKR{O} -\SKL{\tilde{O}}\right)\,, \nonumber \\
   \gradcomm{\Qzero}{\SKG{O}} = 0 \,,\qquad & \gradcomm{\Qzero}{\SKGb{O}} = 0 \,.
   \label{eq:Q0defOrig}
 \end{align}
Translating to the thermally adapted advanced-retarded basis we can simplify this action, for using our earlier definitions we find:
\begin{equation}
\gradcomm{\Qzero}{\SKRet{O}} =  \gradcomm{\Qzero}{\SKG{O}} = \gradcomm{\Qzero}{\SKGb{O}} = 0\,,
\qquad \gradcomm{\Qzero}{\SKAdv{O}} = \delKMS\SKRet{O} \,.
\label{eq:Q0ar}
\end{equation}
These definitions are fixed by demanding the diagram \eqref{eq:QzeroDiag0} up to overall normalization.\footnote{  The operator $\Qzero$ was called $i{\cal Q}_0$ in \cite{Haehl:2015foa}. As  with the KMS supercharges, this change is motivated by naturalness in superspace, cf., footnote \ref{foot:conventions}.} Our conventions are chosen to be natural from a superspace point of view as will become clear when we give a complete explanation of this structure from an equivariant cohomology point of view in \cite{Haehl:2016uah}.

Let us take stock of the various operators that we have defined on the Schwinger-Keldysh multiplet of fields
$\SF{\Op{O}}$. We have a total of six operators: the Schwinger-Keldysh supercharges $\{\QSK,\QSKb\}$, the KMS supercharges $\{\QKMS, \QKMSb\}$  and Grassmann-even generators $\{\Qbeta,\Qzero\}$. Given the commutation relations in \S\ref{sec:Qsk}, \S\ref{sec:Qkms} and \S\ref{sec:quadruplet} it is a simple matter to check that these supercharges give rise to a closed algebra, which is very reminiscent of supersymmetric structures:
\begin{gather}
\QSK^2 =\QSKb^2 = \QKMS^2=\QKMSb^2 = 0\ , \nonumber\\
\gradcomm{\QSK}{\QKMS} =  \gradcomm{\QSKb}{\QKMSb} = \gradcomm{\QSKb}{\QSK} =
\gradcomm{\QKMS}{\QKMSb} = 0\ ,\nonumber \\
\gradcomm{\QSK}{\QKMSb} =  \gradcomm{\QSKb}{\QKMS} = \Qbeta\,, \label{eq:Qkmsalg}\\
\gradcomm{\QKMS}{\Qzero} = \gradcomm{\QKMSb}{\Qzero} = 0 \,,\nonumber\\
\gradcomm{\QSK}{\Qzero} =  \QKMS\,, \qquad \gradcomm{\QSKb}{\Qzero} =- \QKMSb\,.\nonumber
\label{eq:kmsalg}
\end{gather}
All of these relations follow from the definitions given, as can be verified in short order. We will refer to this structure as the {\it SK-KMS superalgebra}. We now turn to its more compact and efficient description in superspace.

\subsection{SK-KMS superalgebra in superspace}
\label{sec:kmsalg}

The superspace representation of $\{\QSK\,\QSKb\}$ was given in \S\ref{sec:superfields}. As we saw there, it is very natural to think about the these operations as being represented on superspace as translation generators:
\begin{equation}
\QSK\simeq\partial_{\bar\theta}\,,\qquad \QSKb\simeq\partial_\theta\,,
\end{equation}
 which act genuinely as derivations on superspace, see also Eq.\ \eqref{eq:Qderact3}. By the symbol ``$\simeq$" we mean equality of operators with the understanding that $\{\QSK,\QSKb\}$ act component-wise, while the right hand sides act as super-derivations.  We now wish to give a similar representation of the KMS supercharges $\{\QKMS,\QKMSb,\Qbeta,\Qzero\}$.

In order to find the right superspace representation of the KMS supercharges, we recollect the following observation: the four KMS supercharges form a supermultiplet as in the diagram \eqref{eq:QzeroDiag0}. Taking the action of $\{\QSK,\QSKb\}$ as Grassmann-odd derivatives seriously, we would thus like to define operator valued superfields of thermal translations whose derivatives give the correct multiplet structure. This is easily achieved by the following linear combinations:
\begin{equation}\label{eq:SKsuperops}
\begin{split}
   \IKMSzero &\equiv \Qzero + \thb \, \QKMS - \theta \, \QKMSb + \thb \theta \, \Qbeta \,,\\
   \IKMS &\equiv \QKMS + \theta \, \Qbeta \,,\\
   \IKMSb &\equiv \QKMSb + \thb \, \Qbeta \,,\\
   \LKMS &\equiv  \Qbeta \,.
\end{split}
\end{equation}
Upon restriction to ordinary space ($\theta=0=\thb$), these operators reduce to the quadruplet of thermal translations. Hence they sensibly generalize the latter to superspace.

Our notation is chosen such as to suggest that $\LKMS$ acts as a super-Lie derivative operation, while $\{\IKMSzero,\IKMS,\IKMSb\}$ act as {\it interior contractions}. These operations should be familiar to readers from differential geometry, where given various differential forms on a manifold, we can Lie drag them along some vector field, or contract their indices against the same. We clearly do not have a manifold structure, but rather an operator super-algebra that acts on an enlarged Schwinger-Keldysh Hilbert space. The above notions of Lie derivation and interior contraction play a natural role in certain algebraic constructions of cohomology, which go by the moniker of equivariant cohomology. The notion of equivariance in this context refers to the set of algebraic structures that are compatible with (i.e., commute across) a group action. We will shortly unveil the group action we have for the SK-KMS algebra; it will turn out to non-trivially involve the KMS transformation. Intuitively one can think of the mathematical framework as a means to build covariant structures under this group action, as we do in gauge theories. A detailed review and exploration of equivariant cohomology and its relevance for the problem at hand is given in our companion paper \cite{Haehl:2016uah}.

For the present purposes, let us record some basic facts and continue with the algebra at hand. We have extended the six generator of the SK-KMS algebra to have natural actions on superspace. One salient feature of the combinations defined in \eqref{eq:SKsuperops} is that they help build covariant objects. We require the interior contraction operations to annihilate any super-operator that transforms covariantly:
\begin{equation}
\IKMSzero \, \SF{\mathbb{O}} = \IKMS \, \SF{\mathbb{O}} = \IKMSb \, \SF{\mathbb{O}}= 0 \,.
\end{equation}
While it appears that we could a-priori have made other choices compatible with the basic ghost number assignments, this choices enables direct contact with the language of equivariant cohomology.

Let us recast the SK-KMS algebra in terms of the super-operations $\IKMSzero$, $\IKMS$,
$\IKMSb$, $\LKMS$. We already know that the BRST charges act as super-derivations
$\QSK \simeq \partial_\thb$ and $\QSKb \simeq \partial_\theta$. It is then easy to infer their action on \eqref{eq:SKsuperops} and see that the structure \eqref{eq:QzeroDiag0} is reproduced directly:
\begin{equation}
\begin{tikzcd}
&\IKMSzero \arrow{ld}{\partial_\thb} \arrow{rd}[below]{\partial_\theta}    &   \\
\IKMS\arrow{rd}{\partial_\theta} & &  -\IKMSb \arrow{ld}[above]{\!\!\!\!\!\!\!\!\!\!\!\!\!\!-\partial_\thb}\\
&   \LKMS &
\end{tikzcd}
\label{eq:QzeroDiag}
\end{equation}
An equivalent  way to encode this multiplet structure is by the super-commutation rules:
\begin{equation}\label{eq:Sops3}
 \gradcomm{\QSK}{{\bf I}} = \partial_\thb {\bf I} \,,\qquad \gradcomm{\QSKb}{{\bf I}} = \partial_\theta {\bf I} \,,\qquad \gradcomm{{\bf I}}{{\bf I}'} = 0 \,,
\end{equation}
for any superoperators ${\bf I},{\bf I}'\in \{\IKMSzero,\IKMS,\IKMSb,\LKMS\}$.

While \eqref{eq:Sops3} is the most compact and elegant writing of the SK-KMS superalgebra, let us for sake of clarity expand out the relations encoded therein to write a superspace analog of Eq.\ \eqref{eq:kmsalg}:
\begin{gather}
\partial_\thb^2 =\partial_\theta^2 = (\IKMS)^2=  (\IKMSb)^2= 0\ , \nonumber\\
\gradcomm{\QSK}{\IKMS} =  \gradcomm{\QSKb}{\IKMSb} = \gradcomm{\QSKb}{\QSK} = \gradcomm{\IKMS}{ \IKMSb}= 0\ ,\nonumber \\
\gradcomm{\QSK}{ \IKMSb} =  \gradcomm{\QSKb}{\IKMS} = \LKMS \,, \label{eq:kmsalg2}\\
\gradcomm{\LKMS}{ \IKMSzero} = \gradcomm{ \IKMSb}{ \IKMSzero} = 0 \,,\nonumber\\
\gradcomm{\QSK}{ \IKMSzero} = \IKMS\,, \qquad \gradcomm{\QSKb}{ \IKMSzero} =  -\IKMSb\,.\nonumber
\end{gather}
In the following we give a brief discussion of this algebra which we argue is a particular case of well-known construction in algebraic topology.

\subsection{Thermal equivariant cohomology}
\label{sec:tech}

The SK-KMS algebra represented as abstract operators \eqref{eq:Qkmsalg} or in terms of super-derivations
\eqref{eq:kmsalg2} can be understood as an extended equivariant cohomology algebra. As explained briefly above, these constructions are relevant when there is a action of a group on some algebraic structure, and we seek to define objects invariant under the group action. The SK-KMS algebra was uncovered in our earlier work
\cite{Haehl:2015foa} where we argued that it has been encountered before in the string theory literature. In particular,  the six SK-KMS operators  generate a so-called  $\mathcal{N}_\smallT =2$  extended equivariant cohomology algebra. The notation following \cite{Dijkgraaf:1996tz} is meant to suggest that we have two topological (BRST) supercharges; here they are just $\QSK$ and $\QSKb$, which are CPT conjugates of each other.
We will explain the details of equivariant cohomology in the companion paper \cite{Haehl:2016uah}.
The key point to note is that such topological/cohomological structures naturally appear in the study of gauge theories, where one studies objects compatible with the action of a gauge group. This is the main idea embodied in the term `equivariance'.

Let us try to physically motivate this structure, in particular, the hitherto unexplained aspect of the mysterious gauge transformations that result in this equivariance. We can get a hint by examining the Lie derivative operator,
for the action of a gauge transformation on configurations is through a Lie drag along a group generator. Thus the Lie derivation can be viewed as the generator of infinitesimal gauge transformations. In the present case, the Lie derivative acts on an SK super-operator $\SF{\Op{O}}$ as a KMS deviation, viz.,
\begin{equation}
 \LKMS \SF{\Op{O}} \equiv \gradcomm{\Qbeta}{\SF{\Op{O}}} = \delKMS \SF{\Op{O}} \,.
\end{equation}

Let us try to unpack this. We started out by looking at the KMS condition as a discrete transformation around the thermal circle. The basic assertion was that in equilibrium an operator and its KMS conjugate are equivalent within correlation functions. As we have remarked in \S\ref{sec:kms}, the operator $\delKMS$ and its infinitesimal avatar
$\deltaB$  are best thought of state-dependent thermal translations along a thermal vector $\Kref^a$. We were there primarily concerned with a transformation that took us once around a thermal period: $\Qbeta$ only acts to map $\Op{O}(t) \mapsto \Op{O}(t-i\beta)$. However, if we open up the imaginary time direction and view the Euclidean time direction as being latticized ${\bf S}^1_\Kref = {\mathbb R}/{\mathbb Z}$, then we can extend our consideration to operators located at various lattice points. For instance we can consider  $\Op{O}(t-i\, m\, \beta)$ with $m \in {\mathbb Z}$ as being arbitrary KMS translates of the operator $\Op{O}(t)$. This structure would suffice for global equilibrium on Minkowski spacetime, but more generally when discussing thermal field theory on curved spacetime backgrounds following the arguments in \S\ref{sec:skthemal}, we would be led to upgrading $m\mapsto m(x)$ an integer valued function on the background geometry.

Allowing arbitrary thermal translations has two important consequences. Firstly, this implies that we can grade the KMS charges by another integer, which tells us how many thermal periods we have traversed.  Secondly, once we consider thermal translations by periods that depend on the spacetime location, we have to face up to the non-trivial fact that two successive thermal translations do not commute once we introduce inhomogeneities. This can be immediately inferred by noting that $\Qbeta$ for instance acts by Lie dragging the operator around the thermal circle, so two Lie drags by $m(x) \Kref^\mu(x)$ and $n(x)\Kref^\mu(x)$ will have a non-trivial commutator. We can check that the resulting behaviour of the commutator of two such transformations is along the `thermal commutator' $(m(x), n(x))_\Kref = m(x) \lieD_\Kref\, n(x) - n(x)\, \lieD_\Kref\,m(x)$.

While it would be interesting to understand the full algebra of discrete thermal translations, we found it easier to make progress in the continuum limit. Let us argue for this not just in equilibrium, but also extend considerations to near-equilibrium settings such as those relevant for low energy hydrodynamic effective field theories. The low energy description is valid on scale large compared to the thermal length scale, so $\omega \, T \ll1 $ and $k \,T \ll 1$. In this limit, we can effectively think of the thermal circle as being infinitesimally small for $\beta \to \infty$.

 Under these circumstances we may view the thermal circle as being fibered over the entire spacetime manifold (the background of our quantum field theory), and consider uplifting discrete thermal translations to local (i.e., spacetime dependent) continuous spacetime thermal translations.


Picking a  gauge parameter $\Lambda(x)$ we consider infinitesimal gauge transformations of the form
\begin{equation}\label{eq:OtrfUT}
  \SF{\Op{O}}\; \mapsto \;  \SF{\Op{O}} +  \Lambda\, \delKMS  \SF{\Op{O}}
\end{equation}
for an operator $\SF{\Op{O}}$ in the fundamental representation of the gauge symmetry.
The algebra generated by these transformations is the gauge algebra we seek. Computing successive gauge transformations and taking a commutator, we obtain a gauge transformation along the commutator involving a \emph{thermal Lie bracket}
\begin{equation}\label{eq:LLgtrf}
(\Lambda_1 ,\Lambda_2)_\Kref  = \Lambda_1 \delKMS \Lambda_2 - \Lambda_2 \delKMS \Lambda_1 \,.
\end{equation}
More generally, we can lift this discussion to superspace and  take the gauge parameter to be an adjoint superfield itself, call it $\SF{\Lambda}$. This leads us to postulate super-gauge transformations as in \eqref{eq:OtrfUT} with superfield gauge parameter $\SF{\Lambda}$. Note that $\SF{\Op{O}}$ in \eqref{eq:OtrfUT} was fundamental with respect to the symmetry. For fundamental fields, we can thus define the super-gauge transformation with the following thermal bracket:
\begin{equation}\label{eq:UTtrf0}
  \SF{\Op{O}} \; \mapsto \;  \SF{\Op{O}} + ( \SF{\Lambda} ,  \SF{\Op{O}} )_\Kref
  \equiv \SF{\Op{O}} + \SF{\Lambda} \delKMS \SF{\Op{O}} \,.
\end{equation}
 More generally, we could have, for example, adjoint fields $\SF{\mathfrak{F}}$.
An example for an adjoint field would be the gauge parameter $\SF{\Lambda}$ itself. Its gauge transformation should be schematically as in \eqref{eq:LLgtrf}. That is, a super-gauge transformation of a generic adjoint superfield $\SF{\mathfrak{F}}$ would read
\begin{equation}\label{eq:UTtrf}
  \SF{\mathfrak{F}} \; \mapsto \;  \SF{\mathfrak{F}} + ( \SF{\Lambda} ,  \SF{\mathfrak{F}} )_\Kref
  \equiv \SF{\mathfrak{F}} + \left\{ \SF{\Lambda} \delKMS \SF{\mathfrak{F}} - \SF{\mathfrak{F}} \delKMS \SF{\Lambda} \right\} \,.
\end{equation}
For explicit examples of this, we refer the reader to the companion paper \cite{Haehl:2016uah} (see also \cite{Haehl:2015uoc}).

The group of transformations inherits various features from diffeomorphisms along a circle, and appears to be a particular deformation of $\text{Diff}({\bf S}^1)$. Since it originates from the KMS condition and involves infinitesimal diffeomorphisms around the thermal circle, we refer to this gauge group as the $\UT$ \emph{KMS gauge symmetry}.

We first encountered this $\UT$ symmetry of gauged thermal translation in the context of hydrodynamics. Our initial postulate for this symmetry was motivated on phenomenological grounds as the missing ingredient in the construction of effective actions for adiabatic transport \cite{Haehl:2014zda,Haehl:2015pja}. In \cite{Haehl:2015uoc} it has been made more precise how to intertwine $\UT$ thermal diffeomorphisms with the supersymmetric structure of Schwinger-Keldysh theories. Earlier analyses of Lorentz anomalies in thermal field theory, also revealed a strong hint of such a symmetry operation \cite{Jensen:2012kj}. While we refer to the symmetry as
$\UT$ since it is a deformation of a circle diffeomorphism, the resulting group has non-abelian characteristics (as is easy enough to check by computing commutators). In fact it bears a strong resemblance to non-commutative gauge theories obtained via deformation quantization.

In \cite{Haehl:2016uah} we study this structure in much more detail and explain how to understand this in the standard
 language of equivariant cohomology. There, we construct this gauge theory of thermal translations in detail, starting from the above algebraic structures. A detailed analysis then reveals  that the continuum version of the discrete symmetries introduced via KMS supertranslations can indeed be consistently described by a group action of thermal diffeomorphisms as sketched above.

\subsection{Physical origins of the supercharges}
\label{sec:details}

Let us pause for a moment to reflect on the algebraic structures derived so far from a physical point of view.
For a Schwinger-Keldysh path-integral with an arbitrary initial state, we have  $\SKR{O}-\SKL{O}=0$ at the initial time. We assume w.l.o.g.\ that the initial state is prepared in some fashion; either by slicing open a Euclidean path integral or simply by giving appropriate data to pick out a desired density matrix. Further, the Schwinger-Keldysh
 boundary condition enforces $\SKR{O}-\SKL{O}=0$ at the final time. This shows that the Schwinger-Keldysh boundary conditions do not activate the difference operator. Further, the evolution of the system as defined by \eqref{eq:ZSKdef} only probes these operator in conjunction with the average operators. The upshot of these statements is that Schwinger-Keldysh construction has an invariance which respects the triviality (in a cohomological sense) of the difference operators. This we have rephrased in terms of an invariance under the supercharges $\{\QSK, \QSKb\}$ for an \emph{arbitrary initial state}.

If we choose the Schwinger-Keldysh action and measure to be $\QSKb$-closed, then there naturally
is a nice $\{\QSK, \QSKb\}$ -cohomological structure to the Schwinger-Keldysh correlators. This is a universal cohomological structure for an arbitrary Schwinger-Keldysh path integral, which as we motivate in \S\ref{sec:frbrst} is inherited from the field redefinition symmetries of the construction. The rationale for this structure, in particular the requirement that  the Schwinger-Keldysh action and measure be closed under $\{\QSK, \QSKb\}$ ensures some fundamental identities of time-ordered correlators.\footnote{ As explained later in \S\ref{sec:cpt} the two charges are related by CPT and thus their presence is natural in a CPT invariant theory. We will also see an explicit implementation of CPT with these two charges in \S\ref{sec:superrules}. }

The primary motivation for our construction is to arrive at the vanishing of difference correlation functions \eqref{eq:diff0}
which we view to be a remarkable and important statement. It holds for all operators and is absolutely agnostic to actual dynamics of the theory. Physically such a statement ought to follow from some underlying invariance, which should be a powerful one, since the statement holds for arbitrary operator insertions. More pertinently it is  independent of the insertion points and thus of the background on which the field theory is defined. The canonical manner in which this is achieved in known examples is the existence of a cohomological structure. We are simply observing that such an interpretation suffices to extract all the physical consequences of the Schwinger-Keldysh construction in arbitrary initial states. Closure of the action and invariance of the measure under $\{\QSK, \QSKb\}$ simply serve to enforce these in a dynamics agnostic fashion. We refer the reader to \S\ref{sec:superrules} for further discussion on the measure and implications for  super-correlation functions.

It is also worth clarifying the origin of two BRST charges: the charges $\{\QSK, \QSKb\}$ are a BRST anti-BRST pair, which is the standard structure in any cohomological setting. For instance in the geometric context of de Rham cohomology the supercharges may be viewed to be simply the exterior derivative $d$ and its adjoint $ d^\dag$, which are both nilpotent. In the physical context, consider  gauge theories where we can employ standard arguments to motivate  the BRST symmetry \cite{Becchi:1974md,Tyutin:1975qk}. While usual discussions of BRST invariance only focus on a single topological charge, the Faddeev-Popov construction implies  the existence of a pair of supercharges, cf., \cite{AlvarezGaume:1981zr}.  The anti-BRST charge in gauge theories usually does not lend a great deal of fundamental insight, but one may view its presence as restoring symmetry between the ghost and anti-ghost and thus maintains CPT invariance. Basically if we view the Nakanishi-Lautrup field as the top component in the superfield language, there should be two directions of descent down from the physical field. In one direction we increase the ghost number and in the other we decrease the ghost number -- the operators implementing this are the two BRST charges.

The presence of the two SK-supercharges, whilst implying the basic relations we expect for the Schwinger-Keldysh path integral, does not suffice to constrain the theory sufficiently. We will discuss how this might be alleviated in \S\ref{sec:entanglement} where we consider potential use of the modular Hamiltonian. For now however, we restrict attention to the case of thermal (or near-thermal) density matrices $\rhoT$ whence we have some mileage to extract from the KMS condition. Since the latter asserts that thermally translated operators are equivalent to the original ones, we have a second set of vanishing correlation functions \eqref{eq:diff1}. In \S\ref{sec:Qkms} we have seen that running the arguments of Schwinger-Keldysh field redefinitions once again with operators replaced by their KMS conjugates we are led to a second set of topological BRST symmetries  $\{\QKMS,\QKMSb\}$.

The combined structure of the four BRST charges turns out to be quite powerful in constraining the structure of thermal Schwinger-Keldysh theory. Now, while the statements  made above are but a rephrasing of known facts of the Schwinger-Keldysh construction, we want to argue that their presence can be put to good use in the construction of low energy Wilsonian dynamics in the Schwinger-Keldysh formalism. Microscopically, given a presentation of the Schwinger-Keldysh theory and the initial conditions, we can compute all the relevant correlation functions without ever making much mileage of the BRST structures. However, in trying to constrain the low energy theory without running into tension with microscopic unitarity, we will find the underlying topological structure extremely helpful.

It is very important to realize that our interest is in understanding the physical theory; the topological symmetries are but a helpful crutch in achieving this goal. The infra-red effective field theories we care about are most certainly not topological.
The underlying SK-KMS topological symmetries serve to constrain terms in the low energy dynamics. One should always be sensitive to the fact  that the effective field theories might involve dynamical degrees of freedom which are not  manifestly present in the microscopic theory, but rather are emergent in the infra-red. A-priori this makes constructing the dynamics in mixed states complicated. As indicated in \S\ref{sec:intro} an important issue involves figuring out how the influence functionals are constrained.

 The primary thesis of our discussion thus far is that the Schwinger-Keldysh construction has enough structure to help us in fixing the influence functionals to be consistent with requirements of microscopic unitarity. In near-thermal field theories such as hydrodynamics, the topological symmetries are not only sufficient to constrain these explicitly, but we also end up with low energy dynamics which is perfectly in synchrony with sensible physical expectations encoded in phenomenological formulations. As explained briefly in \cite{Haehl:2015uoc}  (see also \cite{Crossley:2015evo}) one ends up deriving the phenomenological theory of dissipative hydrodynamics which justifies the reformulation of the Schwinger-Keldysh formalism in terms of the BRST symmetries.  Our hope with the current discussion is that one may employ the general principles outlined herein to address questions in other physical contexts. A list of situations which would be fascinating from our perspective are discussed in \S\ref{sec:applications}.

\section{CPT symmetries}
\label{sec:cpt}

Let us now turn to examine the CPT properties of Schwinger-Keldysh path integrals. This discussion is not only useful to see how the discrete symmetries operate, but as we indicate below, they are also important in understanding the emergence of dissipation and arrow of time in the low energy theory.

We will first begin by lifting the usual notion of CPT from single copy field theory into  Schwinger-Keldysh and study its implications. As we will describe below, this crucially involves CPT conjugating
the initial state. We will then describe an alternate way of implementing CPT which exploits the fact that the two contours of Schwinger-Keldysh path integrals are time-reversed
copies of each other to implement time-reversal.

Let us start with the initial state $\rhoi$ and define $\rhoi^{\CPT}$ to be the CPT conjugated initial state. Existence of such a state is guaranteed in a local quantum field theory by the CPT theorem. Let us illustrate this point with a simple example from low energy dynamics. We consider a quantum field theory in $d$-dimensions in near-thermal equilibrium, where the dynamics is well approximated by a hydrodynamic effective field theory. Configurations in the latter can be thought of as fluid dynamical states characterized by a small number of intensive variables.

Consider then the example of a fluid state in flat spacetime specified by a velocity field $u^\sigma=\frac{(1,v_i)}{\sqrt{1-v^2}}$, temperature field $T$ and  chemical potential field(s) $\mu$. Its CPT conjugate state is given by $\{u^\sigma,T,\mu\}^{\CPT}$ with
\begin{equation}
\begin{split}
(u^\sigma)^{\CPT}(x^0,x^1,y) &\equiv \frac{(1,v_1(-x^0,-x^1,y),-v_y(-x^0,-x^1,y))}{\sqrt{1-v_1^2(-x^0,-x^1,y)-v^2_y(-x^0,-x^1,y)}}\ , \\
T^{\CPT}(x^0,x^1,y) &\equiv T(-x^0,-x^1,y)  \ , \\
\mu^{\CPT}(x^0,x^1,y)&\equiv-\mu(-x^0,-x^1,y)\ .
\end{split}
\end{equation}
Here $\{x^1,y\}$ are spatial Cartesian coordinates and $x^0$ denotes the time coordinate. Thus, we reverse  all of the chemical potentials, all but one velocities
along with a reversal of the flow in time and a reflection about one spatial direction. Note that both parity and time-reversal  flip the overall sign of the velocity $v_1$,
thus CPT does not flip its overall sign. In addition, time-reversal  flips the overall sign of the velocities $v_y$ and charge conjugation flips
the overall sign of the chemical potentials thus leading to the signs above.

Sometimes, it is more convenient  to add in a rotation which also flips all the $y$ directions (in even spacetime dimensions) or one that flips    all the $y$ directions
as well as the $x^1$ direction  (in odd spacetime dimensions). We get
\begin{equation}
\begin{split}
(u^\sigma)^{\CPT}(x^0,x^1,y) &\equiv \frac{(1,(-)^{d}v_1(-x^0,(-)^{d-1} x^1,-y),v_y(-x^0,(-)^{d-1}x^1,-y))}{\sqrt{1-v_1^2(-x^0,(-)^{d-1}x^1,-y)-v^2_y(-x^0,(-)^{d-1}x^1,-y)}}\ , \\
T^{\CPT}(x^0,x^1,y) &\equiv T(-x^0,(-)^{d-1}x^1,-y)  \ , \\
\mu^{\CPT}(x^0,x^1,y)&\equiv-\mu(-x^0,(-)^{d-1}x^1,-y)\ .
\end{split}
\end{equation}
In $d=4$, for example, this is the commonly used definition of CPT.

Whichever definition one may prefer, the basic principles of QFT ensure that the above state is an admissible initial state in the underlying quantum field theory. However,
we note a well-known subtlety: the macroscopic fluid equations  do not seem to have any such symmetry. This in turn means that
while there is  nothing problematic about performing the above transformation for the \emph{initial state}, one is not allowed to perform such
a transformation on a whole fluid solution over a period of time. This can also be seen by the fact that the above transformations naively
turn a dissipative configuration into an anti-dissipative configuration (and hence an admissible solution seems to be taken to an inadmissible solution).
This is a well-known feature of dissipation crucial to the conceptual foundations  of statistical mechanics: dissipative systems should be thought
of as systems where microscopic CPT is spontaneously broken.  This spontaneous breaking leads to various interesting consequences, including
fluctuation relations \cite{Gaspard:2012la}. With  this subtlety in mind, we will assume  for now that we can always CPT conjugate any given initial state.
As we will see later, there is an \emph{alternate} and a more convenient notion of CPT present in Schwinger-Keldysh formalism which sidesteps this subtlety.

In order to work out the CPT conjugate of the Schwinger-Keldysh path integral defined in \eqref{eq:ZSKdef},viz.,
\begin{equation}
\mathscr{Z}_{SK}[{\cal J}_\skR,{\cal J}_\skL] \equiv \Tr{\ U[{\cal J}_\skR]\ \rhoi\ (U[{\cal J}_\skL])^\dag\ } \, ,
\end{equation}
we will also need to CPT conjugate the left and the right sources $\{{\cal J}_\skR,{\cal J}_\skL\}$. This can be done by defining\footnote{ For a clear discussion of time reversal in Schwinger-Keldysh formalism, we refer the reader to  \cite{Sieberer:2015hba}. We discuss the connection to our implementation at the end of this section.\label{fn:sieb}} 
\begin{equation}
\begin{split}
{\cal J}_\skR^{\CPT}(x^0,x^1,y) &\equiv \sigma^{\CPT}_{{\cal J}_\skR} {\cal J}^*_\skR(-x^0,-x^1,y) , \\
{\cal J}_\skL^{\CPT}(x^0,x^1,y) &\equiv \sigma^{\CPT}_{{\cal J}_\skL} {\cal J}^*_\skL(-x^0,-x^1,y) .
\end{split}
\end{equation}
where $\sigma^{\CPT}_{{\cal J}_\skR}$ and   $\sigma^{\CPT}_{{\cal J}_\skL}$ are the appropriate intrinsic CPT parity operators and $^*$ denotes
the appropriate conjugation. In addition, for the CPT transformation to be anti-linear we also take $i \to (-i)$ after conjugation. 

If we denote by $\rhoi^{\CPT}$ the CPT conjugate state to the initial state, we then obtain the CPT conjugate SK path integral as
\begin{equation}
\begin{split}
\left(\mathscr{Z}_{SK}[{\cal J}_\skR,{\cal J}_\skL] \right)^{\CPT} &\equiv \Tr{\ U[{\cal J}^{\CPT}_\skR]\ \rhoi^{\CPT}\ (U[{\cal J}^{\CPT}_\skL])^\dag\ } \,,
\end{split}
\end{equation}
where we have left the anti-linear transformation implicit.  The CPT-invariance of the underlying microscopic theory is the assertion that
\begin{equation}
\begin{split}
 \mathscr{Z}_{SK}[{\cal J}_\skR,{\cal J}_\skL] = \left(\mathscr{Z}_{SK}[{\cal J}_\skR,{\cal J}_\skL] \right)^{\CPT}
 \,.
\end{split}
\end{equation}
If $\rhoi$ is thermal and without chemical potentials or charges, we can then write $\rhoi^{\CPT} = \rho^*$ and impose the above equation as a symmetry of Schwinger-Keldysh path integral. This is then the direct Schwinger-Keldysh counterpart of CPT in the usual path integrals which is however difficult to implement in an effective field theory because of the CPT conjugation on the
initial state. 

In Schwinger-Keldysh path-integrals, there is however an alternate way to implement CPT by  exploiting the fact that the two contours of Schwinger-Keldysh path integrals are time-reversed
copies of each other. So, one could intertwine usual CPT with an exchange of contours in order to get a new CPT transformation. We remind the reader that,
in general, when a theory  has some global and gauge symmetries, CPT can be defined as any one of the various  anti-unitary symmetries of the  path integral.
Different choices of CPT then merely differ by global and gauge symmetries of the theory and are all equally valid. Thus, our aim would be to find the simplest
anti-unitary symmetry of the Schwinger-Keldysh path integral. To this end consider the complex conjugate of Schwinger-Keldysh path integral:
 \begin{equation}
\begin{split}
 \mathscr{Z}_{SK}[{\cal J}_\skR,{\cal J}_\skL]^* &= \Tr{\ U[{\cal J}_\skR]\ \rhoi\ (U[{\cal J}_\skL])^\dag\ }^* =
 \Tr{\ U[{\cal J}_\skL^*]\ \rhoi^\dag\ U^\dagger[{\cal J}_\skR^*] \ } \\
 &=  \Tr{\ U[{\cal J}_\skL^*]\ \rhoi\ U^\dagger[{\cal J}_\skR^*]\ }  \, ,\\
 &=  \mathscr{Z}_{SK}[{\cal J}_\skL^*,{\cal J}_\skR^*]\,.
\end{split}
\end{equation}

Thus, the Schwinger-Keldysh path integral obeys the following reality condition: it is equal to its complex conjugate with right and left sources exchanged. Since complex conjugation is
explicitly an anti-unitary transformation, we then have a simple candidate for implementing CPT conjugation. This also allows us to sidestep the issue of CPT-conjugating
the initial state. Note that, in the version of Schwinger-Keldysh that occurs in the context of cutting rules a la Veltman \cite{tHooft:1973pz}; this is exactly the CPT action on the S-matrix
which exchanges ingoing and outgoing states.

We will now turn to how this CPT conjugation is implemented at the level of Schwinger-Keldysh superspace. Note that under the exchange of right and left theory, the average operators
map to themselves whereas the difference operators map to minus of themselves:
\begin{equation}
  (\SKAv{O})^{\CPT} = \SKAv{O} \,,\qquad (\SKDif{O})^{\CPT} = - \SKDif{O} \,.
\end{equation}
This implies that for the superfield to be covariant under CPT, we should have
$(\thetab\theta)^{\CPT} = - \thetab\theta$.
If we also demand that CPT exchanges the ghosts and the anti-ghosts,
\begin{equation}
(\SKG{O})^{\CPT} = \SKGb{O} \,,\qquad (\SKGb{O})^{\CPT} = \SKG{O} \,,
\end{equation}
 this implies
\begin{equation}
\thetab^{\CPT} = \theta \,, \qquad \theta^{\CPT} = \thetab \,,
\end{equation}
i.e., CPT is implemented as an R parity
on the superspace. From here on, we will use this definition of CPT and insist that the Schwinger-Keldysh path integrals be invariant under an anti-linear symmetry which exchanges right and left
fields and  exchanges   $\thetab$ and  $\theta$. We will exemplify this explicitly in the next section where we give a superspace version of the Keldysh rules.

Finally, let us see how our implementation relates to standard discussion of the discrete symmetries in the Schwinger-Keldysh formalism. As mentioned in footnote \ref{fn:sieb} a good discussion can be found in 
\cite{Sieberer:2015hba} Section V.B. In particular, in Eq.~(41) of {\cite{Sieberer:2015hba}, the authors  define the time reversal operation (called ${\sf T}'$ there) which reverses the time argument, and acts anti-linearly, whilst keeping the contour ordering intact. However, they also introduce a linear operation  ${\sf T}$ defined in Eq.~(40) of {\cite{Sieberer:2015hba} , which exchanges the left and right contours, with ${\sf T'}$ above. To be specific:
\begin{equation}
\begin{split}
 \text{Linear} \;\; & {\sf T}:  
\bigg( \begin{array}{c} \SKR{O}(t) \\ \SKL{O}(t)  \end{array} \bigg)  \mapsto \bigg(\begin{array}{c} \SKL{O}^*(-t) \\
\SKR{O}^*(-t)  \end{array} \bigg)   \\
\text{Anti-linear}\;\;   & {\sf T}':  \prn{\begin{array}{c} \SKR{O}(t) \\ \SKL{O}(t)  \end{array}}  \mapsto \prn{\begin{array}{c} \SKR{O}(-t) \\ 
\SKL{O}(-t)  \end{array}}   \\
\text{Anti-linear} \;\; & {\sf TT}':  \prn{\begin{array}{c} \SKR{O}(t) \\ \SKL{O}(t)  \end{array}}  \mapsto \prn{\begin{array}{c} \SKL{O}^*(t) \\ 
\SKR{O}^*(t)  \end{array}}  
\end{split}
\label{eq:sieb}
\end{equation}
In our implementation of  CPT, we invoke the anti-linear operation  ${\sf TT}'$ as corresponding to the time-reversal.

\section{Superspace Keldysh rules}
\label{sec:superrules}

Given the elegance of the Schwinger-Keldysh superspace formalism, it is desirable to understand better how the time ordering prescription of Schwinger-Keldysh contours is compatible with it. Ideally we would like to derive from the superfields the Keldysh bracket rules that give us an algorithm to convert the Schwinger-Keldysh two-sided correlation functions into the single-copy correlation functions. Along the way we should also determine all correlation functions involving the ghost fields we introduced to complete the Schwinger-Keldysh multiplet. All told we expect there should be a natural superspace prescription for determining all ghost correlators in terms of the physical single-copy correlators (nested commutators and anti-commutators). We now describe in some detail how this works.

\subsection{Correlation functions in Schwinger-Keldysh superspace}
\label{sec:supcorr}

Let us study a superspace $n$-point function with a super-SK time ordering $\SF{{\cal T}}_{SK}$ which we need to determine. We use the following notation for such super-correlators:
\begin{align}
\vev{\SF{{\cal T}}_{SK}\, \SF{\Op{A}}_1 \, \SF{\Op{A}}_2 \, \cdots \SF{\Op{A}}_n }
= \vev{ \SF{{\cal T}}_{SK}\, \prod_{k=1}^n
\, \left(\SKAv{A}^k + \theta_k\, \SKGb{A}^k + \bar{\theta}_k \,\SKG{A}^k+ \bar{\theta}_k\,\theta_k\, \SKDif{A}^k\right)} \,.
\label{eq:GeneralSKScorr}
\end{align}
We can expand the left hand side in the Grassmann odd coordinates, such that each component will then involve various combination of the physical fields and the ghost partners.
To keep the equations readable we will make some notational simplifications by writing:
\begin{align}
\avA_i \equiv \SKAv{A}^i  \,,
\quad \ghbA_i \equiv \SKGb{A}^i \,,
\quad \ghA_i \equiv \SKG{A}^i\,,
\quad \difA_i \equiv \SKDif{A}^i
\label{eq:skredefs}
\end{align}

The above correlation functions should satisfy some basic requirements. For instance, the fact that the Schwinger-Keldysh path integral  is invariant under $\QSK$ and $\QSKb$ implies that the  correlation functions have supertranslational invariance  in the Grassmann-odd directions. Recall here that the BRST operators act as derivations  $\partial_\theta$ and $\partial_{\thb}$ in superspace.   However, this statement will only give non-trivial answers for observables once we determine an appropriate measure for the path integral.

As is well known in topological field theories, the integration over the Grassmann-odd directions often involves zero modes for the ghost fields. If we do not soak up these ghost zero modes we will end up with a trivial correlation function. Before getting into the details, let us motivate a concern: the naive superspace correlation function \eqref{eq:GeneralSKScorr} will end up being trivial unless we determine the correct measure. This in particular entails that we ascertain the correct set of admissible boundary conditions.

Firstly, let us note that the rationale for introducing the Schwinger-Keldysh supercharges $\QSK$ and $\QSKb$ was to ensure that difference operators $\SKDif{O}$ are BRST exact. This guarantees that the correlation functions of only difference operators vanish as noted earlier. The correlation functions should also ensure another important fact visible in the Schwinger-Keldysh construction, the largest time equation \eqref{eq:LargestTime}, viz., that difference operators cannot be future-most, see \S\ref{sec:LargestTime}.

While these statements in terms of difference operators are manifest for the purposes of the Schwinger-Keldysh path integral, in superspace we need to decide whether the future-most state is annihilated by $\QSK$ or by $\QSKb$. Demanding supertranslational invariance with respect to both of these charges is tantamount to requiring that the future-most state is annihilated by both. We remind the reader that the Schwinger-Keldysh construction projects the left-right evolution onto the maximally entangled state at the future-most point.

These are the only requirements that we need to impose on any superspace rules. We must not demand any specific action on the initial state, since we are free to pick any initial density matrix $\rhoi$, which may have a-priori exhibit any pattern of entanglement between the left and right fields. In particular, we should not be asking for the difference operators to have any particular action on the initial state generically.

It turns out that one is unable to give a consistent prescription that determines all the superspace correlation functions subject to the requirements detailed above, without soaking up some ghost zero modes. In other words
we have to decide how to treat the ghosts vis a vis the initial and final states. Demanding that the final maximally entangled state be annihilated by both the ghost and the anti-ghost will turn out to be too constraining, unless we source background ghost zero modes from the initial state. We will demonstrate this explicitly below without derivation.

 For now based on the above discussion  let us postulate that the superspace correlation function should allow for the insertion of at least one ghost zero mode.\footnote{ An inspiration for this proposal is the manner in which world-sheet vertex operators in string perturbation theory require ghost zero mode dressing. } Let $\SF{\Op{A}}_0$ be a reference background superfield of zero modes which we take to be of the form:
\begin{equation}
\SF{\Op{A}}_0 =
\mathbb{1} + \theta_0 \,  \ghbA_0  + \thetab_0 \,  \ghA_0 +  \thetab_0 \theta_0 \; \difA_0 \,.
\label{}
\end{equation}
We now proceed to demonstrate that the supercorrelation functions are given by demanding $\bar{\theta}$ and $\theta$ translational invariance of the defining object:
\begin{align}
\Tr{\rhoi \, \SF{\Op{A}}_1 \, \SF{\Op{A}}_2 \, \cdots \SF{\Op{A}}_n }\equiv
\vev{\SF{{\cal T}}_{SK}\, \SF{\Op{A}}_1 \, \SF{\Op{A}}_2 \, \cdots \SF{\Op{A}}_n \SF{\Op{A}}_0}\,.
\label{eq:ScorrTot}
\end{align}

The superspace expansion of such a correlator will lead to various terms involving the Grassmann coordinates
$\bar{\theta}_i$ and $\theta_j$. Imposing supertranslation invariance will then lead to relations between these components of the correlator. It is clear is that any such  relation can only involve terms with the same number of $\bar{\theta}_i\theta_j$ pairs. Another physical consequence of supertranslation invariance is that only correlation functions of operators with total ghost number charge zero can be non-vanishing. This gives a nice superselection rule for our correlators. In particular, note that a non-vanishing correlator not only should be balanced in the occurrence of ghost, anti-ghost fields, but should also respect the rule that a pair of ghost-anti-ghost fields can be traded for an average-difference pair of fields (essentially by $\QSK$ or $\QSKb$ actions).

These observations imply that any relation we obtain between components  in the superspace expansion of \eqref{eq:ScorrTot} will be at a given number of $\bar{\theta}_i\theta_j$ pairs. As a result we will break up any given supercorrelation function into {\it levels} based on the number of these pairs. We refer to the superspace expansion of the $n$-point correlation function that has $n_d$ pairs of $\bar{\theta}_i\theta_j$ as the level $n_d$ correlator of $n$ fields. We further denote the set of $n$-point functions at level $n_d$ as ${}^n{\bf L}_{n_d}$. It is clear that a correlator of type ${}^n{\bf L}_{n_d}$ contains at most $n_d$ difference fields; in fact the ghost-free correlator in this family has precisely $n_d$ difference and $n_a \equiv n-n_d$ average fields. As noted above a pair of ghosts, $\ghA_i \ghbA_j$, can be counted equivalently as an average-difference pair, $\difA_i \avA_j$, justifying our classification. We now demonstrate how this works in practice with some low point correlation functions explicitly, before giving the general result.

\subsection{Two-point functions}
\label{sec:2ptSuper}

A generic superfield two-point function reads
\begin{equation}
\begin{split}
\Tr{\rhoi \, \SF{\Op{A}}_1 \, \SF{\Op{A}}_2  }&\equiv
\vev{\SF{{\cal T}}_{SK}\, \SF{\Op{A}}_1 \, \SF{\Op{A}}_2 \, \SF{\Op{A}}_0}\,,
 \\
&= \vev{\SF{{\cal T}}_{SK}\,
\left(\avA_1+ \theta_1\, \ghbA_1+ \bar{\theta}_1\,\ghA_1+ \bar{\theta}_1\,\theta_1\, \difA_1\right)
 \left(\avA_2 + \theta_2\, \ghbA_2 + \bar{\theta}_2 \,\ghA_2+ \bar{\theta}_2\,\theta_2\, \difA_2\right)
\\
&  \hspace{1.5cm} \times
 \left(\mathbb{1} + \theta_0 \,  \ghbA_0  + \thetab_0 \,  \ghA_0 +  \thetab_0 \theta_0 \, \difA_0 \right)
 }\,.
\end{split}
\label{eq:2ptExpand}
\end{equation}
By superspace translation invariance, it is immediately clear that all correlators with non-zero net ghost number have to vanish.
We can thus proceed to expand out the r.h.s. and collect the non-trivial terms of vanishing ghost number at various levels:
\begin{equation}
\begin{split}
^2{\bf L}_0:& \quad \vev{\SF{{\cal T}}_{SK}\,\avA_1\, \avA_2}
\\
^2{\bf L}_1:& \quad -\vev{\SF{{\cal T}}_{SK}\, \bar{\theta}_1 \, \theta_2 \;\ghA_1\, \ghbA_2} -\vev{\SF{{\cal T}}_{SK}\, \bar{\theta}_2 \, \theta_1 \;\ghA_2\, \ghbA_1}
-\vev{\SF{{\cal T}}_{SK}\, \bar{\theta}_1 \, \theta_0 \;\ghA_1\, \ghbA_0}
-\vev{\SF{{\cal T}}_{SK}\, \bar{\theta}_2 \, \theta_0 \;\ghA_2\, \ghbA_0}
\\
& \qquad
-\vev{\SF{{\cal T}}_{SK}\, \bar{\theta}_0 \, \theta_1 \;\ghbA_1\, \ghA_0}
-\vev{\SF{{\cal T}}_{SK}\, \bar{\theta}_0 \, \theta_2 \;\ghbA_2\, \ghA_0}
+\vev{\SF{{\cal T}}_{SK}\, \bar{\theta}_0 \, \theta_0 \; \avA_1 \,\avA_2\, \difA_0}
\\
& \qquad
 +\vev{\SF{{\cal T}}_{SK}\, \bar{\theta}_1 \, \theta_1 \;\avA_2\, \difA_1}
  +\vev{\SF{{\cal T}}_{SK}\, \bar{\theta}_2 \, \theta_2 \;\avA_1\, \difA_2}
  \\
^2{\bf L}_2:& \quad
- \vev{\SF{{\cal T}}_{SK}\, \bar{\theta}_1 \bar{\theta}_2 \, \theta_0 \, \theta_1\;
\ghA_2\, \ghbA_0 \, \difA_1}
+ \vev{\SF{{\cal T}}_{SK}\, \bar{\theta}_1 \bar{\theta}_2 \, \theta_0 \, \theta_2\;
\ghA_1\, \ghbA_0 \, \difA_2}
\\
& \qquad
- \vev{\SF{{\cal T}}_{SK}\, \bar{\theta}_0 \bar{\theta}_1 \, \theta_1 \, \theta_2\;
\ghA_0\, \ghbA_2 \, \difA_1}
+ \vev{\SF{{\cal T}}_{SK}\, \bar{\theta}_0 \bar{\theta}_2 \, \theta_1 \, \theta_2\;
\ghA_0\, \ghbA_1 \, \difA_2}
\\
& \qquad
- \vev{\SF{{\cal T}}_{SK}\, \bar{\theta}_0 \bar{\theta}_1 \, \theta_0 \, \theta_1\;
\difA_0 \, \difA_1}
- \vev{\SF{{\cal T}}_{SK}\, \bar{\theta}_0 \bar{\theta}_2 \, \theta_0 \, \theta_2\;
\difA_0\,  \difA_2}
\\
& \qquad
+\vev{\SF{{\cal T}}_{SK}\, \bar{\theta}_0 \bar{\theta}_1 \, \theta_0 \, \theta_2\;
\difA_0 \, \ghA_1\, \ghbA_2}
+\vev{\SF{{\cal T}}_{SK}\, \bar{\theta}_0 \bar{\theta}_2 \, \theta_0 \, \theta_1\;
\difA_0\,\ghbA_1\, \ghA_2}
\\
& \qquad
- \vev{\SF{{\cal T}}_{SK}\, \bar{\theta}_1 \bar{\theta}_2 \, \theta_1 \, \theta_2\;
\difA_1 \, \difA_2 }
\end{split}
\label{}
\end{equation}
We now can analyze each level in turn and learn what the consequences of requiring supertranslational invariance is.

\begin{itemize}
\item $^2{\bf L}_0$: At level 0, there are no $\thetab$s or $\theta$s, so we do not obtain any constraints from supertranslation invariance. This is as it should be as the correlation function of all average operators is a symmetrized function. We therefore can write the general expression for a two point function of superspace average opeators (with $i_1,i_2 \in \{1,2\}$ and $i_1 \neq i_2$)
\begin{equation}
 \quad \vev{\SF{{\cal T}}_{SK}\,\avA_{i_1}\, \avA_{i_2}} =
\vev{{\cal T}_{SK}\,\avA_{i_1}\, \avA_{i_2}}  = \vev{\gradAnti{\OpH{A}_{i_1}}{\OpH{A}_{i_2}}}\,.
\label{}
\end{equation}
\item $^2{\bf L}_1$: Things are lot more interesting at level 1. Here we have to impose invariance with respect to shifts of $\bar{\theta}$s and $\theta$s respectively. This way we find six relations, which can be written succintly as two relations up to index permutations and ghost, anti-ghost exchange:
\begin{equation}
\begin{split}
\vev{{\cal T}_{SK} \ \ghA_{i_2} \, \ghbA_0 \, \avA_{i_1} } +
\vev{{\cal T}_{SK} \ \ghA_{i_2} \, \ghbA_{i_1}} - \vev{{\cal T}_{SK} \ \avA_{i_1} \, \difA_{i_2} } &=0 \,,
\\
 \vev{ {\cal T}_{SK} \ \ghA_0 \, \ghbA_{i_2} \, \avA_{i_1}} + \vev{{\cal T}_{SK} \ \ghA_{i_1} \, \ghbA_{i_2}}
- \vev{{\cal T}_{SK} \ \avA_{i_1} \, \difA_{i_2} }  &=0\,,
\\
\vev{{\cal T}_{SK} \ \ghA_{i_1} \, \ghbA_0 \,\avA_{i_2}} + \vev{{\cal T}_{SK} \ \ghA_{i_2} \, \ghbA_0 \,\avA_{i_1}}
-\vev{{\cal T}_{SK} \ \difA_0\,\avA_{i_1} \,\avA_{i_2}} &= 0  \,,
\\
\vev{{\cal T}_{SK} \ \ghA_0 \, \ghbA_{i_1} \,\avA_{i_2}} + \vev{{\cal T}_{SK} \ \ghA_0 \, \ghbA_{i_2 }\,\avA_{i_1}}
- \vev{{\cal T}_{SK} \ \difA_0\,\avA_{i_1} \,\avA_{i_2}} &= 0 \,.
\end{split}
\label{}
\end{equation}
The equations as written are pairwise CPT conjugate of each other; CPT exchanges a ghost for an anti-ghost. Explicitly writing these out with choices of $i_1, i_2 \in \{1,2\}$ the two equations gives four relations, while the second pair gives rise to two relations.
We can solve this system of equations to determine the ghost anti-ghost correlation functions in terms of the physical average-difference correlators. However, it turns out that the solution is not unique: there exists a one-parameter ambiguity, which we will parameterize by an arbitrary constant $\al{1}{2,1}$. The superscript refers to the level and the subscript indexes the ambiguity.

This example illustrates the need for the ghost zero mode insertion into the correlation function. If we set $\SF{\Op{A}}_0 \mapsto \mathbb{1}$, then we would be led to a contradiction. The ghost anti-ghost correlations
$\vev{{\cal T}_{SK} \ \ghA_{i_2} \, \ghbA_{i_1}}$ from the first equation together with its CPT conjugate would demand that we set
$\vev{{\cal T}_{SK} \ \avA_{i_1} \, \difA_{i_2} } =\vev{{\cal T}_{SK} \ \avA_{i_2} \, \difA_{i_1} } $.
We however know that the difference operator cannot be future-most; each of the above is non-vanishing for a  complementary temporal ordering. We would then be forced to conclude that we need to set both of these correlators to zero. This however is unphysical and suggests that we have not accounted for
 the ghost charges of the initial state correctly.

We then learn that the set of correlators are determined to be
\begin{equation}
\begin{split}
\vev{{\cal T}_{SK} \ \ghA_{i_1}\, \ghbA_{i_2}}   &=
    \al{1}{2,1}\left( -\vev{{\cal T}_{SK} \ \avA_{i_1} \, \difA_{i_2} } + \vev{{\cal T}_{SK} \ \difA_{i_1} \, \avA_{i_2} } \right)
\\
& =
    - \al{1}{2,1} \vev{\gradcomm{\OpH{A}_{i_1}}{\OpH{A}_{i_2}}}\,,
\\
\vev{{\cal T}_{SK} \ \ghA_{i_1} \, \ghbA_0 \,\avA_{i_2}}
& =
    \vev{{\cal T}_{SK} \ \avA_{i_2} \, \difA_{i_1}} + \al{1}{2,1}\left(  \vev{{\cal T}_{SK} \ \avA_{i_1} \, \difA_{i_2} } -\vev{{\cal T}_{SK} \ \avA_{i_2} \, \difA_{i_1}}\right) \\
&=
    \stepFn{\Op{A}_{i_2} \Op{A}_{i_1}} \vev{\gradcomm{\OpH{A}_{i_2}}{\OpH{A}_{i_1}} } + \al{1}{2,1} \vev{\gradcomm{\OpH{A}_{i_1}}{\OpH{A}_{i_2}}}\,,
\\
\vev{{\cal T}_{SK} \ \ghbA_{i_1} \, \ghA_0 \,\avA_{i_2}}
& =
    -\vev{{\cal T}_{SK} \ \avA_{i_2} \, \difA_{i_1}} +\al{1}{2,1}\left(  \vev{{\cal T}_{SK} \ \avA_{i_1} \, \difA_{i_2} } -\vev{{\cal T}_{SK} \ \avA_{i_2} \, \difA_{i_1}}\right) \\
&=
    -\stepFn{\Op{A}_{i_2} \Op{A}_{i_1}} \vev{\gradcomm{\OpH{A}_{i_2}}{\OpH{A}_{i_1}} } + \al{1}{2,1} \vev{\gradcomm{\OpH{A}_{i_1}}{\OpH{A}_{i_2}}}\,,
\\
\vev{{\cal T}_{SK} \ \difA_0 \,\avA_{i_1}\,\avA_{i_2}}
& =
    \vev{{\cal T}_{SK} \ \avA_{i_1} \, \difA_{i_2}} + \vev{{\cal T}_{SK} \ \avA_{i_2} \, \difA_{i_1}}
\\
&=
    \left(\stepFn{\Op{A}_{i_1} \Op{A}_{i_2}} -\stepFn{\Op{A}_{i_2} \Op{A}_{i_1}} \right) \vev{\gradcomm{\OpH{A}_{i_1}}{\OpH{A}_{i_2}} } \,.
\end{split}
\label{}
\end{equation}
Note that for the canonical choice $\al{1}{2,1}=0$, the solution is very simple. The Schwinger-Keldysh partner ghost anti-ghost correlator vanishes, and we are only left with correlators involving the background ghosts.

\item $^2{\bf L}_2$: For two-point functions level two is the highest achievable level, despite the background ghost superfield $\SF{\Op{A}}_0$ insertion. It is a simple matter to learn that supertranslational invariance demands the expected answer:
\begin{equation}
\vev{{\cal T}_{SK} \ \difA_{i_1} \, \difA_{i_2}} =  \vev{{\cal T}_{SK} \ \ghA_{i_1} \, \ghbA_0\, \difA_{i_2}} =  \vev{{\cal T}_{SK} \ \ghbA_{i_1} \, \ghA_0\, \difA_{i_2}} =0 \,.
\label{}
\end{equation}
\end{itemize}

Having explicitly solved the two-point function relations from supertranslation invariance by hand, let us now give a more general perspective for generating the solutions. We have seen that levels 0 and 2 are trivial, so they do not give much information. However, we can package the solution for level 1 in a very compact fashion
by writing down the most general supertranslational invariant ansatz.  Since we want to express the final result in terms of the average-difference correlators which are correctly time-ordered, consider then the ansatz:
\begin{align}\label{eq:Phi21}
 ^2\Phi_1 =  \sum_{i,j=1}^2  \left[\thetab_{0j}\theta_{0j}  +\frac{1}{2} \,\al{1}{2,1} \left(\thetab_{0i}\theta_{0j} +\thetab_{0j} \theta_{0i} \right) \right] \stepFn{ij} \, \langle \avA_i \, \difA_j  \rangle\,.
\end{align}
Here, we use the supertranslation invariant combinations
\begin{equation}\label{eq:TransInv}
 \thetab_{ij} \equiv \thetab_i - \thetab_j \,,\qquad  \theta_{ij} \equiv \theta_i - \theta_j \,.
\end{equation}
It is a simple matter to check that expanding out $^2\Phi_1$ and matching it with the supercorrelator \eqref{eq:2ptExpand} yields precisely the solutions to the various ghost correlators which we enlisted above. The secret reason for this is that we constructed the solution \eqref{eq:Phi21} as the most general linear combination of $\langle \avA_i \, \difA_j  \rangle$ correlators, which satisfies the following necessary conditions:
\begin{enumerate}
 \item Supertranslation invariance: Grassmann coordinates can only occur in the translation invariant combinations \eqref{eq:TransInv}.
  \item A-priori there are $n+1$ allowed combinations $\thetab_{ij}$, but only $n$ of them are linearly independent (similarly for $\theta_{ij}$). Without loss of generality, we can hence assume that ghost zero-modes are part of each translation invariant combination, i.e., the only allowed Grassmann-odd factors are $\thetab_{0i}$ and $\theta_{0j}$.
 \item Net ghost number should be zero: that is, Grassmann coordinates can only occur in the product form $\thetab_{0i} \theta_{0j}$.
 \item CPT invariance: this requires that to any combination $\thetab_{0i_1}\theta_{0j_1}\cdots \thetab_{0i_r}\theta_{0j_r}$, we need to add its CPT conjugate $\thetab_{0j_1}\theta_{0i_1}\cdots \thetab_{0j_r}\theta_{0i_r}$. For instance, this explains the form of the bracketed ambiguity term in \eqref{eq:Phi21}.
\end{enumerate}

With these rules one can check that the ansatz \eqref{eq:Phi21} exhausts all possibilities. The crucial thing to keep in mind for the future analysis is that we will isolate the contributions of the form $\bar{\theta}_{0j} \,\theta_{0j}$ and deem the correlator to be primarily determined by this combination. The remaining combinations like the coefficient of the $\al{1}{2,1}$ will be referred to as ambiguities.

We will henceforth use the notation ${}^n \Phi_{n_d}$ to denote an expression analogous to \eqref{eq:Phi21} that encodes the solutions and ambiguities in the ghost correlators for an $n$-point function at level $n_d$. To complete the two-point function discussion, we note that
\begin{equation}
{}^2\Phi_0 =  \vev{ {\cal T}_{SK} \ \avA_{i_1} \, \avA_{i_2} }\,, \qquad {}^2\Phi_2 = 0 \,,
\label{}
\end{equation}
 encode the obvious solutions at lowest and highest level, as described above.

\subsection{Three-point functions}
\label{sec:3ptSuper}

There are many relations at the three point function level which have to be unpacked.
\begin{itemize}
\item $^3{\bf L}_0$: As in the two-point function case this is simple. We get a correlation function of three average operators, $\vev{\SF{{\cal T}}_{SK}\,\avA_{i_1}\, \avA_{i_2}\, \avA_{i_3}}$ which is unconstrained and fully symmetric, and was determined in \eqref{eq:3ptCorrA}. Equivalently we can simply write:
\begin{equation}
{}^3\Phi_{0} = \vev{ {\cal T}_{SK} \ \avA_{i_1} \, \avA_{i_2} \, \avA_{i_3} }\,.
\end{equation}
\item $^3{\bf L}_1$: It is useful to perform a count of the possible operators. We have
3 correlators which are of the form $\vev{{\cal T}_{SK} \ \avA_{i_1} \, \avA_{i_2} \,\difA_{i_3}}$.
There are likewise 6 correlators with a single $\ghbA_0$ or $\ghA_0$, viz.,
$\vev{{\cal T}_{SK} \ \ghA_{i_1} \, \ghbA_0\, \avA_{i_2} \, \avA_{i_3}} $ and their CPT conjugates
$\vev{{\cal T}_{SK} \ \ghbA_{i_1} \, \ghA_0\, \avA_{i_2} \, \avA_{i_3}} $.
However, now there are many more ghost anti-ghost correlation functions. With the choice of two positions to fill with a ghost or an anti-ghost, we find 6 correlators of the form $\vev{{\cal T}_{SK} \  \ghA_{i_1}\, \ghbA_{i_2} \,\avA_{i_3}} $ and 1 correlator $\vev{{\cal T}_{SK} \  \difA_0 \,\avA_{i_1}\,\avA_{i_2}\,\avA_{i_3}} $. This yields a total of 16 correlators, only the first 3 of which are physical (viz., do not contain any ghosts).

There are however  4 $\bar{\theta}$ and 4 $\theta$ translations  at our disposal, so we only obtain 8 relations.
These 8 relations allow us to determine the 13 correlators involving ghosts in terms of the average-difference correlators. The relations we find from supertranslation invariance modulo index permutations and CPT conjugation are:
\begin{equation}
\begin{split}
& \vev{{\cal T}_{SK}\ \ghA_{i_3}\, \ghbA_{i_1}\, \avA_{i_2} } +
 \vev{{\cal T}_{SK}\ \ghA_{i_3} \,\ghbA_{i_2} \,\avA_{i_1} } +
  \vev{{\cal T}_{SK}\ \ghA_{i_3}\, \ghbA_0\, \avA_{i_1} \,\avA_{i_2} } -
   \vev{{\cal T}_{SK}\ \difA_{i_3} \, \avA_{i_1} \, \avA_{i_2} }
   =0 \,,\\
  & \vev{{\cal T}_{SK}\ \ghA_{i_1}\, \ghbA_{i_3}\, \avA_{i_2}} +
   \vev{{\cal T}_{SK}\ \ghA_{i_2} \,\ghbA_{i_3}\,\avA_{i_1} } +
   \vev{{\cal T}_{SK}\ \ghA_0\, \ghbA_{i_3}\, \avA_{i_1} \,\avA_{i_2}} -
     \vev{{\cal T}_{SK}\ \difA_{i_3} \, \avA_{i_1}\, \avA_{i_2} }
   =0 \,,\\
 &\vev{{\cal T}_{SK}\ \ghA_0 \, \ghbA_{i_1} \, \avA_{i_2} \, \avA_{i_3}}
 + \vev{{\cal T}_{SK}\ \ghA_0 \, \ghbA_{i_2} \, \avA_{i_1} \, \avA_{i_3}}
 + \vev{{\cal T}_{SK}\ \ghA_0 \, \ghbA_{i_3} \, \avA_{i_1} \, \avA_{i_2}}
- \vev{{\cal T}_{SK}\ \difA_0 \, \avA_{i_1}\, \avA_{i_2} \, \avA_{i_3}}
 = 0 \,, \\
&   \vev{{\cal T}_{SK}\ \ghA_{i_1} \, \ghbA_0 \, \avA_{i_2} \, \avA_{i_3}}
+  \vev{{\cal T}_{SK}\ \ghA_{i_2}\, \ghbA_0 \, \avA_{i_1} \, \avA_{i_3}}
+  \vev{{\cal T}_{SK}\ \ghA_{i_3} \, \ghbA_0 \, \avA_{i_1}\, \avA_{i_2} }
-\vev{{\cal T}_{SK}\ \difA_0 \, \avA_{i_1}\, \avA_{i_2} \, \avA_{i_3}}   =0\,.
\end{split}
\end{equation}
 These relations can again be solved for the ghost correlators in terms of the average-difference ones. The general solution in this case is parameterized by 6 ambiguities $\al{k}{3,1}$ with $k=1,\ldots,6$.

Instead of writing all these solutions explicitly, we use the more abstract construction from the end of the previous subsection. The principles enumerated there lead us to construct the following ansatz which encodes all three-point functions at level 1:
\begin{equation}\label{eq:Phi31}
\begin{split}
 ^3\Phi_1 &=  \sum_{i_1,i_2,j_1=1}^3  \bigg\{ \,
 \Big[\thetab_{0j_1}\theta_{0j_1}  + \al{1}{3,1} \left(\thetab_{0j_1}\theta_{0i_1} +\thetab_{0i_1}\theta_{0j_1}\right) + \al{2}{3,1} \left(\thetab_{0j_1}\theta_{0i_2} +\thetab_{0i_2}\theta_{0j_1}\right)  \\
  &\qquad\qquad\qquad\quad + \al{3}{3,1} \left(\thetab_{0i_1}\theta_{0i_2} +\thetab_{0i_2}\theta_{0i_1}\right)  \bigg] \stepFn{i_1 j_1 i_2}\\
  &\qquad\qquad\;\;\, +  \Big[\thetab_{0j_1}\theta_{0j_1}  + \al{4}{3,1} \left(\thetab_{0j_1}\theta_{0i_1} +\thetab_{0i_1}\theta_{0j_1}\right) + \al{5}{3,1} \left(\thetab_{0j_1}\theta_{0i_2} +\thetab_{0i_2}\theta_{0j_1}\right) \\
  &\qquad\qquad\qquad\quad + \al{6}{3,1} \left(\thetab_{0i_1}\theta_{0i_2} +\thetab_{0i_2}\theta_{0i_1}\right)  \bigg] \stepFn{i_1 i_2 j_1}\; \bigg\} \;\langle \avA_{i_1} \, \avA_{i_2} \, \difA_{j_1}  \rangle\,.
\end{split}
\end{equation}
This expression deserves some explanation. First, note that there are two possible time orderings, corresponding to the allowed positions of the difference operator $\difA_{j_1}$ subject to the requirement that it cannot be the future-most insertion. For each such time ordering there exists one basic solution which as advertised earlier we take to be given by the combination of $\bar{\theta}_{0j} \theta_{0j}$; its normalization is fixed to unity by  matching with Eq.~\eqref{eq:GeneralSKScorr}). In addition we have a 3-parameter family of ghost correlator ambiguities for each time ordering which have been parameterized by the arbitrary coefficients $\{\al{k}{3,1}\}_{k=1,2,3}$ and  $\{\al{k}{3,1}\}_{k=4,5,6}$, respectively.

By matching the superspace expansions of the generic three-point correlator, Eq.~\eqref{eq:GeneralSKScorr} with $n=3$, with ${}^3\Phi_1$, all ghost correlators are determined. For sake of completeness, we list them in Appendix \ref{app:SuperCorr}, including all ambiguities. Here, for brevity we only note the basic solution for the canonical case where all ambiguities $\al{k}{3,1}$ are set to zero:
\begin{equation}
\begin{split}
 \vev{\mathcal{T}_{SK}\ \ghA_{i_1} \, \ghbA_{i_2} \, \avA_{i_3}} &= 0 \,,\\
 \vev{\mathcal{T}_{SK}\   \ghA_j \, \ghbA_0\,\avA_{i_1}\, \avA_{i_2}}
  &= \vev{\mathcal{T}_{SK}\   \ghA_0 \,\ghbA_j\, \avA_{i_1}\, \avA_{i_2}}  =
   \vev{ \mathcal{T}_{SK} \ \difA_j \, \avA_{i_1}\,\avA_{i_2} }\,,
  \\
    \vev{\mathcal{T}_{SK}\ \difA_0  \, \avA_{i_1} \, \avA_{i_2}\, \avA_{i_3}} &  = \vev{ \mathcal{T}_{SK} \ \difA_{i_1} \, \avA_{i_2}\,\avA_{i_3} }  + \vev{ \mathcal{T}_{SK} \ \avA_{i_1} \, \difA_{i_2}\,\avA_{i_3} }  + \vev{ \mathcal{T}_{SK} \ \avA_{i_1} \, \avA_{i_2}\,\difA_{i_3} }  \,,
\end{split}
\end{equation}
where explicit expressions for average-difference correlators on the right hand sides can be found in \S\ref{sec:keldysh}.

\item $^3{\bf L}_2$: The situation here is similar to the level 2 result of the two-point functions, despite an increase in number of potential terms. Now there are
3 correlation functions of the form $ \vev{{\cal T}_{SK} \ \avA_{i_1} \, \difA_{i_2} \, \difA_{i_3}}$. There are however 6 correlators of the type $\vev{{\cal T}_{SK} \ \ghA_{i_1} \, \ghbA_{i_2} \, \difA_{i_3}}$. Finally, with background field insertions we have 6 of the form $\vev{{\cal T}_{SK} \ \ghA_{i_1} \, \ghbA_0 \,\avA_{i_2} \difA_{i_3}}$, another 6 of the form $\vev{{\cal T}_{SK} \ \ghA_0 \, \ghbA_{i_1} \,\avA_{i_2} \difA_{i_3}}$, and similarly 3 of the form
$\vev{{\cal T}_{SK} \ \ghA_{i_1} \, \ghA_{i_2} \,\ghbA_0\, \ghbA_{i_3}}$ and 3 as $\vev{{\cal T}_{SK} \ \ghbA_{i_1} \, \ghbA_{i_2} \,\ghA_0\, \ghA_{i_3}}$. Finally, there are 6 correlators with two background ghosts, namely 3 of the form $\vev{{\cal T}_{SK} \ \difA_0\,\avA_{i_1}\, \avA_{i_2} \, \difA_{i_3}}$ and 3 of type $\vev{{\cal T}_{SK} \ \difA_0 \,\ghA_{i_1}\, \ghbA_{i_2} \, \avA_{i_3}}$. This yields a total of 27 three-point functions of level 2.

We find 54 relations among them, not all of which are independent; these can be solved to determine the ghost correlations as before. We find that there exists again a unique fundamental solution along with a 2-parameter family of ambiguities $\{\al{k}{3,2}\}_{k=1,2}$. Doing this exercise brute force obviously becomes increasingly tedious. However, there is still a very compact way of communicating the full solution, including the two ambiguities, by writing the solution superspace expression as we did before. It should now involve only one time ordering, since two out of three operator insertions will be difference operators, which can never be future-most. Indeed, we find the following expression by following the principles outlined at the end of the previous subsection:
\begin{equation}
\begin{split}
 {}^3\Phi_2 &= \sum_{i_1,j_1,j_2=1}^3 \Big\{ \thetab_{0j_1}\theta_{0j_1}\thetab_{0j_2}\theta_{0j_2} + \al{1}{3,2} \left(  \thetab_{0j_1}\theta_{0j_1} \thetab_{0j_2}\theta_{0i_1} + \thetab_{0j_1}\theta_{0j_1} \thetab_{0i_1}\theta_{0j_2} \right)\\
 &\qquad \qquad\quad + \al{2}{3,2} \left(  \thetab_{0j_2}\theta_{0j_2} \thetab_{0j_1}\theta_{0i_1} + \thetab_{0j_2}\theta_{0j_2} \thetab_{0i_1}\theta_{0j_1} \right) \Big\} \, \stepFn{i_1 j_1 j_2} \, \vev{ {\cal T}_{SK}\ \avA_{i_1} \, \difA_{j_1} \, \difA_{j_2} } \,.
 \end{split}
 \label{eq:Phi32}
\end{equation}
Again, this compact expression encodes the full solution of ghost correlators with ambiguities. For sake of completeness we present all of them in Appendix \ref{app:SuperCorr}, and list here simply the solution where ambiguities are chosen to vanish:
\begin{equation}
\begin{split}
 &\vev{\mathcal{T}_{SK}\ \difA_{j_1} \, \ghbA_{j_2} \, \ghA_{j_3}}  = \vev{\mathcal{T}_{SK}\ \ghbA_0\, \ghA_{j_1} \, \ghA_{j_2} \, \ghbA_{j_3}} = \vev{\mathcal{T}_{SK}\ \ghA_0\, \ghbA_{j_1} \, \ghbA_{j_2} \, \ghA_{j_3}} = 0 \,,
 \\
 &\vev{\mathcal{T}_{SK}\ \ghA_0\,\difA_{j_1} \, \ghbA_{j_2} \, \avA_i  } =  \vev{\mathcal{T}_{SK}\ \difA_{j_1} \, \ghA_{j_2} \, \avA_i \, \ghbA_0 } =
 \vev{\mathcal{T}_{SK}\ \difA_0 \, \ghA_{j_1} \, \ghbA_{j_2} \, \avA_i }
 = \vev{ {\cal T}_{SK} \ \difA_{j_1} \, \difA_{j_2} \, \avA_i } \,,
 \\
 &\vev{{\cal T}_{SK}\ \difA_0 \, \difA_j \, \avA_{i_1} \, \avA_{i_2} }   =  \vev{ {\cal T}_{SK} \ \difA_j \, \difA_{i_1} \, \avA_{i_2} } +  \vev{ {\cal T}_{SK} \ \difA_j \, \avA_{i_1} \, \difA_{i_2} } \,.
\end{split}
\end{equation}

\item $^3{\bf L}_3$: This is the highest level for three-point functions and the only consistent solution for the correlators at this level is:
\begin{equation}
\begin{split}
0&= \vev{{\cal T}_{SK} \ \difA_{j_1} \, \difA_{j_2} \, \difA_{j_3}}=  \vev{{\cal T}_{SK} \ \ghA_{j_1} \, \ghbA_0\, \difA_{j_2} \, \difA_{j_3}}
 = \vev{{\cal T}_{SK} \ \ghbA_{j_1} \, \ghA_0\, \difA_{j_2} \, \difA_{j_3}} \\
  &= \vev{{\cal T}_{SK} \ \difA_0\, \avA_{j_1} \difA_{j_2} \, \difA_{j_3}} =  \vev{{\cal T}_{SK} \ \difA_0\, \ghA_{j_1} \ghbA_{j_2} \, \difA_{j_3}} \,.
\end{split}
\end{equation}
In terms of the superfield solution ansatz, this can be stated as $^3\Phi_3 = 0$.
\end{itemize}

\subsection{n-point functions}
\label{sec:nptSuper}


The complexity grows rapidly owing to increased set of permutations. Before ascertaining the number of ambiguities in the solution $^n\Phi_{n_d}$, let us write the general expression for the basic solution whose normalization is fixed:
\begin{equation}\label{eq:BasicSoln}
\begin{split}
{}^n\Phi_{n_d}  =   \sum_{\substack{i_1,\ldots,i_{n_a}=1\\ j_1,\ldots,j_{n_d}=1}}^n
\left( \prod_{k=1}^{n_d} \thetab_{0j_k}\theta_{0j_k} \right)
  \left( \stepFn{i_1 \ldots i_{n_a} j_1  \ldots j_{n_d}} + \text{perms.} \right)
  \vev{ {\cal T}_{SK} \ \avA_{i_1} \cdots \avA_{i_{n_a}} \, \difA_{j_1} \cdots \difA_{j_{n_d}} }\,,
 \end{split}
\end{equation}
where $n_a \equiv n-n_d$. The indicated permutations refer to all permutations of labels which are such that
\begin{itemize}
\item no $j$-index is ever future-most and
\item  we do not consider permutations of $i$-type (or $j$-type) indices among each other, i.e., we preserve the order of $i$-type indices and the order of $j$-type indices.
\end{itemize}
This constitutes the minimal solution to all the constraints on $n$-point functions at level $n_d$. Its normalization is fixed by matching with \eqref{eq:GeneralSKScorr}. In principle this solution suffices for our purposes. According to the principles we have identified so far, all further ambiguities can be chosen freely, and hence we can always make the minimal choice of setting them to zero.

For sake of full generality, however,  let us now turn to a counting of ambiguity terms that one is allowed to add to this basic solution. We denote the number of ambiguities in $^n\Phi_{n_d}$ as
$$ A[n,n_d] \equiv \text{number of ambiguities in fixing ghost correlators of type } {}^n{\bf L}_{n_d}\,. $$
Two trivial observations are that in general we have
 \begin{equation}\label{eq:GeneralThThb}
  {}^n\Phi_{0} = \vev{ {\cal T}_{SK} \ \avA_1 \cdots \avA_n } \qquad \text{and} \qquad
 {}^n\Phi_n = 0 \,,
 \end{equation}
and therefore $A[n,0] = A[n,n] = 0$.

 For values $0< n_d <n$, the counting of solutions is less trivial. Let us start by counting the number of different allowed time orderings occurring in $^n\Phi_{n_d}$. We split this into two stages. In the first instance note that there are ${n-1 \choose n_d}$ choices for inserting the difference operators, as no difference operator can be inserted in the future-most position. This exhausts all potential temporal permutations;  we conclude that there are ${n-1 \choose n_d}$ time orderings in $^n\Phi_{n_d}$ (up to permutations of labels).

We now turn to the counting of ambiguities for any given time ordering. Note that each ambiguity is characterized by a CPT invariant combination
\begin{equation}
\thetab_{0x_1}\theta_{0y_1} \cdots \thetab_{0x_{n_d}}\theta_{0y_{n_d}} + \thetab_{0y_1}\theta_{0x_1} \cdots \thetab_{0y_{n_d}}\theta_{0x_{n_d}}
\end{equation}
 for some allowed choice of $x_i,y_i \in \{ i_1,\ldots,i_{n_a},j_1,\ldots,j_{n_d}\}$. We obviously need $x_i \neq x_j$ and $y_i \neq y_j$ for any $i\neq j$, since the term would be zero otherwise. For simplicity, let us for the moment ignore the fact that difference operator insertions cannot be future-most. Most other constraints are then already manifestly upheld by the way we formulated the counting problem. The only thing remaining, which we need to be careful about, is the fact that we should not choose the entire set of $\{x_1,\ldots,x_{n_d}\}$ to be the same as the set of $\{y_1,\ldots,y_{n_d}\}$. This unique special choice would correspond to the basic solution \eqref{eq:BasicSoln} which we should not count as an ambiguity. Other than that the choices of $x_i$ and $y_i$ are unconstrained, and hence there are
\begin{equation}\label{eq:Fnnd}
F(n,n_d) \equiv \frac{1}{2} {n \choose n_d} \left[ {n \choose n_d} - 1 \right] =  {{n\choose n_d} \choose 2}
\end{equation}
choices for the two unequal lists of indices $\{x_1,\ldots,x_{n_d}\}$ and $\{y_1,\ldots,y_{n_d}\}$.\footnote{ The right hand side of Eq.\ \eqref{eq:Fnnd} denotes a double Binomial coefficient. It should be read as: ``There are ${n \choose n_d}$ distinct lists of $n_d$ indices. Out of the set of all such lists, we choose two: one for $\{x_1,\ldots,x_{n_d}\}$ and one for $\{y_1,\ldots,y_{n_d}\}$".} The factor $\frac{1}{2}$ in the above expression comes from the fact that we double count everything due to the CPT symmetrization in \eqref{eq:GeneralThThb}, i.e., for every choice of indices there is an inequivalent choice which actually leads to the same superspace expression after CPT symmetrization.

In the above counting, we disregarded the fact that difference operators can never be inserted at the future-most time. A difference operator $\difA_{j}$ will be present if the superspace expression \eqref{eq:GeneralThThb} contains the pair $\thetab_{0j}\theta_{0j}$. Making sure that such a term never carries the $i$-type index of the futuremost average operator (call it $i_f$) is tantamount to requiring that this particular index $i_f$ cannot appear in both lists $\{x_1,\ldots,x_{n_d}\}$ and $\{y_1,\ldots,y_{n_d}\}$ at the same time. In the counting of the previous paragraph, we did count such configurations, so for a correct result we should subtract them now. Fortunately, these configurations are easy to count: they can be described as those terms where the first pair $\thetab_{0x_1} \theta_{0y_1} = \thetab_{0i_f} \theta_{0i_f}$ and all the remaining indices $\{x_2,\ldots,x_{n_d},y_2,\ldots,y_{n_d}\}$ are chosen freely as described above. The number of such terms can then be counted the same way and is given by $F(n,n_d-1)$.

We conclude that the number of ambiguities in the general solution $^n\Phi_{n_d}$ is given by the difference $F(n,n_d) - F(n,n_d-1)$ multiplied by the number of distinct time orderings as counted above:
\begin{equation}
A[n,n_d] = {n-1 \choose n_d} \big(F(n,n_d) - F(n,n_d-1)\big) = {n-1 \choose n_d} \left[{{n\choose n_d} \choose 2}  - {{n\choose n_d-1} \choose 2} \right]\,.
\end{equation}
Up to six-point functions, the numbers of ambiguities $A[n,n_d]$  are explicilty enumerated in Table ~\ref{tab:Acount}.
\begin{table}[h]
\centering
\begin{tabular}{|| c || c | c | c | c | c | c | c||}
\hline\hline
  \shadeB{$n_d$} & \shadeB 0 & \shadeB 1 &\shadeB 2 &\shadeB 3 &\shadeB 4 &\shadeB 5 & \shadeB 6 \\
  \hline\hline
$A[1,n_d]$ & 0 & 0 & & & & & \\
$A[2,n_d]$ & 0 & 1 & 0 & & & & \\
$A[3,n_d]$ & 0 & 6 & 2 & 0 & & & \\
$A[4,n_d]$ & 0 & 18 & 36 & 3 & 0 & & \\
$A[5,n_d]$ & 0 & 40 & 234 & 120 & 4 & 0 &\\
$A[6,n_d]$ & 0 & 75 & 950 & 1450 & 300 & 5 & 0\\
\hline\hline
\end{tabular}
\caption{The number of ambiguities in the consistent solution of ghost correlators in 1- to 6-point functions. The level $n_d$ refers to the number of difference operator insertions (any ghost-anti-ghost insertion counts as an average-difference pair for this purpose).}
\label{tab:Acount}
\end{table}

This then concludes the explicit enumeration of the ambiguities at level $n_d$ for $n$ point functions. The total ambiguity is rather large, so it quickly becomes formidable to parameterize them. We have explicitly checked the results up to four point functions for all level and some of the non-trivial levels of the five point function, confirming the data presented in Table~\ref{tab:Acount}.

It is actually an interesting exercise to try to match these ambiguities to superspace analysis of effective actions directly; we will comment upon these briefly in \S\ref{sec:veltman}, but will leave a detailed analysis for future investigation.

\section{Timefolds and out-of-time-order observables}
\label{sec:oto}

We now turn to an interesting extension of the Schwinger-Keldysh formalism to more complicated contours. For the most part we will focus on the generalities of this construction and draw out an observation of larger BRST algebras that are present in these constructions. The physical motivation for the extension as we first explain arises from out-of-time order observables that have been considered in recent discussions of scrambling and chaos. Our preliminary analysis is aimed at highlighting the general principles which promises to open up new insights into non-equilibrium quantum dynamics.

\subsection{$k$-OTO contours: physical motivation}
\label{sec:koto}

A very interesting class of observables that has recently come to prominence are the so-called out-of-time-order correlation functions in a QFT. The interest in this set of observables arose from holographic considerations of trying to engineer situations where the entanglement between two (sub)systems can be disrupted in a suitable manner.  The canonical example of such observables are correlation functions of the form $\vev{\comm{\OpH{A}(t)}{\OpH{B}(0)}^2}$, i.e, squares of commutators of local Heisenberg picture operators. Expanding out the commutator we find contributions of the form
$\vev{\OpH{A}(t) \OpH{B}(0) \OpH{A}(t) \OpH{B}(0)}$ where the operator insertions are clearly out-of-time order (OTO). Such observables were first explored in \cite{Larkin:1969aa} and interest in them was revived by recent work of Kitaev \cite{Kitaev:2015aa}.

To explain their significance, in the first instance \cite{Shenker:2013pqa} studied the behaviour of such observables as a diagnostic of quantum chaos. The initial explorations were in the context of black hole physics and holography, aimed at understanding how black holes scramble information. These analyses then inspired an interesting bound on the Lyapunov exponent, which is defined by examining the intermediate time behaviour of the commutator. More recently, \cite{Maldacena:2015waa} argued for a fundamental bound on quantum processing leading to an upper bound on the Lyapunov exponent $\lambda_L \leq 2\pi\,\beta$, when evaluated in an initial thermal state (inverse temperature $\beta$). This bound is saturated by holographic field theories dual to Einstein gravity and also by an interesting quantum mechanical model of free fermions which has come to be known as the Sachdev-Ye-Kitaev  (SYK) model \cite{Sachdev:1992fk,Kitaev:2015aa} -- we refer the reader to \cite{Maldacena:2016hyu} for a comprehensive discussion of the model and its solution.

Recent explorations of this model \cite{Jensen:2016pah,Maldacena:2016upp} have unveiled an interesting structure of the   infra-red physics in this SYK model. While the theory has a large ground state degeneracy, the low energy dynamics is dominated by a single mode which remains gapless. Its dynamics is governed by an emergent $SL(2)$ symmetry, which in turn, leads to the bound on the out-of-time-order correlation function. It has been pointed out in the aforementioned references that the effective action for this low energy mode is qualitatively similar to that of hydrodynamic effective actions (to be specific, \cite{Jensen:2016pah} has argued that the effective action can be brought by a field redefinition to the Landau-Ginzburg Class L effective action of \cite{Haehl:2015pja}).

We would like to argue that this may not be entirely coincidental, and that the class of effective actions of interest in hydrodynamics might have some close connection with those of the SYK model with an emergent (nearly broken) conformal symmetry in the infra-red. Let us argue for this from our Schwinger-Keldysh contour, or generalizations thereof. Realize first of all that we cannot compute expectation values of commutators-squared using the standard Schwinger-Keldysh construction; by definition the latter is meant to give us a handle on computing time-ordered multi-point functions, which all can be written as sums of nested commutators and anti-commutators with suitable time-ordering. At no point in our explicit analysis did we encounter products of commutators. To attain the latter, we need a generalization of the Schwinger-Keldysh contour to include two more horizontal legs, i.e., we need a path integral contour in complex time domain which has two forward and two backward legs. Such contours have come to be known as timefolds; see \cite{Heemskerk:2012mn} where the authors motivate the construction as a generalization of Schwinger-Keldysh.\footnote{ As this paper was in preparation we became aware of the recent work \cite{Aleiner:2016eni} who explore generalizations of the Schwinger-Keldysh contour to compute OTO correlators.}  Similar contours also appear in the computation of R\'enyi entropies using the replica trick in Lorentzian spacetime \cite{Dong:2016hjy}.

\subsection{OTO contours}
\label{sec:otoc}

With the above motivation, let us define the out-of-time-order generating functions.
Based on the discussion above, to compute such observables, we need to introduce
timefolds into the path integral, necessarily involving a sequence of forward/backward evolution. Every forward segment will involve a unitary operator $U$ with some source deformations, while each backward segment will be represented by conjugate $U^\dag$ also with sources. Given such a timefolded contour we can compute out-of-time-order correlation functions.

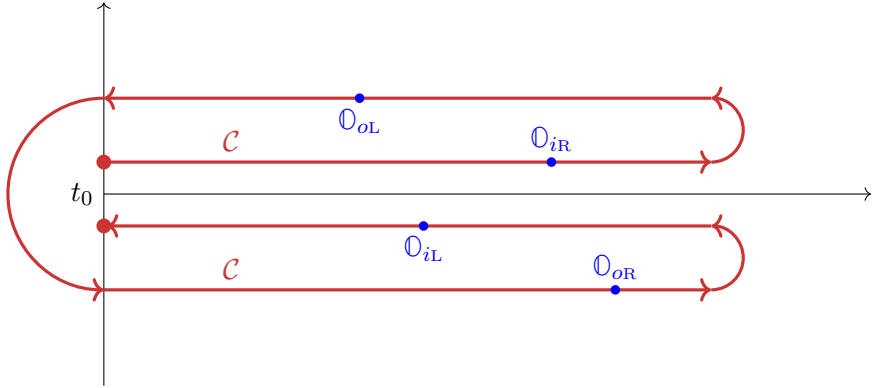
\begin{figure}[h!]
\begin{center}
\begin{tikzpicture}[scale=0.85]
\draw[thin,color=black,->] (-5,0) -- (7,0);
\draw[thin, color=black,->] (-5,-3) -- (-5,0) node [left] {$t_0$} --  (-5,3);
\draw[very thick,color=rust,->] (4.5,1.5) -- (-5,1.5);
\draw[very thick,color=rust,->] (-5,0.5) -- (-3,0.5) node [above] {$\color{rust}\mathcal{C}$} -- (4.5,0.5);
\draw[very thick,color=rust,->] (4.5,-0.5) -- (-4.95,-0.5);
\draw[very thick,color=rust,->] (-5,-1.5) -- (-3,-1.5) node [above] {$\color{rust}\mathcal{C}$}  -- (4.5,-1.5);
\draw[very thick,color=rust,->] (4.5,0.5) arc (-90:90:0.5);
\draw[very thick,color=rust,->] (4.5,-1.5) arc (-90:90:0.5);
\draw[very thick,color=rust,<-] (-5,-1.5) arc (-90:-270:1.5);
\draw[thick,color=rust,fill=rust] (-5,0.5) circle (0.6ex);
\draw[thick,color=rust,fill=rust] (-5,-0.5) circle (0.6ex);
 \draw[thick,color=blue,fill=blue] (-1,1.5) circle (0.35ex) node [below] {$\Op{O}_{o\skL}$} ;
\draw[thick,color=blue,fill=blue] (2,0.5) circle (0.35ex) node [above] {$\Op{O}_{i\skR}$};
\draw[thick,color=blue,fill=blue] (0,-0.5) circle (0.35ex) node [below] {$\Op{O}_{i\skL}$};
 \draw[thick,color=blue,fill=blue] (3,-1.5) circle (0.35ex)  node [above] {$\Op{O}_{o\skR}$};
\end{tikzpicture}
\caption{The 2-OTO contour computing the correlation functions with operators inserted out of the conventional Schwinger-Keldysh time-ordering, cf. Eq.~\eqref{eq:2oto}. As usual the initial state is prepared at time $t_0$ and the latest operator insertion happens at time $t$. The indicated operator insertions correspond to the correlation function $\vev{\Op{O}_{i\skR}(t_1)  \Op{O}_{o\skL}(t_2) \Op{O}_{o\skR}(t_3) \Op{O}_{o\skL}(t_4)}$.}
\label{fig:oto}
\end{center}
\end{figure}

In particular, for  computing squares of commutators etc., the generating function we need involves two timefolds. The generating function may be written down explicitly with the time-evolution implementing unitaries as
\begin{equation}
\mathscr{Z}_{2-oto}[\mathcal{J}_{o\skR},  \mathcal{J}_{i\skR}; \mathcal{J}_{o\skL},  \mathcal{J}_{i\skL}]=
\Tr{ (U[\mathcal{J}_{o\skL}])^\dag  U[\mathcal{J}_{i\skR}] \;\rhoi \; (U[\mathcal{J}_{i\skL}])^\dag
U[\mathcal{J}_{o\skR}]} \,.
\label{eq:2oto}
\end{equation}
Pictorially we can represent this contour with a series of switchbacks as in Fig.~\ref{fig:oto}. We will refer to this class of generating functions as the $2$-OTO generating function, since it allows us to compute a two-fold out-of-time-order correlation function.

One can similarly extend this to defining $k$-OTO generating functions whose generating function can be expressed in the form
\begin{equation}
\mathscr{Z}_{k-oto}[\mathcal{J}_{\alpha\skR},  \mathcal{J}_{\alpha\skL}]=
\Tr{   \cdots
U[\mathcal{J}_{3\skR}]
(U[\mathcal{J}_{2\skL}])^\dag  U[\mathcal{J}_{1\skR}] \;\rhoi \; (U[\mathcal{J}_{1\skL}])^\dag
U[\mathcal{J}_{2\skR}] (U[\mathcal{J}_{3\skL}])^\dag \cdots} \,.
\label{eq:koto}
\end{equation}
with $\alpha \in \{1,2,\cdots, k\}$. Pictorially these would be represented as in  Fig.~\ref{fig:koto}.
Our nomenclature is meant to suggest that the $0$-OTO is computed by the  standard Feynman path integral, while $1$-OTO corresponds to the Schwinger-Keldysh contour, etc..
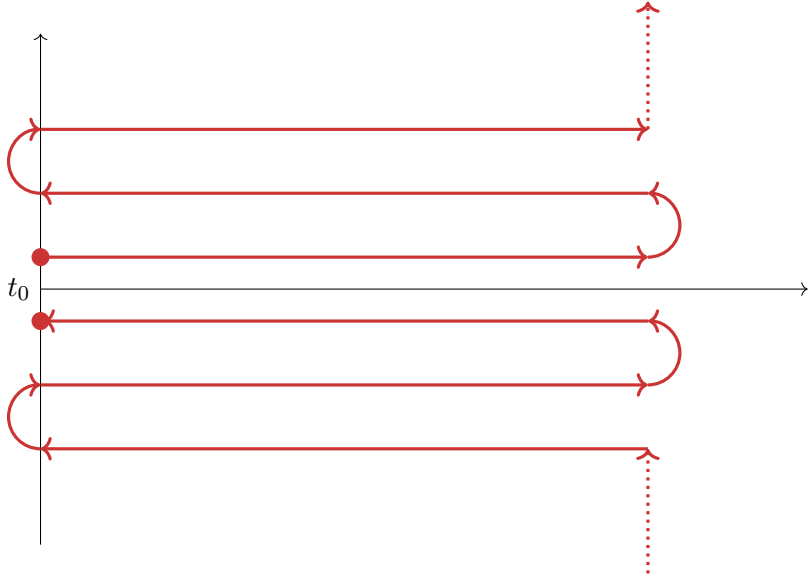
\begin{figure}[t!]
\begin{center}
\begin{tikzpicture}[scale=0.85]
\draw[thin,color=black,->] (-5,0) -- (7,0);
\draw[thin, color=black,->] (-5,-4) -- (-5,0) node [left] {$t_0$} --  (-5,4);
\draw[very thick,color=rust,->,dotted] (4.5,2.5) -- (4.5,4.5);
\draw[very thick,color=rust,<-,dotted] (4.5,-2.5) -- (4.5,-4.5);
\draw[very thick,color=rust,{}->] (-5,2.5) -- (4.5,2.5);
\draw[very thick,color=rust,->] (4.5,1.5) -- (-5,1.5);
\draw[very thick,color=rust,->] (-5,0.5)  -- (4.5,0.5);
\draw[very thick,color=rust,->] (4.5,-0.5) -- (-4.95,-0.5);
\draw[very thick,color=rust,->] (-5,-1.5)   -- (4.5,-1.5);
\draw[very thick,color=rust,->] (4.5,-2.5)   -- (-5,-2.5);
\draw[thick,color=rust,fill=rust] (-5,0.5) circle (0.75ex);
\draw[thick,color=rust,fill=rust] (-5,-0.5) circle (0.75ex);
\draw[very thick,color=rust,->] (-5,1.5) arc (-90:-270:0.5);
\draw[very thick,color=rust,->] (4.5,0.5) arc (-90:90:0.5);
\draw[very thick,color=rust,->] (4.5,-1.5) arc (-90:90:0.5);
\draw[very thick,color=rust,->] (-5,-2.5) arc (-90:-270:0.5);
\end{tikzpicture}
\caption{The k-OTO contour computing the out-of-time-ordered correlation functions encoded in the generating functional \eqref{eq:koto}.}
\label{fig:koto}
\end{center}
\end{figure}

Let us record some general features of these contours at the outset. Clearly the computation of an $n$-point function from the $k$-OTO generating function results in $(2k)^n$ potential possibilities. By switching off or aligning some of the sources, we can collapse some of the timefolds. One should anticipate that the $k$-OTO generating functional would collapse in an alignment limit of sources to a $j$-OTO with $j\leq k$. Furthermore, in an $n$-point function,  $\vev{\OpH{O}(t_1) \OpH{O}(t_2) \cdots \OpH{O}(t_n)}$, all possible time orderings are attained by restricting attention to $\left[\frac{n+1}{2}\right]$-OTO contours.\footnote{ $[x]$ denotes the integer part of $x$.} This observation simply follows by viewing the operators as Heisenberg operators and drawing the time-evolution contours that compute the Wightman correlator. Conversely, $k$-OTO contours with $k>[\frac{n+1}{2}]$ give no new information for $n$-point functions: $1$-point functions are computed by the Feynman path integral, $2$-point functions are computed by Schwinger-Keldysh contours, while it is only $3$- and $4$-point functions, where we encounter some data which requires $2$-OTO contours. $3$-OTO contours become relevant for $5$- and $6$- point functions, and so on.

\subsection{Unitarity and localization in $2$-OTO contours}
\label{sec:u2oto}

To get a feeling for  the $k$-OTO correlation functions, let us examine the case $k=2$. We have four independent legs of the contour in Fig.~\ref{fig:oto}. On each leg we can insert operators, so we naively expect there to be $4^n$ $n$-point correlation functions for any given set of $n$ operators. Clearly, based on  our experience with the Schwinger-Keldysh contour,  we do not expect these correlators to be independent; the question is what is the useful way to encode the relations imposed by unitarity.

To address this issue, let us introduce a set of 2-OTO difference operators:
\begin{equation}
\begin{split}
\Op{A}_{d}^{ii} &= \Op{A}_{i\skR} - \Op{A}_{i\skL} \,, \\
\Op{A}_{d}^{oo} &= \Op{A}_{o\skR} - \Op{A}_{o\skL} \,, \\
\Op{A}_{d}^{oi} &= \Op{A}_{o\skR} - \Op{A}_{i\skL} \,, \\
\Op{A}_{d}^{io} &= \Op{A}_{i\skR} - \Op{A}_{o\skL} \,.
\end{split}
\label{eq:2diff}
\end{equation}
The four operators are not independent but rather satisfy a linear relation
\begin{equation}
\Op{A}_{d}^{ii} + \Op{A}_{d}^{oo} = \Op{A}_{d}^{oi} + \Op{A}_{d}^{io}\,.
\label{eq:2otorel}
\end{equation}
In terms of these we can write down the relations we get upon alignment of the various background sources directly from the generating functional \eqref{eq:2oto}. We have two different topological limits, where the $2$-OTO contour collapses to the initial state:
\begin{itemize}
\item  Aligning $\mathcal{J}_{o\skL}  =\mathcal{J}_{o\skR}$ and  $\mathcal{J}_{i\skL}  =\mathcal{J}_{i\skR}$ collapses the $2$-OTO generating functional to the initial state:
\begin{equation}
\mathscr{Z}_{2-oto} [\mathcal{J}_{o\skL}  =\mathcal{J}_{o\skR},\mathcal{J}_{i\skL}  =\mathcal{J}_{i\skR}]= \Tr{\rhoi}\,.
\end{equation}
We will encode this relation as asserting that $n$-point correlation functions of inner ($i$) difference operators  and outer ($o$) difference operators vanish as a consequence of unitarity. For example, at the level of two-point functions we have:
\begin{equation}
\vev{\Ttwo\Op{A}_d^{oo} \, \Op{B}_d^{oo}}=\vev{\Ttwo \Op{A}_d^{ii} \, \Op{B}_d^{ii}}=\vev{\Ttwo\Op{A}_d^{oo} \, \Op{B}_d^{ii}}=\vev{\Ttwo\Op{A}_d^{ii} \, \Op{B}_d^{oo}} = 0 \,,
\label{}
\end{equation}
where $\Ttwo$ denotes time ordering along the 2-OTO contour of Figure \ref{fig:oto}.
This straightforwardly generalizes to a set of $n^2$ vanishing $n$-point functions of the same type.
 \item Setting $\mathcal{J}_{o\skL}  =\mathcal{J}_{i\skR}$ and  $\mathcal{J}_{i\skL}  =\mathcal{J}_{o\skR}$ leads to a second localization to the initial state. Now the operators with vanishing correlation functions are those involving a different set of difference operators, viz., $\Op{A}_d^{io}$ and $\Op{A}_d^{oi}$. At the two-point function level, this leads to:
\begin{equation}
\vev{\Ttwo\Op{A}_d^{io} \, \Op{B}_d^{io}}=\vev{\Ttwo\Op{A}_d^{oi} \, \Op{B}_d^{oi}}=\vev{\Ttwo\Op{A}_d^{io} \, \Op{B}_d^{oi}}=\vev{\Ttwo\Op{A}_d^{oi} \, \Op{B}_d^{io}} = 0 \,.
\end{equation}
Again, there are $n^2$ relations of this type at the level of $n$-point functions.
\end{itemize}
We will refer to these limits as {\it full localization}. Similarly, there are limits of {\it  1-OTO localization}, where the $2$-OTO contour does not quite collapse to the initial state, but instead to a $1$-OTO Schwinger-Keldysh contour. The partial localizations can be described as follows:
\begin{itemize}
\item There are three different limits where we align sources such that the $2$-OTO generating functional collapses to a standard Schwinger-Keldysh theory. They take the form:
\begin{equation}
\begin{split}
\mathscr{Z}_{2-oto}[\mathcal{J}_{o\skR}=\mathcal{J}_{o\skL},  \mathcal{J}_{i\skR},  \mathcal{J}_{i\skL}] &=
\mathscr{Z}_{1-oto}[ \mathcal{J}_{i\skR},  \mathcal{J}_{i\skL}]\,,\\
\mathscr{Z}_{2-oto}[\mathcal{J}_{o\skR}= \mathcal{J}_{i\skL},  \mathcal{J}_{o\skL},\mathcal{J}_{i\skR}]&=
\mathscr{Z}_{1-oto}[\mathcal{J}_{i\skR},\mathcal{J}_{o\skL}]\,,\\
\mathscr{Z}_{2-oto}[\mathcal{J}_{o\skL}=\mathcal{J}_{i\skR},  \mathcal{J}_{i\skL},  \mathcal{J}_{o\skR}]&=
\mathscr{Z}_{1-oto}[   \mathcal{J}_{o\skR},  \mathcal{J}_{i\skL}]\,.
\end{split}
\label{eq:2otoSK1}
\end{equation}
\end{itemize}
It is clear that any partial localization can be extended into one of the two full localization limits by aligning the remaining pair of sources in each case.

Finally, there are limits which are similar to partial localization, but have further interesting features. We refer to these as {\it timefolded 1-OTO localization}:
\begin{itemize}
\item Let us consider correlation functions such as
$\vev{\Ttwo\Op{A}_{i\skR}  \, \Op{B}_{o\skR} }$, $\vev{\Ttwo\Op{A}_{i\skL}  \, \Op{B}_{o\skL} }$, or
$\vev{\Ttwo\Op{A}_{o\skR} \, \Op{B}_{o\skL} }$. These correlators can be computed without loss of generality in the limits ${\cal J}_{i\skL} = {\cal J}_{o\skL}=0$, ${\cal J}_{i\skR} = {\cal J}_{o\skR}=0$, and ${\cal J}_{i\skR} = {\cal J}_{i\skL}=0$, respectively. A quick inspection of the generating function should convince the reader that while a pair of sources are set to zero, the non-zero sources are separated from the initial density matrix. For instance:
\begin{equation}
\begin{split}
\vev{\Ttwo\Op{A}_{i\skL}  \, \Op{B}_{o\skL} }  &= \frac{\delta^2}{\delta \mathcal{J}_{o\skL, \Op{B}} \;
\delta \mathcal{J}_{i\skL,\Op{A}}}
\Tr{ (U[\mathcal{J}_{o\skL,\Op{B}}])^\dag\, U[0] \;\rhoi \;(U[\mathcal{J}_{i\skL,\Op{A}}])^\dag \,U[0] }
\\
&\equiv \frac{\delta^2}{\delta \mathcal{J}_{o\skL, \Op{B}} \;
\delta \mathcal{J}_{i\skL,\Op{A}}}
 \Tr{ \widetilde{U}[\mathcal{J}_{o\skL,\Op{B}}] \; \rhoi\; (U[\mathcal{J}_{i\skL,\Op{A}}])^\dag}
\end{split}
\label{}
\end{equation}
where we have tried to reduce the computation to the form of a Schwinger-Keldysh, or $1$-OTO observable, by redefining the source.  Our definition for the composite evolution operator $\tilde{U}$ can be read off to be
\begin{equation}
\widetilde{U}[\mathcal{J}]   \equiv U[0] \, U[\mathcal{J}]^\dag \,  U[0]  \equiv U[{\mathcal{J}}^{(\hookleftarrow)}]\,,
\label{eq:preU}
\end{equation}
where $U[0]$ is the standard Hamiltonian evolution in the absence of any external sources.  This definition clearly satisfies $\widetilde{U}[0] = U[0]$. One can equivalently imagine that the action of $\widetilde{U}[\mathcal{J}]$ can be understood as a new source deformation by formally writing the expression as $U[{\mathcal{J}}^{(\hookleftarrow)}]$. Of course, obtaining a precise expression for ${\mathcal{J}}^{(\hookleftarrow)}$ involves explicitly evaluating the sequence of evolutions in the given time-order.\footnote{ It is instructive to draw the contour for the above evolution to see how insertion of the operator coupling to ${\mathcal{J}}^{(\hookleftarrow)}$ can be viewed as a forward temporal evolution interrupted by a time-reversed evolution to insert an operator at an earlier time. This is the reason for interpreting these correlation functions as precursor or timefolded correlators. }
We can summarize these three limits as follows:
\begin{equation}
\begin{split}
\mathscr{Z}_{2-oto}[{\cal J}_{i\skL} = {\cal J}_{o\skL}=0,  \mathcal{J}_{i\skR},  \mathcal{J}_{o\skR}] &=
\mathscr{Z}_{1-oto}[{\mathcal{J}}_{i\skR}, {\mathcal{J}}^{(\hookleftarrow)}_{o\skR}]\,, \\
\mathscr{Z}_{2-oto}[{\cal J}_{i\skR}={\cal J}_{o\skR}=0,  \mathcal{J}_{i\skL},  \mathcal{J}_{o\skL}] &=
\mathscr{Z}_{1-oto}[{\mathcal{J}}^{(\hookleftarrow)}_{o\skL},{\mathcal{J}}_{i\skL}]\,, \\
\mathscr{Z}_{2-oto}[{\cal J}_{i\skR} = {\cal J}_{i\skL}=0,  \mathcal{J}_{o\skR},  \mathcal{J}_{o\skL}] &=
\mathscr{Z}_{1-oto}[{\mathcal{J}}^{(\hookleftarrow)}_{o\skL}, {\mathcal{J}}_{o\skR}^{(\hookleftarrow)}]\,.
\end{split}
\label{eq:2otoSK2}
\end{equation}
\end{itemize}
These correlators have been described as precursor/postcursor correlators in literature, owing to the fact that the action of the sources is to insert operators at an earlier/later time to achieve a particular effect at a particular moment of interest.\footnote{ Note that the generating functionals also suggest that we could consider a further generalization of
\eqref{eq:preU} where instead of switching off the source we consider aligned non-vanishing sources; this would lead $\widetilde{U}[\mathcal{J}_{\mathcal{J}_{c}}] =
U[\mathcal{J}_{c}] \,U[\mathcal{J}]^\dag \, U[\mathcal{J}_{c}]$ where $\mathcal{J}_{c}$ is the common source which is aligned.}

The observations made above make it clear that there are non-trivial relations to be obtained between the various observables. Per se, this is not surprising, since the definition of the $2$-OTO contour naturally comes with further redundancy. The question we face is one of understanding how to encode all of these redundancies succinctly, taking on board the lessons learnt from the Schwinger-Keldysh construction.
As the reader might anticipate we are aiming to argue for a generalization of the Schwinger-Keldysh BRST symmetries which would implement the various localizations of the OTO contours.

Let us try to address the various relations systematically and obtain the set of independent correlation functions. Rather than working out the story for general $n$-point functions, we focus on the case of $2$-point functions for simplicity to build up intuition. While we know that the set of arbitrary $2$-point functions can be computed from $1$-OTO or Schwinger-Keldysh path integrals, it is instructive to go through the exercise in some detail to see how things work.  This will suffice to make the point we wish to convey in the present discussion.

\paragraph{$2$-OTO two-point functions:}
Naively there are $4^2=16$ two-point functions obtained by inserting operators $\Op{O}_{o\skR}$, $\Op{O}_{i\skR}$, $\Op{O}_{o\skL}$, and $\Op{O}_{i\skL}$ respectively. By examining the structure of $\mathscr{Z}_{2-oto}$ we can make the  sequence of inferences described below. We are essentially going to group the set of 16 correlators into sets, depending on whether they involve a Feynman contour, a Schwinger-Keldysh contour, and more specifically in the latter case whether the operators are usual Heisenberg operators or precursors. In each case we will also relate the answer to a well-understood combination of Schwinger-Keldysh two-point functions that we have discussed hitherto.\footnote{ While we choose a convenient basis of Schwinger-Keldysh correlators below, the reader can verify that all 16 two-point functions can be expressed in terms of retarded, advanced and Keldysh correlators as defined in \S\ref{sec:skreview}, if the operators are the same. For instance, note that for a real bosonic operator $\OpH{A}$, we have
\begin{equation}
\begin{split}
 \langle {\cal T} \,\OpH{A}(x) \OpH{A}(x')\rangle &= \frac{i}{2} \left( G_K(x,x') + G_{ret}(x,x') + G_{adv} (x,x') \right) \,,\\
  \langle \overline{\cal T} \,\OpH{A}(x) \OpH{A}(x')\rangle &= \frac{i}{2} \left( G_K(x,x') - G_{ret}(x,x') - G_{adv} (x,x') \right) \,,\\
 \langle \left[\OpH{A}(x) ,\OpH{A}(x')\right]\rangle &= i \left(  G_{ret}(x,x') -G_{adv} (x,x') \right) \,.
\end{split}
\end{equation}
}

\begin{enumerate}
\item There are two time-ordered correlation functions which are obtained when $\mathscr{Z}_{2-oto}$ collapses onto the usual Feynman contour for a single copy. This is attained in the alignment limit
$\mathcal{J}_{o\skL} = \mathcal{J}_{i\skL} =0$ with either $\mathcal{J}_{o\skR} =0 $ or $\mathcal{J}_{i\skR} =0$. The standard $0$-OTO correlator obtained this way is simply the usual time-ordered correlation function. One can thus write the identity:
\begin{equation}
\vev{\mathcal{T} \;\OpH{A} \,\OpH{B}} =   \vev{\Ttwo\Op{A}_{i\skR}\Op{B}_{i\skR}} = \vev{\Ttwo\Op{A}_{o\skR} \Op{B}_{o\skR}}  \,,
\label{eq:2oto0}
\end{equation}
where ${\cal T}$ denotes usual (Feynman) time ordering.
\item Similarly, the alignment limits where $\mathcal{J}_{o\skR} = \mathcal{J}_{i\skR} =0$ with either $\mathcal{J}_{o\skL} =0 $ or $\mathcal{J}_{i\skL} =0$ gives correlation functions of left operators, which are simply anti-time ordered:
\begin{equation}
\begin{split}
\vev{\overline{\mathcal{T}} \;\OpH{A} \,\OpH{B}}
&=
	 \vev{\Ttwo\Op{A}_{i\skL} \Op{B}_{i\skL}} = \vev{\Ttwo\Op{A}_{o\skL} \Op{B}_{o\skL}}  \,,
\end{split}
\end{equation}
where $\overline{\cal T}$ denotes usual anti-time ordering.
\item There are three different limits where the $2$-OTO generating functional collapses to a usual Schwinger-Keldysh contour. These three limits correspond to $\mathcal{J}_{o\skL} = \mathcal{J}_{o\skR}$, $\mathcal{J}_{o\skL} = \mathcal{J}_{i\skR}$, and $\mathcal{J}_{i\skL}=\mathcal{J}_{o\skR}$.
This reduction of the $2$-OTO generating function to the $1$-OTO generating function fixes the following three combinations of two-point functions:
\begin{equation}
\begin{split}
\vev{\comm{\OpH{A}}{\OpH{B}}}
& =
	\vev{\Ttwo\Op{A}_{i\skL}\Op{B}_{i\skR} } - \vev{\Ttwo\Op{A}_{i\skR} \Op{B}_{i\skL}}
	= \vev{\Ttwo\Op{A}_{i\skL}\Op{B}_{o\skR}} - \vev{\Ttwo\Op{A}_{o\skR} \Op{B}_{i\skL}} \\
&	= \vev{\Ttwo\Op{A}_{o\skL}\Op{B}_{i\skR}} - \vev{\Ttwo\Op{A}_{i\skR}\Op{B}_{o\skL}} 	 \,.
\end{split}
\label{eq:2oto1}
\end{equation}
These are the three separate instances of the Schwinger-Keldysh identity \eqref{eq:2ptadda}, corresponding to the three $1$-OTO reductions.
\item In a similar fashion one can establish that six other correlation functions get related to the linear combination of time-ordered and anti-time-ordered Schwinger-Keldysh correlators:
\begin{equation}
\begin{split}
\vev{\overline{\mathcal{T}} \;\OpH{A} \,\OpH{B}}  +
\vev{\mathcal{T} \;\OpH{A} \,\OpH{B}}
&
	= \vev{\Ttwo\Op{A}_{i\skR}\Op{B}_{i\skL}} + \vev{\Ttwo \Op{A}_{i\skL}\Op{B}_{i\skR}}
	= \vev{\Ttwo\Op{A}_{o\skR} \Op{B}_{i\skL}} + \vev{\Ttwo \Op{A}_{i\skL}\Op{B}_{o\skR}} \\
	&
	= \vev{\Ttwo\Op{A}_{i\skR}\Op{B}_{o\skL}} + \vev{\Ttwo\Op{A}_{o\skL}\Op{B}_{i\skR}}
	= \vev{\Ttwo\Op{A}_{o\skR} \Op{B}_{o\skL}} + \vev{\Ttwo\Op{A}_{o\skL}\Op{B}_{o\skR}} 	\\
&
	= \vev{\Ttwo\Op{A}_{o\skR} \Op{B}_{i\skR}} + \vev{\Ttwo\Op{A}_{o\skL} \Op{B}_{i\skL}}
	= \vev{\Ttwo\Op{A}_{i\skR}\Op{B}_{o\skR}} + \vev{\Ttwo\Op{A}_{i\skL}\Op{B}_{o\skL}} \,.
\end{split}
\label{}
\end{equation}
\item The above relations account for 13 of the 16 $2$-OTO correlators which we have seen reduce  to well known combination usual Schwinger-Keldysh observables. The new element in $\mathscr{Z}_{2-oto}$ are the  precursor or timefolded correlation functions. We find it convenient to pick the following linear combinations as the representatives of these observables:
\begin{equation}
\begin{split}
\vev{\comm{\OpH{A}}{\OpH{B}}}
&= \vev{\Ttwo\Op{A}_{i\skL}\Op{B}_{o\skL}} - \vev{\Ttwo\Op{A}_{i\skR} \Op{B}_{o\skR}} \\
&= \vev{\Ttwo\Op{A}_{o\skR} \Op{B}_{i\skR}} -\vev{\Ttwo\Op{A}_{o\skL}\Op{B}_{i\skL}} \\
&= \vev{\Ttwo\Op{A}_{o\skR}\Op{B}_{o\skL}} -\vev{\Ttwo\Op{A}_{o\skL}\Op{B}_{o\skR}} \\
\end{split}
\label{eq:2o2p}
\end{equation}
In writing the expression for the timefolded correlators in terms of the standard single-copy correlation functions, we have also invoked the largest time equation which prevents difference operators from being futuremost. While the combinations in the r.h.s. \eqref{eq:2o2p} reduce to \eqref{eq:2oto0} we present them independently to emphasize the precursor nature of  the operators.
\end{enumerate}
We note for completeness that in order to see genuine 2-OTO correlators (i.e., ones which cannot be expressed as combinations of Schwinger-Keldysh correlation functions), we need to consider at least 3-point functions. As an example, consider
$\vev{\gradAnti{\gradAnti{\OpH{A}(t_A)}{\OpH{B}(t_B)}}{\OpH{C}(t_C)}}$, which for arbitrary time orderings is manifestly not a combination of Schwinger-Keldysh three-point functions. We leave a full exploration of higher point $2$-OTO correlators to the future, and content ourselves with the following general remarks.

\paragraph{BRST symmetries of $2$-OTO contours:}

The analysis for $2$-point functions should make it amply clear that the general structure of the $2$-OTO generating function will involve new field redefinition BRST symmetries which enforce the various relations obtained above. If we view the construction of $\mathscr{Z}_{2-oto}$ as a nested sequence of Schwinger-Keldysh constructions, it becomes clear that we can upgrade the operator algebra to an extended operator superalgebra to make manifest the structure. One can  speculate that\footnote{ We would like to thank Michael Geracie and David Ramirez for discussions on the precise number of BRST symmetries that are necessary for the full set of localizations  of the OTO contours.} 
a quartet of BRST charges $\{\QSK^o, \QSKb^o, \QSK^i, \QSKb^i \}$ arising from the set of outer and inner contours respectively.  We will refer to this structure as the $\mathcal{N}_\smallT =4$ superstructure following \cite{Dijkgraaf:1996tz}.

An efficient way to encode the operation of these BRST charges is work in superspace. Having four BRST charges implies that we need two sets of Grassmann variables, so the superspace will be generated by $\{\theta_i, \thb_i, \theta_o ,\thb_o\}$. The charges act effectively as translation generators along these Grassmann-odd directions.
Thus we should embed every operator $\OpH{O}$ into an extended super-operator
$\SF{\SF{\Op{O}}}$ which will be an $\mathcal{N}_\smallT =4$ superfield with $16$ components. The components comprise of linear combinations of $\{\Op{O}_{o\skR}, \Op{O}_{o\skL}, \Op{O}_{i\skR}, \Op{O}_{i\skL}\}$ together with the $\mathcal{N}_\smallT =4$ ghost partners.

While we can naively write down a suitable expression (for instance by working with a sequence of nested $\mathcal{N}_\smallT=2$) superfields, it is helpful to a-priori understand the action of discrete symmetries such as CPT in building a useful representation. We defer  this construction for a separate discussion. Nevertheless, it should be clear that the structure will capture all the relations we obtain in the various alignment limits. A related question is to understand the interplay of these generators with the KMS generators in the context of a thermal initial density operator. The Grassmann-even KMS translation operator $\Qbeta$ will pick up $15$ superpartners to form a $16$-component KMS-superoperator. The full algebra including the KMS charges may provide a realization of $\mathcal{N}_\smallT =4$ thermal equivariant cohomology algebra, which we believe should shed light on the chaos bound obtained in \cite{Maldacena:2015waa}.

\paragraph{BRST symmetries of $k$-OTO contours:} Finally, while we have extensively discussed the case of $2$-OTO, the general structure should now be clear. The $k$-OTO generating function will have $2k$ field redefinition BRST symmetries, which we speculate are naturally realized by an action on an $\mathcal{N}_\smallT =2k$ superspace.  These charges constrain the alignment limits or localizations of external sources, relating $k$-OTO correlation functions to  $j$-OTO and timefolded $j$-OTO correlators with $j<k$.
In the context of thermal field theory, the story is certainly a lot richer with potentially a large $\mathcal{N}_\smallT =2k$ thermal equivariant cohomology algebra constraining various correlators.
We hope to report on these structures and implications for various physical questions involving scrambling and chaos in the near future.

\section{Applications to physical problems}
\label{sec:applications}

We have given a detailed explanation of the BRST supersymmetry inherent in the Schwinger-Keldysh formalism. It is useful to take stock of these symmetries as applied to various physical problems. In the current section we give a brief overview of various areas of physics where this symmetries can help elucidate the basic physical ideas.

\subsection{Stochastic dynamics}
\label{sec:stochastic}

It is well known that stochastic dynamics admits a BRST-supersymmetric formulation. The prototype example of interest in this context is Langevin dynamics of a Brownian particle, which thanks to the Martin, Siggia, Rose  (MSR) \cite{Martin:1973zz} construction is known to be obtained from an effective action with the BRST symmetries. A general connection between stochastic time evolution and supersymmetry was spelt out by Parisi and Sourlas in  \cite{Parisi:1982ud}.  These developments are well reviewed in classic references on topological field theories \cite{Birmingham:1991ty} and critical phenomena \cite{ZinnJustin:2002ru}.

The basic starting point in all this discussion is to note that the natural way to write an action whose equations of motion are the stochastic differential equation is to enlarge the variables to a quartet of fields. Given a random variable $\phi(t)$ which satisfies the stochastic equation of motion say $E(\phi) = 0$, the trick involves using a Lagrange multiplier field $\tilde{\phi}$ to exponentiate it. The idea is to essentially treat $E(\phi) =0$  as a gauge fixing condition, which is imposed as a delta-functional constraint on the configuration space parameterized by the $\phi$ field. One then uses the standard Faddeev-Popov trick to exponentiate the delta function and introduce Grassmann partners $\phi_\psi$ and $\phi_{\bar \psi}$ to account for the measure. Note that the latter involves the Jacobi functional $\frac{\delta E(\phi)}{\delta \phi}$, which forms the kinetic term for the Grassmann partners.

It should be clear from the enumeration of the fields involved that the natural language for writing the action of the stochastic differential equation requires a quartet of fields. A quick check of the symmetries, especially the ghost number makes it clear that we are talking about an $\mathcal{N}_\smallT= 2$ multiplet as discussed in the text. The occurrence of this topological symmetry algebra has been noted in the literature \cite{Gozzi:1989vv,Cattaruzza:2013iua}. These references were interested in providing a path integral formulation of classical mechanics as a counterpart to the  operator formalism developed by Koopman and von Neumann  \cite{Koopman:1931aa,Neumann:1932aa}.

However, this standard discussion eschews the full power of the $\mathcal{N}_T=2$ symmetries. They manifestly work with the so-called Cartan charges as symmetry generators. As we elaborate on in \cite{Haehl:2016uah}, we can decompose those into the fundamental Weil charges and interior contraction operations making the algebraic structure manifest. It is easy to check that the general structure we have described in the main part of the text continues to apply in this case. We have given a preliminary treatment for the case of Langevin dynamics in the Appendix of \cite{Haehl:2015foa}; the reader can find further details in \cite{Haehl:2016uah}. In particular, we show there how to embed the standard construction into a gauged worldline theory for a point (super)-particle. Whilst simple it provides insight into the workings of more complicated theories such as hydrodynamics which is our next point of contact.

\subsection{Hydrodynamics}
\label{sec:hydro}

As we described in the introduction \S\ref{sec:intro}, the primary motivation for our foray into a discussion of equivariant cohomology was to better understand the construction of hydrodynamic effective field theories, and in particular to argue for constraints on the class of influence functionals that are admissible in the low energy theory. Much of this discussion treats the hydrodynamic system of interest as a field theory in an approximately Gibbs state. The intensive parameters of this density matrix are allowed to vary on macroscopic length scales, and one is interested in the low energy dynamics of the collective degrees of freedom.

It is well known that the phenomenological axioms of hydrodynamics require that the low energy theory capture only the dynamics of conserved currents subject to the requirement of non-negative definite entropy production. The dynamical part of the theory is intuitive: in systems that relax back to thermal equilibrium, the short lived high-energy modes relax exponentially fast. The only perturbations that survive to late time and long distances are the conserved currents, which persist owing to their local conservation. In \cite{Haehl:2014zda,Haehl:2015pja} we have given a complete classification of all solution to the phenomenological axioms. Our eightfold classification in particular constrains extensively the form of influence functionals that could arise in any effective field theory of dissipative hydrodynamics.

Inspired by the prospect of confronting potential effective actions with the eightfold classification, we have embarked on the construction of topological sigma models for dissipative hydrodynamics \cite{Haehl:2015uoc}. The philosophy we followed was espoused in \cite{Haehl:2015foa} and involves certain crucial ingredients gleaned from our classification scheme.
In a nutshell, we had learnt that there is a $\UT$ KMS gauge symmetry which couples to the entropy current and that an effective action for the non-dissipative, adiabatic sector, of hydrodynamics involves a doubling of degrees of freedom. The key observation of our analysis was to note a structural similarity between the adiabatic effective action and the construction of MSR \cite{Martin:1973zz} (which was also used by \cite{Kovtun:2014hpa}).

The main new ingredient is to upgrade the MSR construction to a gauged version to account for the $\UT$ symmetry. The manner in which we initially inferred this symmetry was driven by phenomenology, but in the end it was rather satisfying to see its microscopic origins in the Schwinger-Keldysh formalism and the KMS condition.  In \cite{Haehl:2015pja} we argued that the adiabatic sector of hydrodynamics, must satisfy an important first law type constraint, dubbed adiabaticity equation, owing to the lack of entropy production. We showed that this constraint naturally follows as a consequence of the $\UT$ gauge symmetry. In other words the entropy current is the Noether current for the $\UT$ symmetry. A further striking feature of the way the symmetry acted on the physical hydrodynamic fields was that it naturally explained the construction in \cite{Jensen:2013rga} who in fact were the first to exploit it to construct equilibrium partition functions for field theories with mixed flavour/gravitational anomalies.

Thus far our construction in \cite{Haehl:2015uoc} only confronts three of the eight classes of admissible transport. We show there that the formalism of the $\mathcal{N}_\smallT=2$ topological symmetry enables us to write down an effective action that captures precisely the dissipative part of transport in addition to the adiabatic classes corresponding to Landau-Ginzburg functionals (which comprises of both equilibrium partition function data and non-equilibrium hydrodynamic terms).  The construction of an effective action that encompasses all eight classes is still underway. The structural similarity of the topological sigma model construction with the adiabatic action lends us confidence that the eightfold classes (and no more) will be attained by exploiting the topological symmetries. Nevertheless the true test of our formalism is to make this explicit.

Crucially, the construction captures not just the dissipative terms that help revert the system back to equilibrium, but it also simultaneously  fixes the fluctuation terms that are missing in classical formulations of hydrodynamics. This point has also been appreciated in the parallel development of \cite{Crossley:2015evo} who have encountered similar structures as us, but do not explicitly make use of the $\mathcal{N}_\smallT=2$ algebra. Instead they work with a single BRST charge associated with the field redefinition symmetry. A second charge is argued to arise in the low energy, near thermal dynamical sector; together with a ${\mathbb Z}_2$ action of the KMS condition, they construct an effective action for dissipative fluids. As noted in \S\ref{sec:intro} despite various differences, the symmetries they invoke to constrain the hydrodynamic effective action bear close resemblance (in the high temperature limit) with our proposal. A detailed connection between the formalisms will be made elsewhere.

A natural corollary of the topological symmetries is a useful derivation of the second law of thermodynamics. We see from our analysis that a BRST symmetry Ward identity for spontaneous CPT breaking, leads to a fundamental identity: the Jarzynski relation \cite{Jarzynski:1997aa,Jarzynski:1997ab}  and its reformulation as the Crooks relation \cite{Crooks:1999fk}. These identities  form the generalized fluctuation-dissipation relations that are valid in out-of-equilibrium settings. We have argued in \cite{Haehl:2015uoc} that they follow from our topological sigma model. This is indeed what one should expect, since we have a gauge symmetry for the entropy current. The detailed balance statement following the Jarzynski relation leads to a statement of the second law of thermodynamics for the dissipative sector.

\subsection{Entanglement and the modular superalgebra}
\label{sec:entanglement}

Let us now return to general density matrices; we have seen in that the Schwinger-Keldysh BRST charges, as in any topological field theory constrain the BRST exact operators  to be trivial. In the Schwinger-Keldysh construction, we have demonstrated that the difference operators belong to this class. The supercharges ensure that the Ward identities \eqref{eq:diff0} are satisfied explicitly, independent of the microscopic dynamics of the quantum system under consideration.

The most useful physical lesson one can extract from these considerations is that the Schwinger-Keldysh ghost operators and the associated BRST symmetries ensure microscopic unitarity of the theory. One can intuit this, given the close similarity of our discussion of field redefinition BRST charges with the well-known Faddeev-Popov ghosts of gauge theories. As long as the BRST symmetries are non-anomalous in the quantum theory, we would have reason to believe that the ghosts are doing the correct job in ensuring that the physical Hilbert space only comprises of positive norm states.

There is one further consequence of the topological structure of the Schwinger-Keldysh path integral, which bears detailed scrutiny. Let us focus on the topological sector of the theory: as noted when the sources for the left and right degrees of freedom are aligned in the Schwinger-Keldysh path integral ${\cal J}_\skL = {\cal J}_\skR$, we have from \eqref{eq:ZSKdef}, a collapse of the generating function onto the theory  of initial conditions, viz.,
\begin{equation}
 \mathscr{Z}_{SK}[{\cal J}_\skR={\cal J}_\skL] =\Tr{ \rhoi} \,.
\label{eq:rhoiloc}
\end{equation}

One usually tends to normalize the initial density matrix. This is equivalent to subtracting out the topological contributions, since these are all that survives when we turn on identical sources as in  \eqref{eq:rhoiloc}. Often however it is convenient to not normalize $\rhoi$, but rather let the trace $\Tr{\rhoi}$ capture the entanglement inherent in the initial state. A simple example to keep in mind is the thermal density matrix $\rhoT$, which, if left unnormalized, computes for us the thermal partition function $\mathscr{Z}_{_T}(\beta)$ of the theory.

Let us try to extract the entanglement built into $\rhoi$, by computing the von Neumann entropy of this initial state, which is given by
\begin{equation}
S_\text{initial} = -\Tr{\rhoi \log \rhoi}
\label{eq:Sidef}
\end{equation}
It is useful however, to think in terms of the \emph{modular Hamiltonian} \cite{Haag:2012aa}
\begin{equation}
\Ki = - \log \rhoi \,.
\label{eq:modularH}
\end{equation}
which is a state-dependent non-linear operator in the theory.  It depends on the chosen state as it actively involves taking the logarithm of the density matrix operator. With its introduction we can write the Schwinger-Keldysh path integral with equal left-right sources as a modular free energy:
\begin{equation}
 \mathscr{Z}_{SK}[{\cal J}_\skR={\cal J}_\skL] =\Tr{ e^{-\Ki}} \,.
\label{eq:}
\end{equation}
The analogy with thermal partition functions is apparent and suggests that the density matrix trace is obtained by modular evolution, i.e., evolution  by the operator $\Ki$ for a unit distance in an imaginary direction.

To obtain the entanglement inherent in the density matrix, let us introduce the notion of R\'enyi entropies, which are obtained from the moments of the density matrix:
\begin{equation}
S^{(q)}(\rhoi) = \frac{1}{1-q}\, \log \, \Tr{(\rhoi)^q} = \frac{1}{1-q}\, \log \Tr{e^{-q\,\Ki}} \,,
\label{eq:Renyi}
\end{equation}
where the standard prefactor of $(1-q)^{-1}$ is introduced for convenience in defining the entanglement entropy. The latter is defined as the
von Neumann entropy  of the initial state, viz.,
\begin{equation}
S(\rhoi) = -\Tr{\rhoi \, \log \rhoi}  = \Tr{\Ki \, e^{-\Ki} } \,.
\label{eq:eei}
\end{equation}
Knowledge of the quantities $S^{(q)}$, which resemble usual thermal partition functions for the `Hamiltonian' $\Ki$ at inverse temperature $q$, up to an inconsequential rescaling by $(1-q)^{-1}$, determines $S(\rhoi)$ via
\begin{equation}
S(\rhoi) = \lim_{q\to 1}\, S^{(q)}(\rhoi) \,.
\label{eq:limq1}
\end{equation}
Another useful quantity which appears to be more natural in holography or gravity is the \emph{modular entropy},  \cite{Dong:2016fnf}, which is defined as the derivative the R\'enyi entropy with respect to its index
\begin{equation}
\tilde{S}^{(q)}(\rhoi) = -q^2\, \frac{\partial}{\partial q} \left(\frac{1}{q} \; \log \Tr{e^{-q\, \Ki}} \right) .
\label{}
\end{equation}
Comparing with thermodynamic formulae, one can be convinced that $\tilde{S}^{(q)}(\rhoi) $ is really the entropy associated with the modular evolution at inverse temperature $q$.

The modular evolution by the operator $\Ki$  for an imaginary time can be viewed as determining moments of the density matrix. This is strictly in analogy with the thermal density matrices where evolution in Euclidean time direction gives us the partition function as a function of the temperature (which is determined by the size of the thermal circle). The main difference from the usual story for thermal density matrices is that the modular Hamiltonian is intrinsically tied to the state of the system $\rhoi$ and is generically a non-local (state-dependent) operator. There are however certain situations where it is well behaved operator acting on the entire Hilbert space. For instance for a QFT on Rindler space, obtained by decomposing ${\mathbb R}^{d-1,1} =
\text{Rindler} \times {\mathbb R}^{d-3,1}$, the reduced density matrix in a single Rindler wedge has as its modular Hamiltonian the boost generator \cite{Bisognano:1975ih,Bisognano:1976za}. Likewise the reduced density matrix obtained by confining the vacuum state of a CFT into  a spherical domain leads to a local modular Hamiltonian \cite{Casini:2011kv}.

For both modular evolution of generic density matrices and the thermal evolution of Gibbs states, the traces are computed by evolution in an imaginary time direction.  One can thus in principle imagine setting up an appropriate Euclidean path integral which computes $\mathscr{Z}_T(\beta)$ or $\mathscr{Z}_{SK}(\rhoi)$ for us. In both cases it is clear that only the information in the initial state is necessary to determine the corresponding partition function. Thus $\mathscr{Z}_{SK}[{\cal J}_\skR = {\cal J}_\skL]$ readily admits a Euclidean path integral representation.

Furthermore, it is tempting to argue that we should involve a set of \emph{modular charges} which implement the Euclidean periodicity of modular evolution. It is not clear to us at present whether these charges, which would be intrinsically non-local be of practical use in understanding the evolution of general density matrix. It however, does appear to be the case, that these modular charges together with the Schwinger-Keldysh charges would generate an $\mathcal{N}_\smallT=2$ extended equivariant cohomology algebra, along the lines of our thermal density matrix discussion.  It appears naively that the discrete version of the modular gauge symmetry, as opposed to the analog of the continuum $\UT$ discussed in the thermal case, should play some role. We think it would be extremely intriguing to see how this construction plays out, and whether it has any lessons to impart for the entaglement/geometry correspondence in holography.

Should this structure pertain, one can use it to argue that the Schwinger-Keldysh path integral localizes onto an appropriate modular partition function at the initial time when the state is prepared. One may infer from this requirement that the topological sector of the Schwinger-Keldysh theory encodes the relevant entanglement structure for the mixed states. We use the phrase `localize' in a precise technical sense used in topological field theories.

 This interpretation allows one to have a clear strategy of defining effective field theories of Schwinger-Keldysh path integrals in the Wilsonian sense. Firstly, one constructs what might be termed as the {\em topological backbone}, viz., a theory that captures appropriate correlations/entanglement of the initial mixed state under study. Once this is achieved,  we may deform away from the ${\cal J}_\skR={\cal J}_\skL$ limit and study the class of mixed states which are continuous deformations of our chosen initial state. These mixed states have similar entanglement structure  to $\rhoi$  and evolve into each other under unitary evolution. The Schwinger-Keldysh topological field theory helps setting up this entanglement pattern. Its efficacy in understanding effective field theories is apparent, for the rigidity of the topological invariance ensures that it is robust against renormalization.

 From a modern quantum information theoretic perspective, one may even go so far as to note that the Schwinger-Keldysh construction employs a topological symmetry to initially set-up a sector of the quantum system which is robust against non-topological perturbations. This topological skeleton may be viewed as the abstract, continuum analog of a tensor network for the initial state $\rhoi$. Splitting the discussion of open quantum systems into a topological theory of initial conditions and dynamics has many advantages. As dynamics is implemented by unitary transformations, it leaves the entanglement structure intact, thus allowing one to decouple the physical consequences of evolution from the manner in which the open system is unitarized in the first instance.

\subsection{Gravitational systems}
\label{sec:grav}

One of our primary interests behind investigating this reformulation of Schwinger-Keldysh formalism, is to understand the lessons it holds for gravitational dynamics, especially in spacetimes with horizons. There are many gravitational questions that naturally fall under the class of questions we have been discussing, having to do with the physics of black hole formation and evaporation, cosmological evolution, etc., which involve systems driven out of equilibrium.

Hydrodynamics is an ideal point of contact, since the fluid/gravity correspondence \cite{Bhattacharyya:2008jc,Hubeny:2011hd} naturally maps the dynamics of (large) black holes in asymptotically AdS spacetimes to fluid dynamics. We have previously outlined in \cite{Haehl:2015foa} elements of our philosophy relating to how the a gravity dual of the dissipative hydrodynamic effective actions could potentially help us understand long-standing issues in black hole physics.

For instance, one natural conjecture is to relate the interior of the black hole with a gauged topological sigma model for dissipative hydrodynamics (dubbed $\UT$ open string theory in the aforementioned reference). Relatedly, the importance of the ghost degrees of freedom, whose condensation can be seen as the origin of dissipation at low energies, we feel holds a crucial clue for understanding how the interior of the black hole should be reconstructed from the dual field theory. As of this writing we do not offer a precise implementation of these ideas. Nevertheless, recent developments in the study of gravity duals of fast scrambling systems (see \S\ref{sec:oto}), appear extremely promising in that they appear to embody some of the basic principles we have explored in this paper, directly in the gravitational realm.

\subsection{Cutting rules and amplitudes}
\label{sec:veltman}

Our discussion so far has focused on the Schwinger-Keldysh formalism aimed at computing time-ordered correlation functions. It is interesting to note that the basic ideas we have described actually apply rather directly to the study of causality constraints on S-matrices. To explain the context, we recall a basic notion of causality formulated by Bogoliubov in the 1950s \cite{Bogoliubov:1959aa} (see also \cite{Peres:1970tr}).

Consider a scattering process where we imaging having a spacetime dependent interaction term that we can
switch on and off at will. Given that the quantum fields undergo a continuous evolution from the initial state defined on past timelike infinity ($I^-$) of Minkowski space to a state on  future timelike infinity ($I^+$), we demand a causality restriction, that the interaction only influences the scattering matrix to its future. The condition is usually written as a variational statement of the S-matrix, with respect to such spacetime dependent coupling functions. Pragmatically, we can intuit this statement in terms of energy flow in the process: if $p$ is to the causal future of $q$, then the Feynman diagrams which involve particles propagating between $q$ and $p$ are such that only those with positive energy flow from $q\to p$ are allowed. One can write this symbolically as:
\begin{equation}
\frac{\delta^2 S}{\delta g(q) \delta g(p)}  \, S^\dagger+ \frac{\delta S}{\delta g(q)}\;
\frac{\delta S^\dagger}{\delta g(p)} =0 \,, \qquad  p^0> q^0\,,
\label{eq:bogc}
\end{equation}
where $g(x)$ is our spacetime dependent coupling.

We are used to an equivalent version of this statement in terms of operators (which are required to commute at spacelike separation), but now we want a direct encoding in terms of the S-matrix, rather than Green's functions.
It was realized by Veltman that a pragmatic way to interpret this statement, is to split up Feynman diagrams that describe a process with particles propagating from $q$ to $p$ in terms of their energy characteristics. As beautifully reviewed in \cite{tHooft:1973pz}, we can take a given Feynman diagram and decide to monitor energy flow through the legs. The simplest way to do this is to examine a propagator that connects two vertices and pick the right Green's function according to the causal ordering.

In order to implement this, the idea of circled and uncircled vertices is introduced into Feynman diagrammatics \cite{tHooft:1973pz}. Every vertex of a Feynman diagram is doubled, with it being either circled, or uncircled.
Each propagator  gets replaced by one of the four choices:
\begin{itemize}
\item propagators linking two uncircled vertices are usual Feynman propagators
\item propagators linking two circled vertices are anti-Feynman propagators
\item propagator from circled to uncircled is given by the positive frequency part of the Feynman propagator
(retarded Green's function)
\item propagator from uncircled to circled is given by the negative frequency part of the Feynman propagator
(advanced Green's function)
\end{itemize}
The reader will immediately recognize these are really the standard Schwinger-Keldysh rules of time ordering!

Based on these rules, a series of Cutkosky cutting rules are derived which encode the causality constraint explicitly.
They can be abstractly summarized as the statement that the sum of a single diagram over all possible circlings vanishes. This is exactly equivalent to the one constraint of the vanishing of the difference operator correlation functions. Convolving this expression against source functions leads to a single Feynman diagram version of Bogoliubov's condition \eqref{eq:bogc}.  In particular, one also finds that the largest time equation is upheld. This asserts the circled vertex cannot be future most.

At a conceptual level the similarity between the implementing causality and unitarity via cutting rules and using the Schwinger-Keldysh formalism to keep track of time-ordering is not surprising. However, given our BRST symmetries it is interesting to ask whether we can employ the superspace techniques developed here to give a complementary picture of causality. Such a development might not only be interesting to do for QFTs but also for string theory, see e.g., \cite{Pius:2016jsl}. Work in this direction is in progress and we hope to report on these applications in the not too distant future.


\newpage
\acknowledgments


It is a pleasure to thank Veronika Hubeny,  Juan Maldacena, and Shiraz Minwalla for useful discussions.

 FH and RL would like to thank the QMAP at UC Davis for hospitality during the course of this project.  FH \& MR would like to thank the Yukawa Institute for Theoretical Physics, Kyoto  and   Perimeter Institute for Theoretical Physics  (supported by the Government of Canada through the Department of Innovation, Science and Economic Development and by the Province of Ontario through the Ministry of Research and Innovation) for hospitality during the course of this project. MR would also like to thank Nordita and Arnold Sommerfeld Center, Munich for hospitality during the concluding states of this project. FH gratefully acknowledges support through a fellowship by the Simons Foundation. RL gratefully acknowledges support from International Centre for
Theoretical Sciences (ICTS), Tata institute of fundamental research, Bengaluru and Ramanujan fellowship from Govt. of India. RL would
also like to acknowledge his debt to all those who have generously supported and encouraged the pursuit of science in India.

\appendix

\section{Complete list of three-point functions}
\label{app:SuperCorr}

This appendix complements section \S\ref{sec:3ptSuper}, where we wrote down compact expressions which determine all three-point ghost correlators in terms of average-difference expectation values. Here, we expand out the solutions explicitly and write down the relations thus obtained.

\paragraph{Level 0:}
The lowest level $^3{\bf L}_0$ does not allow for any ghosts, so there are no new correlators to fix. Here, we only have the obvious
\begin{equation}
{}^3\Phi_{0} = \vev{ {\cal T}_{SK} \ \avA_1 \, \avA_2 \, \avA_3 }
\end{equation}
and its explicit form was given in \eqref{eq:3ptCorrA}.

\paragraph{Level 1:}
At level 1 $^3{\bf L}_1$, the ghost correlators are solved as follows (with a 6-parameter family of ambiguities):
\begin{equation}
\begin{split}
\vev{\mathcal{T}_{SK}\ \ghA_1 \, \ghbA_2 \, \avA_3}  &= \left[  \al{1}{3,1}\stepFn{2 1 3} + \al{2}{3,1} \stepFn{3 1 2} +\al{4}{3,1} \stepFn{2 3 1} +  \al{5}{3,1}\stepFn{3 2 1} \right] \vev{ \mathcal{T}_{SK} \ \difA_1 \, \avA_2\,\avA_3 }
 \\
 &\;\;+\left[ \al{1}{3,1} \stepFn{1 2 3} +  \al{2}{3,1}\stepFn{3 2 1} +\al{4}{3,1} \stepFn{1 3 2} +  \al{5}{3,1}\stepFn{3 1 2} \right]\vev{ {\cal T}_{SK} \ \avA_1 \, \difA_2\,\avA_3 }
 \\
 &\; \; +\left[ \al{3}{3,1} (\stepFn{1 3 2} + \stepFn{2 3 1}) +\al{6}{3,1} (\stepFn{1 2 3} + \stepFn{2 1 3}) \right] \vev{ \mathcal{T}_{SK} \ \avA_1 \, \avA_2\,\difA_3 }
\end{split}
\end{equation}
\begin{equation}
\begin{split}
&\vev{\mathcal{T}_{SK}\ \ghbA_0 \, \ghA_1 \, \avA_2\, \avA_3}  = \vev{\mathcal{T}_{SK}\ \ghbA_1 \, \ghA_0 \, \avA_2\, \avA_3}  = \\
&\quad \qquad\quad=\big[  \left(\al{1}{3,1}+\al{2}{3,1}-1\right) \left( \stepFn{2 1 3}  + \stepFn{3 1 2}\right) \\
 &\qquad\quad\qquad\qquad +\left(\al{4}{3,1}+\al{5}{3,1}-1\right) \left( \stepFn{2 3 1}  + \stepFn{3 2 1}\right) \big] \vev{ \mathcal{T}_{SK} \ \difA_1 \, \avA_2\,\avA_3 } \qquad\qquad
 \\
 &\qquad\qquad\quad+\big[ \left(\al{1}{3,1} + \al{3}{3,1}\right) \stepFn{1 2 3} +  \left(\al{2}{3,1} + \al{3}{3,1}\right) \stepFn{3 2 1}
 \\
 &\qquad\quad\qquad\qquad +\left(\al{4}{3,1} + \al{6}{3,1}\right) \stepFn{1 3 2} +  \left(\al{5}{3,1} + \al{6}{3,1}\right) \stepFn{3 1 2} \big]\vev{ {\cal T}_{SK} \ \avA_1 \, \difA_2\,\avA_3 }
 \\
 &\qquad \qquad\quad+\big[ \left(\al{1}{3,1} + \al{3}{3,1}\right) \stepFn{1 3 2} +  \left(\al{2}{3,1} + \al{3}{3,1}\right) \stepFn{2 3 1}
 \\
 &\qquad\quad \qquad\qquad+\left(\al{4}{3,1} + \al{6}{3,1}\right) \stepFn{1 2 3} +  \left(\al{5}{3,1} + \al{6}{3,1}\right) \stepFn{2 1 3} \big]\vev{ {\cal T}_{SK} \ \avA_1 \, \avA_2\,\difA_3 }
\end{split}
\end{equation}
\begin{equation}
\begin{split}
&\vev{\mathcal{T}_{SK}\ \difA_0 \, \avA_1 \, \avA_2\, \avA_3}  = \big[  \left(1-2\al{1}{3,1}-2\al{2}{3,1}-2\al{3}{3,1}\right) \left( \stepFn{2 1 3}  + \stepFn{3 1 2}\right) \\
 &\qquad\qquad\qquad\qquad\quad +\left(1-2\al{4}{3,1}-2\al{5}{3,1}-2\al{6}{3,1}\right) \left( \stepFn{2 3 1}  + \stepFn{3 2 1}\right) \big] \vev{ \mathcal{T}_{SK} \ \difA_1 \, \avA_2\,\avA_3 }
 \\
 &\qquad\qquad\qquad\quad\; +\big[  \left(1-2\al{1}{3,1}-2\al{2}{3,1}-2\al{3}{3,1}\right) \left( \stepFn{1 2 3}  + \stepFn{3 2 1}\right) \\
 &\qquad\qquad\qquad\qquad\quad +\left(1-2\al{4}{3,1}-2\al{5}{3,1}-2\al{6}{3,1}\right) \left( \stepFn{1 3 2}  + \stepFn{3 1 2}\right) \big] \vev{ \mathcal{T}_{SK} \ \avA_1 \, \difA_2\,\avA_3 }
  \\
 &\qquad\qquad\qquad\quad\; +\big[  \left(1-2\al{1}{3,1}-2\al{2}{3,1}-2\al{3}{3,1}\right) \left( \stepFn{1 3 2}  + \stepFn{2 3 1}\right) \\
 &\qquad\qquad\qquad\qquad\quad +\left(1-2\al{4}{3,1}-2\al{5}{3,1}-2\al{6}{3,1}\right) \left( \stepFn{1 2 3}  + \stepFn{2 1 3}\right) \big] \vev{ \mathcal{T}_{SK} \ \avA_1 \, \avA_2\,\difA_3 }
\end{split}
\end{equation}

\paragraph{Level 2:}

At level 2, there are the following three-point functions found by expanding the solution \eqref{eq:Phi32}:

\begin{equation}
\begin{split}
\vev{\mathcal{T}_{SK}\ \difA_1 \, \ghbA_2 \, \ghA_3}  &= \left( \al{1}{3,2}  \stepFn{3 1 2}  + \al{2}{3,2}  \stepFn{3 2 1} \right)\vev{ \mathcal{T}_{SK} \ \difA_1 \, \difA_2\,\avA_3 }    \\
&\quad + \left( \al{1}{3,2}  \stepFn{2 1 3}  + \al{2}{3,2}  \stepFn{2 3 1} \right)\vev{ \mathcal{T}_{SK} \ \difA_1 \, \avA_2\,\difA_3 }
\end{split}
\end{equation}
\begin{equation}
\begin{split}
\vev{\mathcal{T}_{SK}\ \ghbA_0 \, \ghA_1 \, \ghA_2 \, \ghbA_3}  = \vev{\mathcal{T}_{SK}\ \ghA_0 \, \ghbA_1 \, \ghbA_2 \, \ghA_3} &= \left( \al{1}{3,2} - \al{2}{3,2}   \right) \left(\stepFn{3 1 2} - \stepFn{3 2 1}\right)  \vev{ \mathcal{T}_{SK} \ \difA_1 \, \difA_2\,\avA_3 }    \\
&\quad +\left( \al{1}{3,2} \stepFn{2 1 3} + \al{2}{3,2} \stepFn{2 3 1} \right) \vev{ \mathcal{T}_{SK} \ \difA_1 \, \avA_2\,\difA_3 } \\
&\quad -\left( \al{1}{3,2} \stepFn{1 2 3} + \al{2}{3,2} \stepFn{1 3 2} \right) \vev{ \mathcal{T}_{SK} \ \avA_1 \, \difA_2\,\difA_3 }
\end{split}
\end{equation}
\begin{equation}
\begin{split}
\vev{\mathcal{T}_{SK}\  \difA_1 \, \ghA_2 \,  \avA_3 \, \ghbA_0} &= \vev{\mathcal{T}_{SK}\  \ghA_0 \, \difA_1 \, \ghbA_2 \,  \avA_3 }   =   \left( \al{1}{3,2} \stepFn{2 1 3} + \al{2}{3,2} \stepFn{2 3 1} \right) \vev{ \mathcal{T}_{SK} \ \difA_1 \, \avA_2\,\difA_3 } \\
&\quad\qquad\qquad + \left[ \left(1+\al{1}{3,2}\right) \,\stepFn{3 1 2}  + \left(1+\al{2}{3,2}\right) \,\stepFn{3 2 1} \right] \vev{ \mathcal{T}_{SK} \ \difA_1 \, \difA_2\,\avA_3 }
\end{split}
\end{equation}
\begin{equation}
\begin{split}
\vev{\mathcal{T}_{SK}\  \difA_0 \, \difA_1 \,  \avA_2 \, \avA_3} &  =  \left[ \left(1+2 \al{1}{3,2}\right) \stepFn{3 1 2} + \left( 1+2\al{2}{3,2} \right) \stepFn{3 2 1} \right] \vev{ \mathcal{T}_{SK} \ \difA_1 \, \difA_2\,\avA_3 } \\
&\quad +  \left[ \left(1+2 \al{1}{3,2}\right) \stepFn{2 1 3} + \left( 1+2\al{2}{3,2} \right) \stepFn{2 3 1} \right] \vev{ \mathcal{T}_{SK} \ \difA_1 \, \avA_2\,\difA_3 }
\end{split}
\end{equation}
\begin{equation}
\begin{split}
\vev{\mathcal{T}_{SK}\ \difA_0\, \ghA_1\,  \ghbA_2 \, \avA_3 } &  =  \left( 1+ \al{1}{3,2} + \al{2}{3,2}\right) \left(\stepFn{3 1 2}+\stepFn{3 2 1}\right)   \,\vev{ \mathcal{T}_{SK} \ \difA_1 \, \difA_2\,\avA_3 }  \\
&\quad +\left( \al{1}{3,2}-\al{2}{3,2}\right) \left(\stepFn{1 2 3}-\stepFn{1 3 2}\right)   \,\vev{ \mathcal{T}_{SK} \ \avA_1 \, \difA_2\,\difA_3 }\\
&\quad + \left( \al{1}{3,2}-\al{2}{3,2}\right) \left(\stepFn{2 1 3}-\stepFn{2 3 1}\right)   \,\vev{ \mathcal{T}_{SK} \ \difA_1 \, \avA_2\,\difA_3 }
\end{split}
\end{equation}

\paragraph{Level 3:}
The highest level, $^3{\bf L}_3$ contains a number of ghost correlators, but the only consistent solution is to set them all to zero:
\begin{equation}
\begin{split}
0&= \vev{{\cal T}_{SK} \ \difA_1 \, \difA_2 \, \difA_3}=  \vev{{\cal T}_{SK} \ \ghA_1 \, \ghbA_0\, \difA_2 \, \difA_3}
 = \vev{{\cal T}_{SK} \ \ghbA_1 \, \ghA_0\, \difA_2 \, \difA_3} \\
  &= \vev{{\cal T}_{SK} \ \difA_0\, \avA_1 \difA_2 \, \difA_3} =  \vev{{\cal T}_{SK} \ \difA_0\, \ghA_1 \ghbA_2 \, \difA_3} \,.
\label{}
\end{split}
\end{equation}

\newpage

 \providecommand{\href}[2]{#2}\begingroup\raggedright\endgroup

\end{document}